\documentclass[%
 reprint,
superscriptaddress,
 amsmath,amssymb,
 aps,
]{revtex4-2}

\usepackage{comment}
\usepackage{graphicx}
\usepackage{dcolumn}
\usepackage{bm}
\usepackage{hyperref}
\usepackage{xcolor}
\usepackage{soul}
\usepackage{ mathrsfs }

\begin{document}

\preprint{APS/123-QED}

\title{Near-Term Fermionic Simulation with \\ Subspace Noise Tailored Quantum Error Mitigation}%

\author{Miha Papi\v c}
\email{miha.papic@meetiqm.com}
\affiliation{IQM Quantum Computers, Georg-Brauchle-Ring 23-25, 80992 Munich, Germany}
\affiliation{Department of Physics and Arnold Sommerfeld Center for Theoretical Physics, Ludwig-Maximilians-Universität München, Theresienstr. 37, 80333 Munich, Germany}

\author{Manuel G. Algaba}
\affiliation{IQM Quantum Computers, Georg-Brauchle-Ring 23-25, 80992 Munich, Germany}
\affiliation{PhD Programme in Condensed Matter Physics, Nanoscience and Biophysics, Doctoral School, Universidad Autónoma de Madrid}

\author{Emiliano Godinez-Ramirez}
\affiliation{IQM Quantum Computers, Georg-Brauchle-Ring 23-25, 80992 Munich, Germany}

\author{Inés de Vega}
\affiliation{IQM Quantum Computers, Georg-Brauchle-Ring 23-25, 80992 Munich, Germany}
\affiliation{Department of Physics and Arnold Sommerfeld Center for Theoretical Physics, Ludwig-Maximilians-Universität München, Theresienstr. 37, 80333 Munich, Germany}

\author{Adrian Auer}
\affiliation{IQM Quantum Computers, Georg-Brauchle-Ring 23-25, 80992 Munich, Germany}

\author{Fedor Šimkovic IV}
\affiliation{IQM Quantum Computers, Georg-Brauchle-Ring 23-25, 80992 Munich, Germany}

\author{Alessio Calzona}
\affiliation{IQM Quantum Computers, Georg-Brauchle-Ring 23-25, 80992 Munich, Germany}

\date{\today}

\begin{abstract}
Quantum error mitigation (QEM) has emerged as a powerful tool for the extraction of useful quantum information from quantum devices. Here, we introduce the Subspace Noise Tailoring (SNT) algorithm, which efficiently combines the cheap cost of Symmetry Verification (SV) and low bias of Probabilistic Error Cancellation (PEC) QEM techniques. We study the performance of our method by simulating the Trotterized time evolution of the spin-1/2 Fermi-Hubbard model (FHM) using a variety of local fermion-to-qubit encodings, which define a computational subspace through a set of stabilizers, the measurement of which can be used to post-select noisy quantum data. We study different combinations of QEM and encodings and uncover a rich state diagram of optimal combinations, depending on the hardware performance, system size and available shot budget. We then demonstrate how SNT extends the reach of current noisy quantum computers in terms of the number of fermionic lattice sites and the number of Trotter steps, and quantify the required hardware performance beyond which a noisy device may compete with current state-of-the-art classical computational methods. 
\end{abstract}

\maketitle


\subsection*{\label{sec:intro} Introduction}
The simulation of fermionic quantum systems from condensed matter physics and quantum chemistry is believed to provide some of the most promising applications where quantum computers are expected to eventually outperform their classical counterparts \cite{Georgescu_2014, Fauseweh_2024}. This belief is largely centered around the task of time-evolving quantum systems which is one of the few cases where exponential quantum speedup has been proven \cite{hoefler2023}. This optimism has sparked a series of proof-of-principle experimental realizations on current quantum devices \cite{Kandala_2017,Sagastizabal_2019,smith_2019,google_2020b_chargespin,Google_2020_hartreefock,Dborin_2022,Stanisic_2022,O_Brien_2023, yoshioka_ibm_2024, koyluoglu_2024, nigmatullin2024, vilchezestevez2025,Mildenberger_2025, will2025, will2025, evered2025}, leading to the question of the ultimate reach of near-term, non-error-corrected quantum computations \cite{IBM_2023_utility}. This question is of essential relevance given that, despite steady recent progress and ambitious company road-maps, current quantum-error-correction experiments are still limited to small-distance codes and few logical qubits, and fully fault-tolerant quantum computers will not come into existence for a number of years to come. 

\begin{figure*}[t]
    \centering
    \includegraphics[width=.9\textwidth]{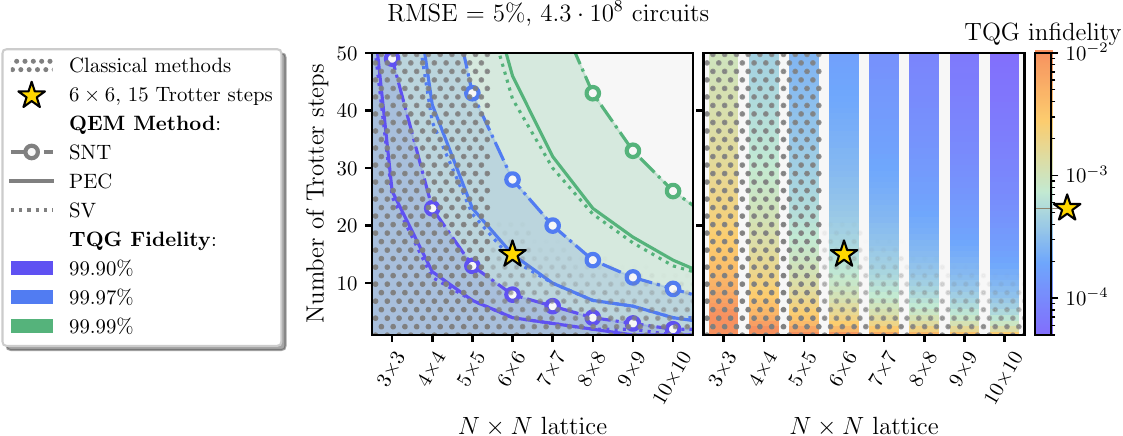}
    \caption{Classical and quantum limits of the simulability of the 2D FHM. \textit{Left}: The maximal number of Trotter steps achievable for a given QEM method at a fixed TQG fidelity, and a fixed $5\%$ root-mean-squared error (RMSE) of the site occupations. For more details see ``Methods''. \textit{Right}: The required TQG fidelity for the simulation of a given FHM with SNT. The dotted region represents the \emph{approximate} reach of classical computations whereas the gradual onset of transparency represents a transitional regime where quantum or classical methods are in close competition, as argued in detail in Supplementary Note 1 \cite{Schollwoeck_2011_dmrg, Paeckel_2019, thompson2025_fermioniq}.}
    \label{fig:1}
\end{figure*}

Recently, effort has been invested into resource estimation for the simulation of fermionic Hamiltonians on quantum hardware in terms of the required circuit depth and gate counts \cite{Cade_2020,Clinton_2021,Algaba_2024,jafarizadeh2024}. It has become increasingly clear that any successful application on current noisy hardware will necessitate the use of Quantum Error Mitigation (QEM) techniques, which reduce the effects of hardware noise at the cost of an exponential increase in the number of circuit executions. A myriad of different QEM approaches has been developed~\cite{Cai_2023_QEMreview}, where different techniques can be characterized by their measurement overhead, referred to as the \textit{cost} of error mitigation, and their accuracy in the limit of infinite resources, referred to as the \textit{bias}. Broadly speaking, approaches with low bias incur higher costs, and vice versa. The community is thus actively exploring error mitigation techniques that strike the right balance between these two factors, with the conjecture that optimal QEM strategies will likely involve hybrid approaches that combine multiple methods, leveraging their complementary strengths~\cite{Cai_2023_QEMreview}. 

One family of commonly utilized QEM techniques is based on symmetry verification (SV) \cite{BonetMonroig_2018,McArdle_2019,McClean_2020,Huggins_2021, Cai_2021}. Given that quantum systems conserve certain quantities, such as the total number of fermions, it is sometimes possible to filter out measurements of a noisy quantum state which fall outside of the correct symmetry-preserving subspace \cite{Google_2020_hartreefock, O_Brien_2023}. Generally, these methods exhibit low cost and high bias, as only a few global symmetries exist in most systems of interest. It is possible to artificially add further symmetries for SV purposes by enlarging the computational space of the system, thus allowing the implementation of SV methods using post-selection (PS) based on the measurement of stabilizer operators, identical to syndrome measurements in quantum-error-correction (QEC) codes \cite{Terhal_2015, BonetMonroig_2018}. Notably, the existence of many local stabilizers is a natural feature of local fermion-to-qubit encodings~\cite{Kitaev_2006, Derby_2021, chien_2022, chien_2023, simkovic_2024, Algaba_2024}, where ancilla qubits are introduced to resolve fermionic commutation relations in a way that avoids high-weight logical operators, which would otherwise appear in standard fermion-to-qubit encodings such as the Jordan-Wigner transformation (JW)~\cite{Jordan1928, bravyi2002fermionic}. This led to stabilizer-based QEM \cite{chien_2022, chien_2023, simkovic_2024} and partial QEC~\cite{jiang2019majorana, setia2019superfast, chen2024error, simkovic_2024} proposals, especially on fermionic systems defined on periodic lattices in two and three dimensions \cite{nigmatullin2024}.

Nonetheless, any symmetry based QEM technique ultimately suffers from a bias due to undetectable errors, which occur within the correct subspace and thus commute with all available stabilizers. In contrast, the probabilistic error cancellation (PEC) method is, at least in principle, able to cancel any type of errors by averaging over many different circuits designed to compensate for previously characterized hardware noise \cite{Temme_2017_PEC, Endo_2018}. However, the overhead associated with a successful PEC implementation is often prohibitively large, up to orders of magnitude larger compared to biased QEM methods \cite{IBM_2023_utility,van_den_Berg_2023_spl_pec}. A naturally arising question is therefore whether PS and PEC can be combined in a way to overcome these challenges and improve the overall performance.

The initial approach of Ref.~\cite{Cai_2021} proposed a scheme where the errors of a two-qubit gate (TQG) were classified as (un)detectable based on total fermion parity conservation. However, a more general fermionic simulation algorithm may contain more than one stabilizer symmetry and it is not necessary that the native entangling operations preserve these symmetries, which means that this approach cannot be straightforwardly applied to exploit all available symmetries of an algorithm. An alternative, presented in Ref.~\cite{Chen_2023}, applies PEC and PS independently, without any noise classification, thus resulting in an unnecessarily high cost -- even higher compared to the costs of applying PEC and PS individually. Building on these insights, it is clear that any practical hybrid of PS and PEC must reduce the QEM cost by: 1. establishing what is the effective form of the noise as the error propagates through the circuit and 2. dividing errors into disjoint sets of detectable and undetectable ones, which are then treated with PS and PEC, respectively.

In this work, we introduce the Subspace Noise Tailoring (SNT) technique, which combines PEC with PS and adheres to the two stated requirements. We show that it is possible to classify Pauli noise appearing at any location in the circuit into detectable and undetectable errors, even for non-Clifford circuits and for any set of stabilizers. Then, by using PEC to cancel only a fraction of all errors, those undetectable to PS, we are able to keep the overall cost close to that of pure PS, while significantly reducing the bias.

We investigate the relative performance of SNT, PEC and SV in terms of their gate fidelity and shot budget requirements for the Trotterized time evolution of the spin-$1/2$ Fermi-Hubbard model (FHM) and find that the relative performance of the three QEM techniques depends largely on the choice of fermion-to-qubit encoding which directly affects the fraction of detectable errors in PS and SNT. Besides JW, we consider four different local fermion-to-qubit encodings \cite{verstraete2005mapping, Kitaev_2006, Derby_2021, chien_2022, Algaba_2024,  simkovic_2024, LE}, and find a rich \emph{state diagram} of optimal encoding plus QEM technique combinations with respect to hardware characteristics, the fermionic lattice size, and dimensionality. 

To delineate the limits of near-term quantum simulation of fermionic many-body systems we consider the time evolution of the 2D FHM on a square lattice whilst allowing for a shot budget corresponding to roughly 12 hours of computation \cite{O_Brien_2023} on superconducting quantum hardware \cite{fruitwala_2024_FPGA_RC}. These results are presented in Fig.~\ref{fig:1}, from which it is clear that SNT greatly extends the reach of noisy hardware compared to its constituents. Fig.~\ref{fig:1} also shows the approximate reach of current classical methods, based on the arguments presented in Supplementary Note 1. Because classical methods are continually evolving and are constrained by different factors than quantum computers, such as e.g. the weight of the observable or the number of fermions in the lattice, these limitations give rise to a transition region rather than a well-defined boundary. To calculate the resource requirements of SNT for a representative point of a problem size in this transition region, we choose a $6\times6$ lattice with 15 Trotter steps (indicated with a star). At this point, SNT requires a TQG fidelity of $99.95\%$ to achieve a root-mean-square-error (RMSE) of $5\%$ (see Fig.~\ref{fig:1}b) for the evaluation of a Pauli observable. This is around $10^6$ times fewer circuit executions compared to PEC and an almost 2 times larger TQG infidelity compared to SV. These results significantly relax the hardware requirements from previous estimates \cite{O_Brien_2023} and may be used to guide future quantum hardware development and experiments.

\section*{Results}

\subsection*{\label{subsec:SNT}Subspace Noise Tailoring}

In this section, we present how PEC can be utilized to cancel a fraction of the errors, which are not detectable by subsequent symmetry verification-based (SV) error mitigation methods. This allows us to perform QEM at a lower cost while maintaining a small bias. We therefore refer to this combination of QEM methods as Subspace Noise Tailoring (SNT) in the remainder of the text. 

\begin{figure}[t]
    \centering
    \includegraphics[width=0.85\linewidth]{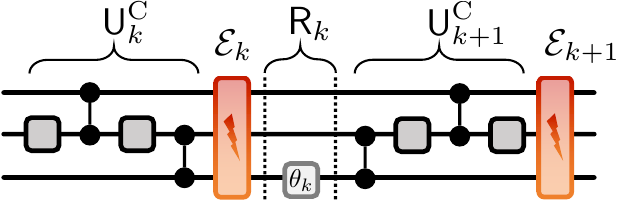}
    \caption{Example of a decomposition of a parameterized multi-qubit Pauli operator $e^{i\theta \mathsf{XZY}}$ into a native gate set, consisting of arbitrary single-qubit rotations and a CZ as the native entangling gate. The red/orange blocks represent the noise channel $\mathcal{E}_k$ associated with implementing the Clifford unitary $\mathsf{U}_k^\mathrm{C}$. The layer of non-Clifford gates is comprised of a single-qubit rotation with angle $\theta_k$.}
    \label{fig:2}
\end{figure}

In order to perform SNT, we rely on the fact that the circuits generated by any fermionic encoding are comprised of a product of exponentials of multi-qubit parameterized Pauli operators of the form $\prod_k e^{-i\theta_k\mathsf{P}_k} $\cite{Algaba_2024}, with the angle $\theta_k$ determined by the Trotter step size and the Pauli operator $\mathsf{P}_k \in \{ \mathsf{I}, \mathsf{X}, \mathsf{Y}, \mathsf{Z} \}^{\otimes N_\mathrm{Q}} \setminus \{\mathsf{I}^{\otimes N_\mathrm{Q}}\}$,  $N_\mathrm{Q}$ being the total number of qubits. Additionally, these circuits preserve a set of stabilizer symmetries $\mathsf{S}_i \in \mathbb{S}$. There exist many ways of decomposing evolution operators into multi-qubit parameterized Pauli generators but most of them consist of two external layers of a linear number of Clifford operators arranged in linear depth, surrounding a central layer of single-qubit non-Clifford gates \cite{Cowtan_2020}, an example of which is shown in Fig.~\ref{fig:2}. More specifically, we utilize the XYZ decomposition for fermionic operators from Refs.~\cite{Sriluckshmy_2023,Algaba_2024} and provide additional derivations for more general decompositions in Supplementary Note 2. After these operators are decomposed into the native gate set, the resulting unitary evolution $\mathsf{U}$ can be divided into $N_L$ layers, of the form
\begin{equation}\label{eq:simple_circuit_unitary}
    \mathsf{U} = \prod_{k=1}^{N_L} \left[\mathsf{U}^\mathrm{C}_k \mathsf{R}_k(\theta_k) \right] \mathsf{U}^\mathrm{C}_0,
\end{equation}
with $\mathsf{R}_k(\theta_k) = e^{-i \frac{\theta_k}{2} \mathsf{M}_k}$, where $\mathsf{M}_k \in \{ \mathsf{X}, \mathsf{Y}, \mathsf{Z}\}$ is a single-qubit Pauli operator, whilst $\mathsf{U_k^\mathrm{C}}$ is a Clifford unitary in layer $k=0,\dots,N_L$. The form of the unitary in Eq.~\ref{eq:simple_circuit_unitary} means that the Clifford unitaries $\mathsf{U_k^\mathrm{C}}$ will contain all of the entangling operations, while $\mathsf{R}_k(\theta_k)$ may be a single single-qubit rotation. However, the reasoning outlined in the remainder of the section also applies to cases such as when several commuting Pauli operators are exponentiated, generating more non-Clifford single-qubit rotations $\mathsf{R}_k(\theta_k)$. It is reasonable to assume that most of the errors in the circuit will appear during the implementation of $\mathsf{U_k^\mathrm{C}}$, and that the noise is Markovian. Furthermore, Randomized Compiling (RC) \cite{hashim_2021_rc} may be used to transform these errors into Pauli errors acting after $\mathsf{U}^\mathrm{C}_k$, as described by the dynamical map:
\begin{equation}\label{eq:pauli_noise_def}
    \mathcal{E}_k[\bullet] = (1 - \sum_i p_i^{(k)}) \bullet + \sum_i p_i^{(k)} \mathsf{P}_i \bullet \mathsf{P}_i,
\end{equation}
where $p_i^{(k)}$ denotes the probability of a Pauli error $\mathsf{P}_i$ appearing in layer $k$ of the algorithm. If the noise of the central non-Clifford rotation cannot be neglected, pseudo-twirling can be used to ensure the noise is still approximately described by a Pauli channel~\cite{santos_2024}. Moreover, Pauli noise can be efficiently characterized using techniques such as Cycle Benchmarking~\cite{Erhard_2019_CB, carignandugas2023} and similar \cite{flammia_2022_ACES,calzona_2024, van_den_Berg_2024,pelaez_2024_ACES, hockings_2024_ACES, chen_2025}, provided that it is sufficiently local. By characterizing the noise of each layer individually, the reconstructed noise model can account for both context- and gate-dependent errors.

\begin{figure*}[t]
    \centering
    \includegraphics[width=0.75\linewidth]{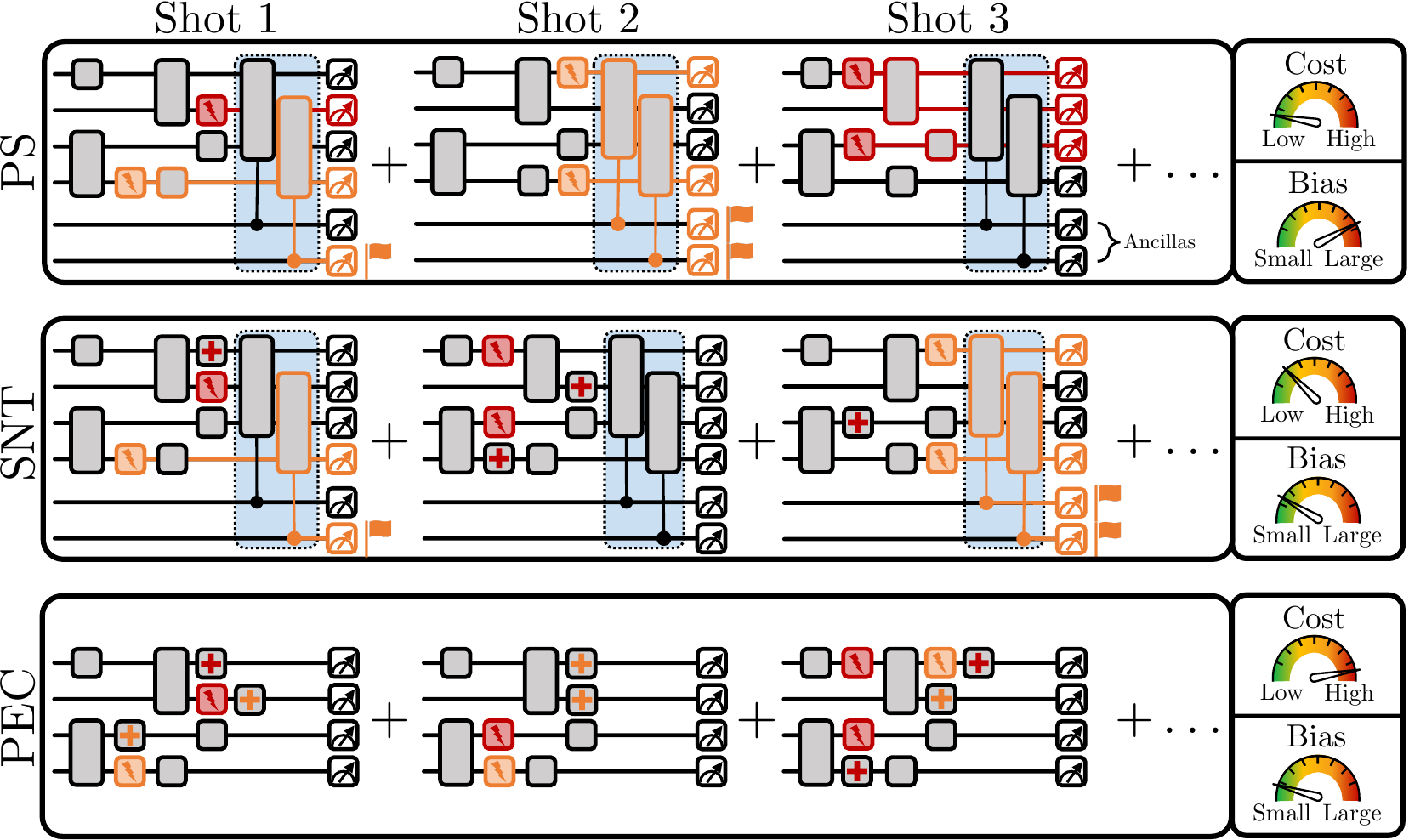}
    \caption{Shot-by-shot representation of three different QEM methods. The (red)orange lightning bolts represent stochastically appearing (un)detectable errors. Gates with (red)orange crosses represent the gates added to the circuit in order to probabilistically cancel the (un)detectable errors using PEC. The parity check layer is highlighted in blue. If a detectable error propagates through the circuits and is subsequently detected via the ancilla measurement, the shot is discarded (indicated by an orange flag). Compared to PS, SNT circuits also utilize extra operations used to cancel undetectable errors, however only undetectable errors are canceled probabilistically while the detectable errors are removed by PS. The qualitative performance of each QEM method in terms of bias and cost is illustrated on the right.}
    \label{fig:3}
\end{figure*}

Given the circuit structure of Eq.~\ref{eq:simple_circuit_unitary}, we prove that if a \emph{single} Pauli error $\mathsf{P}_i$ occurs after the $k$-th Clifford unitary, the error is \emph{undetectable} after layer $l>k$ iff the Pauli operator $\mathsf{Q}_i^{(k:l)} \equiv \mathsf{U}_{l}^\mathrm{C}\dots\mathsf{U}_{k+1}^\mathrm{C}\mathsf{P}_i (\mathsf{U}_{k+1}^\mathrm{C})^\dagger \dots (\mathsf{U}_{l}^\mathrm{C})^\dagger$ commutes with all stabilizers, i.e. iff
\begin{equation}\label{eq:detectable_pauli_condition}
    [ \mathsf{Q}_i^{(k:l)} , \mathsf{S}_j] = 0 \,\, \forall \mathsf{S}_j \in \mathbb{S}.
\end{equation}
Otherwise, the error $\mathsf{P}_i$ is \emph{detectable} and SV performed right after layer $l$ will mitigate its effects. Remarkably, as detailed in ``Methods'', this result holds despite the presence of several non-Clifford rotations $\mathsf{R}_m(\theta_m)$ in the circuit, thus allowing the implementation of SV at arbitrary points in the circuit.

Once the desired SV strategy is set in place, SNT utilizes PEC to cancel the undetectable errors in the circuit. In particular, denoting the set of \emph{undetectable} errors of layer $k$ as $\mathbb{U}_k \subseteq  \{\mathsf{I},\mathsf{X},\mathsf{Y},\mathsf{Z} \}^{\otimes N_\mathrm{Q}} \setminus \{\mathsf{I}^{\otimes N_\mathrm{Q}}\}$, PEC effectively implements the inverse noise map
\begin{equation}\label{eq:1st_order_noise_inverse}
    \mathcal{N}^{-1}_k[\bullet] = (1 + \sum_{i| \mathsf{P}_i \in \mathbb{U}_k} p_i^{(k)}) \bullet - \sum_{i| \mathsf{P}_i \in \mathbb{U}_k} p_i^{(k)} \mathsf{P}_i \bullet \mathsf{P}_i,
\end{equation}
which cancels, up to first order in $\eta_k \equiv \sum_{i|\mathsf{P}_i  \in \mathbb{U}_k} p_i^{(k)}$, the effects of the errors $\mathsf{P}_i \in \mathbb{U}_k$ \cite{Cai_2021,guimaraes_2024} in the dynamical map defined in Eq.~\ref{eq:pauli_noise_def}. By construction, all Paulis in $\mathcal{N}^{-1}_k$ belong to $\mathbb{U}_k$, meaning that the PEC implementation does not alter the set of circuit stabilizers. This ensures full compatibility between SV and PEC, which can be applied after every noisy Clifford layer. Examples of the circuits needed to implement PS, SNT and PEC are illustrated in Fig.~\ref{fig:3}.

While SNT efficiently merges PEC with SV, it nonetheless inherits some of the shortcomings of its constituents. Indeed, the efficacy of PEC is limited by higher-order terms in $\eta_k$, upper bounded by the (entanglement) layer infidelity $\varepsilon_k \equiv \sum_{i| \mathsf{P}_i \in \mathbb{P}_k}p_i^{(k)}$, which is expected to be small. If this is not the case, the layer infidelities can be reduced by splitting layers into their constituent two-qubit gates. Additionally, imperfect noise characterization due to drifts will contribute to the PEC error \cite{govia_2024}. The issue of unlearnable degrees of freedom associated with Pauli noise characterization \cite{Chen2023Jan, calzona_2024, chen_2024} can be avoided by characterizing the noise self-consistently and could even be used to minimize the cost of PEC \cite{chen_2025}. As for SV, it suffers from the imperfect implementation of parity checks and, importantly, from the detrimental effects of multiple errors happening within the causal cone of individual stabilizer parity checks, as a combination of detectable errors may become undetectable. In the remainder of the paper, we use numerical simulations to investigate the bias of SNT, stemming from these residual sources of bias, and provide a deeper understanding of their impact.

Assuming a Poisson distribution of errors, the probability of an error-free circuit execution, also known as the circuit success probability (CSP), is given by $\mathrm{CSP} \equiv \prod_k (1-\varepsilon_k) =  e^{-\lambda}$, where the circuit error rate $\lambda$ indicates the average number of errors occurring in the circuit \cite{Cai_2023_QEMreview}. The cost of any QEM method scales exponentially in $\lambda$ as $C^2 \equiv \mathrm{Var}[\mathsf{O}_\mathrm{est.}]/\mathrm{Var}[\mathsf{O}] \propto e^{2 \beta \lambda}$, where $\mathrm{Var}[\mathsf{O}_\mathrm{est.}]$ and $\mathrm{Var}[\mathsf{O}]$ are the variances of the error mitigated and noisy observable, respectively \cite{Cai_2023_QEMreview}, and $\beta$ is a coefficient specific to the QEM method. The cost $C_{\mathrm{SNT}}$ of implementing SNT is bounded by the costs of SV and PEC \cite{Temme_2017_PEC,Endo_2018,Cai_2023_QEMreview} as
$C_{\mathrm{SV}} \leq C_{\mathrm{SNT}}\leq C_{\mathrm{PEC}}\approx e^{2\lambda}$.
Assuming that SV is performed via PS, we get $C_\mathrm{SV} \leq e^{\lambda/2}$\cite{Cai_2023_QEMreview}, where the inequality is due to the fact that not all of the noise is detectable. In the simple scenario where the ratio of detectable to total noise is known and layer-independent, i.e. $R_k \equiv 1-\eta_k/\varepsilon_k = R \in [0,1]$, the cost of SNT can be approximated (up to lowest-order in $\varepsilon_k$) as: 
\begin{align}\label{eq:cost_definition}
    C_\mathrm{SNT} \sim e^{\beta_\mathrm{SNT}\lambda}&\approx e^{ \frac{R}{2} \lambda } e^{2(1 - R) \lambda} = e^{\frac{4 - 3R }{2} \lambda}.
\end{align}

\subsection*{Simulation of the Fermi-Hubbard model}

\begin{table*}[t] 
\begin{tabular}{|c||c|c|c|c|c|c|c|c|c|c|c|}
\hline
 &  &  &  & Max. &  &  &  & & \multicolumn{3}{c|}{ $\log C^\mathrm{(...)}_\mathrm{SNT}/(1 - \mathcal{F}_\mathrm{TQG})  N_\mathrm{Trotter} N $} \\ \cline{10-12}

Encoding & Type & Dist. & $Q_r$ & Conn. (2D)  & $N_\mathrm{TQG}^\mathrm{1D}$ & $N_\mathrm{TQG}^\mathrm{2D}$ & $R_\mathrm{PS}$ & $R_\mathrm{PP}$ & SV & PEC  & SNT ($\sum$)  \\ \hline\hline
JW & 1D & 1 & 1 & 2 &$6N - 4$ & $4N^{3/2} + 6N - 4 \sqrt{N} - 4$ & 0\% & 64\%  & $2.6\sqrt{N} + 3.9$  & $2.6\sqrt{N} + 3.9$   & $5.2\sqrt{N} + 7.8$  \\ \hline
LE & 1D & 2 & 2  & 3(+2)&$14N - 8$ & $8N^{3/2} + 10N - 4 \sqrt{N} - 4$ & 89\% & 2\%   &  $3.6\sqrt{N} + 4.5 $  & $1.9\sqrt{N} + 2.4$  & $5.5\sqrt{N} + 6.9$  \\ \hline \hline
DK & 2D & $1$ & $1.5$ &  4(+4) & $14N-12$ &  $20N - 36 \sqrt{N} + 18$& 80\% & 3\% & $8.6$ & $7.6$ & $16.2$ \\ \hline
VC & 2D & 1 & 2  & 4(+4) &$14N - 12$ & $22N -20 \sqrt{N} $ & 83\% & 5\% & $10.3 $  & $6.5 $ & $16.8 $ \\ \hline
HX & 2D & 2 & 2  & 3(+3) &$16N - 9$ & $44N -24 \sqrt{N} $ & 91\% & 2\% & $20.1$  & $7.8$  & $28.8$   \\ \hline

\end{tabular}

\caption{Fermionic encodings used in this work and their properties. We report $N_\mathrm{TQG}$ required for simulating the Fermi-Hubbard model on a 1D chain and on a 2D square lattice, highlighting the superlinear scaling of the two 1D encodings in the the number of fermionic sites $N$. Values in brackets within the connectivity column represent the additional connectivity required for the implementation of a parity check. The fractions of noise detected by PS $R_\mathrm{PS}$ and PP $R_\mathrm{PP}$, presented in the last two columns, are computed numerically based on a local noise model, as detailed in ``Methods''. The last three columns represents the scaling of the SV (left) and PEC (middle) contributions to the total SNT cost (right) for the simulation of a 2D FHM. The values are accurate up to $\mathcal{O}(N^{-1/2})$.}

\label{tab:encodings}
\end{table*}

We now demonstrate the performance of SNT with simulations of the 1D FHM, defined in Eq.~\ref{eq:fhm},  on system sizes of 2 and 4 sites and examine fermionic encodings beyond JW. Specifically, we consider another 1D encoding, LE\cite{LE}, and three 2D encodings, VC \cite{Algaba_2024}, DK \cite{Derby_2021} and HX \cite{Kitaev_2006, chien_2022, simkovic_2024}. The main properties of these encodings are summarized in Table~\ref{tab:encodings}, with additional details provided in ``Methods''. While having a larger qubit-to-fermion ratio $Q_r$, those four encodings feature local stabilizers, with weights$w_{\mathsf{S}}$, independent of the number of fermionic sites $N$, and whose number scales as $|\mathbb{S}|\propto N$. This, combined with the higher code distance of LE and HX, enables the detection of more errors compared to JW, with the added benefit of allowing scalable PS based on local parity checks using ancilla qubits. The last key metric to evaluate the performances of these encodings in combination with QEM is the number $N_{\rm TQG}$ of two-qubit gates required to implement a single Trotter step, while assuming a QPU connectivity native to the encoding and that only native CZ entangling gates are available. Indeed, for a fixed gate fidelity, a larger number of gates leads to a lower CSP and thus to a worse RMSE. We consider a finite budget of individual circuits ($N_\mathrm{circuits}$) and total shots ($N_\mathrm{shots}$), i.e. of rounds of final measurements ($N_\mathrm{shots} \geq N_\mathrm{circuits}$).

The FHM also preserves the total spin of the system, which can be exploited for the purposes of QEM. Although the total spin operator is not a Pauli operator, making it incompatible with the stabilizer formalism and thus difficult to handle~\cite{McClean_2020}, the \textit{parity} of the number of up ($\uparrow$) and down ($\downarrow$) spins is a conserved Pauli symmetry for all fermionic encodings~\cite{McClean_2020}. We can thus incorporate the two corresponding operators $\hat{S}_{\uparrow/\downarrow} =  \prod_{i} \hat{n}_{i}^{\uparrow/\downarrow}$ into the set of stabilizers, increasing the error detection capabilities of the system. However, just like the single stabilizer of JW, these two stabilizers are global, with weights $w_{\mathsf{S}_{\uparrow/\downarrow}} = N$. Measuring them with parity checks is therefore not scalable~\cite{Wintersperger_2023}, as it would require entangling one ancilla with $\mathcal{O}(N)$ qubits. While this rules out the possibility of implementing SV based on PS, one can still perform SV for global stabilizers via a post-processing (PP) procedure, as described in Refs.~\cite{BonetMonroig_2018, Huggins_2021, Cai_2021}. The downside of PP-based SV is that it comes with a quadratically worse cost and a constant numerical prefactor, i.e. $C_\mathrm{PP}\leq 1.5 \, e^{\lambda}$, as shown in Supplementary Note 3. Therefore, whenever possible, parity check-based SV is performed beforehand, so that only a small fraction of the noise, detected by the global stabilizers, is removed via the more costly PP. 

The cost of SNT associated with the simulation of the FHM has three main contributions, associated with the costs of PS, PP and PEC: 
\begin{equation}
\label{eq:cost_fhm}
    C_{\rm SNT} = C_{\rm SNT}^{(\rm PS)} C_{\rm SNT}^{(\rm PP)} C_{\rm SNT}^{(\rm PEC)} \approx 1.5 e^{\beta_{\rm SNT} \lambda}.
\end{equation}
While the PEC contribution can be readily computed from the known error characterization as $C_{\rm SNT}^{(\rm PEC)} = \exp(2\sum_k \eta_k)$, the remaining two can be estimated from numerical results, and are listed in Table~\ref{tab:encodings}. Specifically, given the shot rejection probability $\Pi$ associated with the parity checks, one has $C_{\rm SNT}^{(\rm PS)} = 1/\sqrt{1 -\Pi}$. As for PP, we have $C_{\rm SNT}^{(\rm PP)} = 1.5/\langle \mathsf{M}_{\mathbb{S}_{\uparrow/\downarrow}}\rangle$, where $\langle \mathsf{M}_{\mathbb{S}_{\uparrow/\downarrow}}\rangle$ is the expectation value of the subspace projector $\mathsf{M}_{\mathbb{S}_{\uparrow/\downarrow}} = \prod_{\mathsf{S}_i \in \mathbb{S}_{\uparrow/\downarrow}} (\mathsf{I}+\mathsf{S}_i)/2$ of the global stabilizers ($\mathbb{S}_{\uparrow/\downarrow} = \{ \mathsf{S}_\uparrow,\mathsf{S}_\downarrow \}$) on the post-selected shots. The numerical results, summarized in Fig.~\ref{fig:8}, allow us to derive an upper bound on $\beta_{\rm SNT}$ by computing $C_{\rm SNT}$ in the high CSP limit $\lambda \to 0$, as described in ``Methods''.  

Values of $\beta_{\rm SNT}$ for different fermionic encodings are reported in Table~\ref{tab:qem_methods}. SNT based on local encodings features cost coefficients well below $1$, placing it among the most affordable QEM techniques. The reason for this is that SNT offloads the mitigation of the majority of the noise to the low-cost PS, leaving only a small remaining fraction to the more costly PP and PEC parts.

The measurement errors of the ancilla qubits in the parity checks will not contribute to $\beta_\mathrm{SNT}$, but will nonetheless increase the cost up to a factor of 
\begin{equation}\label{eq:meas_err_cost}
    C_\mathrm{SNT}^{(\mathrm{meas.})} \leq 1/(1 - \varepsilon_\mathrm{meas.})^{N_\mathrm{meas.}/2} \approx e^{N_\mathrm{meas.}\varepsilon_\mathrm{meas.}/2},
\end{equation}
with $\varepsilon_\mathrm{meas.}$ being the measurement error probability and $N_\mathrm{meas.}$ representing the number of stabilizer measurements. The equality is valid in the limit of a noiseless circuit, and the measurement error induced cost is smaller for noisier circuits, as confirmed in Supplementary Fig. S1 and Supplementary Note 4. Conversely, measurement errors on the data qubits have no effect on the cost of SNT.

\begin{table*}[] 
\begin{tabular}{|c||c|c|c|}
\hline
 QEM Method & Cost coeff. $\beta$ & Noise characterization? & Ref.  \\
 \hline\hline
 SV (via PS) & $\leq 0.5$ &    & \cite{BonetMonroig_2018,Cai_2023_QEMreview} \\  \hline
 SNT (+LE or HX) & $\approx 0.7$ &  \checkmark  &  Figs.~\ref{fig:4},\ref{fig:8}\\ \hline
 SNT (+VC or DK) & $\approx 0.8$ & \checkmark  &  Figs.~\ref{fig:4},\ref{fig:8}\\ \hline
 SV (via PP) & $\leq 1$ & &\cite{Huggins_2021,Cai_2021,Cai_2023_QEMreview}  \\ \hline
 TEM & $1$ &  \checkmark &\cite{filippov_2023_TEM} \\ \hline
 EV & $1$ &    & \cite{Cai_2023_QEMreview,OBrien_2021_EV,Huo_2022_EV}\\ \hline
 SNT (+JW) & $\approx 1.3$ &  \checkmark &Figs.~\ref{fig:4},\ref{fig:8} \\ \hline
PEC & 2 &   \checkmark & \cite{Temme_2017_PEC,Li_2017_PEC} \\ \hline 
VD ($M$-th order) & $M$ ($\geq 2$)  &   & \cite{czarnik_2021_VD,Huggins_2021_VD} \\ 
\hline
\end{tabular}

\caption{\label{tab:qem_methods}
Comparison of the cost of several QEM methods. The cost coefficient of SNT is derived based on the data in ``Methods'' under ``Cost of SNT''. }
\end{table*}

\begin{figure*}[t]
    \centering
    \includegraphics[width=.9\textwidth]{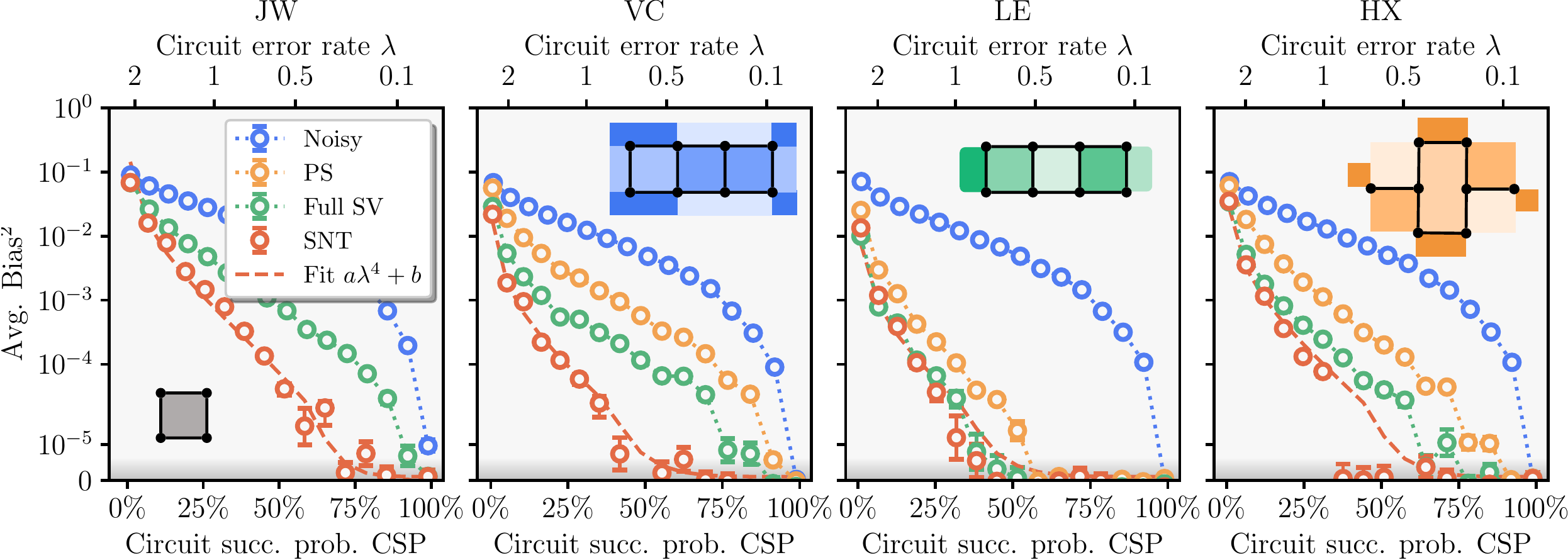}
\caption{Squared bias (averaged over the site occupations) of the time evolution of a FHM with two sites after 10 Trotter steps as a function of CSP and the circuit error rate $\lambda$, for four different encodings and different mitigation schemes: no mitigation (blue), PS on local stabilizers (yellow), full SV including PP based on global stabilizers (green) and SNT (red). The dashed line represents a fit to the SNT data, assuming a second-order-error dominated bias $\propto \lambda^2$. The error bars represent a 1-$\sigma$ uncertainty due to a finite number of shots/circuits, which starts to dominate in the gray shaded area. The noisiness of the circuits is varied by changing the CZ gate fidelity. The insets display the simulated systems, with each node representing a qubit and different shades corresponding to different stabilizers.
}
    \label{fig:4}
\end{figure*}

Our numerical simulations, whose details are presented in ``Methods'', enable us to directly extract the mean-squared-error and variance of the mitigated results. As described in Supplementary Note 5, we can then construct a reliable estimator for the squared bias, quantifying also its statistical uncertainty due to the finite number of shots used in the simulation. The results are presented in Fig.~\ref{fig:4}, where we plot the estimated squared bias, averaged over the site occupations as a function of the CSP, for four fermionic encodings, and with a single round of parity checks. As detailed in ``Methods'', since the circuit depth and two-qubit gate count significantly exceed the number of qubits, we do not expect the choice of observables (or observable weights) to significantly affect the results, as also demonstrated in Supplementary Note 6. The DK encoding cannot be implemented on 2 fermionic sites whilst preserving the same stabilizer weight as in a larger system and was therefore omitted from the results in Fig.~\ref{fig:4}. The squared bias computed from a 4-site Clifford simulation with the DK encoding is available in Supplementary Fig. S2.

Consistent with our expectations, SNT (red markers) yields a smaller bias than SV (green markers), as the additional PEC stage eliminates errors that cannot be detected by the two SV stages, i.e. PP and PS (yellow markers), the latter implemented only for local encodings. This effect is particularly pronounced for distance-1 encodings, JW and VC, where the fraction of errors detectable by SV is smaller. As indicated by the red dashed line, the bias of SNT, scales with $\lambda^2$, and is offset by a constant $b \sim 1/N_\mathrm{circuits}$, representing the finite resolution of the data.

For high CSP, SNT achieves very small squared bias values, below the uncertainty (around $10^{-5}$) associated with the finite number of shots used in the simulations. The fit also demonstrates that the bias behavior follows the simple theoretical predictions, and increases polynomially, even when taking into account the noise in the parity check measurement circuits. Similar behavior of the bias is also observed for the LE encoding in combination with full SV, thus suggesting that in this case, the fraction of detectable noise is high enough that the bias is limited by higher-order detectable errors. However, as the CSP rate decreases, the SNT squared bias increases significantly above the shot resolution of $10^{-5}$. This occurs first for the JW encoding, at around $60\%$ CSP, followed by the three local encodings, at approximately $35\%$ CSP. The reason for this behavior lies in the increasing probability of multiple errors occurring within the causal cones of stabilizers at larger CSP, which renders some combinations of detectable errors undetectable, meaning they are not canceled at the PEC stage. This effect is more pronounced in the JW case due to its few global stabilizers which feature large causal cones \cite{tran_2023_cones,eddins_2024_cones}.

\begin{figure*}[t]
    \centering
    \includegraphics[width=0.7\textwidth]{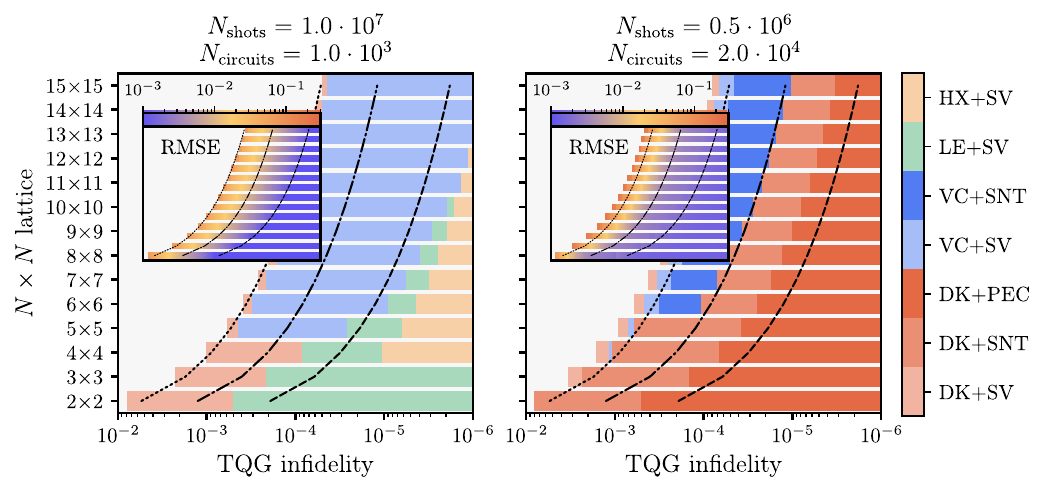}
    \caption{Optimal combinations of encoding and QEM, for the time evolution ($10$ Trotter steps) of a 2D FHM. The three black contour lines represent a CSP of 5\%, 50\% and 90\% (left to right) of the circuit generated by the encoding with the smallest number of TQGs, depending on the system size. The white region in the top left corresponds to low circuit success probabilities, where a single parity check is not sufficient for a significant bias reduction, and the cost of PEC $C_\mathrm{PEC}^2 \gtrsim 10^6$ is too large for the given shot budget. Insets show the RMSE for optimal combinations of QEM and encoding. 
    }
    \label{fig:5}
\end{figure*}

In the presence of measurement errors, readout error mitigation \cite{Bravyi_2021_REM, Cai_2023_QEMreview} can be employed for the data qubits, but not for the ancillas. However, as shown in Supplementary Note 4, the effect of ancilla measurement errors on the bias is second order, i.e. a measurement error must occur on \textit{all} stabilizer measurements that would have detected an error appearing in the circuit. We demonstrate in Supplementary Fig. S1 that with an average of one measurement error on all the ancillas there is no discernible effect on the bias. Even an average of 2.5 ancilla measurement errors, still allow for a significant suppression of the bias compared to a noisy simulation. Additionally, due to the aforementioned mechanism, as the system is scaled up and the number of stabilizers increases, the bias at a fixed measurement error rate will be even less susceptible to ancilla measurement errors.

\subsection*{\label{sec:QEM_fermionic_combo} Error Mitigation and Fermionic encodings: \\ Optimal Strategies for Large-Scale Simulations}

Having evaluated the performance of SNT on smaller system sizes, we now focus on problems at the limit of classical computational abilities - simulating the FHM on up to a $15\times15$ lattice. Initially, we consider the simple scenario where a single parity check round is performed at the end of the circuit. This analysis allows us to determine which combinations of fermionic encodings and mitigation techniques perform best in terms of RMSE in different regimes.

To assess the RMSE of mitigation techniques in this challenging regime, we extrapolate their cost and bias from smaller-scale numerical simulations. We expect the SNT cost coefficient $\beta_\mathrm{SNT}$ to stay largely independent of the circuit size, prompting the use of approximate upper bounds, as reported in Table~\ref{tab:encodings}, regardless of the problem size. This is a consequence of the fact that an (un)detectable error will remain (un)detectable even if a noiseless logical operator is applied afterwards. The detectability of a single error is thus not affected by the number of Trotter steps or system size. The main determining factor of the cost of SNT on an arbitrarily large system is therefore the locality of the hardware noise and the weight of logical operators (due to the errors within the logical operators), which is independent of the system size. This is also numerically demonstrated in ``Methods'' - ``Cost of SNT''. The same applies to the cost coefficient for SV based on PP and PS. 

For the bias, we rely on the interpolation of numerical data from small-scale systems (see Supplementary Fig. S2) to derive an approximate functional dependence of the squared bias on the CSP for each QEM method. This approach is justified by the robustness of the CSP metric, which emerges from the fact that noisy data in Fig.~\ref{fig:4} (blue markers) are largely insensitive to the specific encoding, which is associated with different circuit size and structure. The same encoding-independence holds true for SNT data, albeit clearly restricted only to the local encodings. Moreover, at a fixed CSP, increasing system size (and gate count) leads to a reduction in the gate infidelity, thereby enhancing the performance of the PEC stage of SNT. Additionally, for local encodings, larger system sizes have fixed-weight local stabilizers whose causal cones cover a smaller fraction of the whole circuit. This reduces the likelihood of multiple errors affecting the same stabilizers, which is a key contributor to SNT and SV bias. Overall, we are thus confident that this approximate bias estimation is conservative and valid in the large-scale regime.

We consider the simulation of a 2D Fermi-Hubbard Model with $10$ Trotter steps for various system sizes, CZ fidelities, and shot and circuit budgets. To correctly asses $\mathrm{Var}[\mathsf{O}_\mathrm{est.}]$, we derive analytical expressions for the variance of QEM methods based on circuit sampling for the realistic case of $N_\mathrm{circuits} \leq N_\mathrm{shots} $ in Supplementary Note 7, and show how these results can be used to construct improved sampling strategies in Supplementary Note 8. 
Specifically, we cover two distinct regimes, with the first corresponding to the case where many shots but few circuits are available, and the second where the number of circuits is similar to recent experiments \cite{koyluoglu_2024}. The CSP, needed to estimate the RMSE, is computed for a given CZ fidelity and system size by using the relations in Table~\ref{tab:encodings} to determine the total number of CZ gates $N_\mathrm{TQG}$ in a given circuit. This approach enables a fair comparison of different encodings, taking into account their footprint in terms of the number of required CZ gates.

\begin{figure*}[t]
    \centering
    \includegraphics[width=0.6\textwidth]{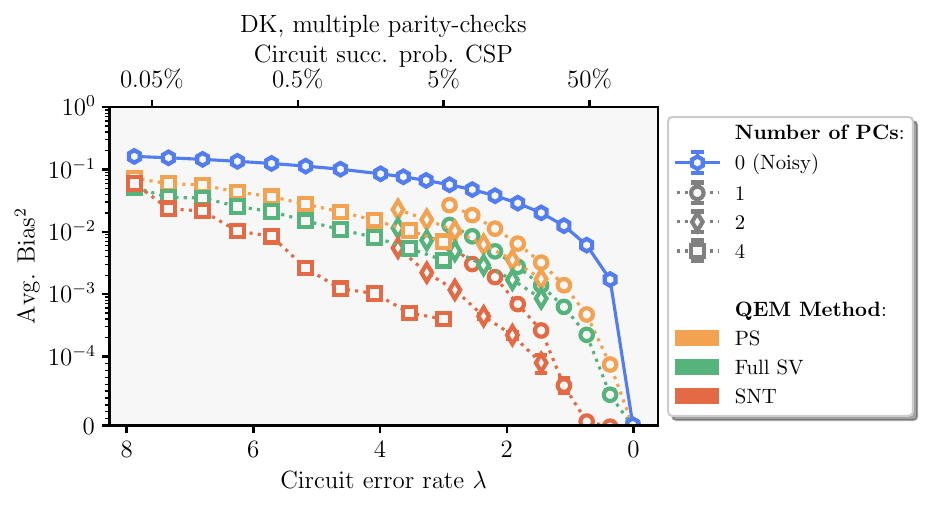}
    \caption{Squared bias with different numbers of parity check rounds (PCs) and the DK encoding for a FHM simulation with four sites and 12 Trotter steps. The parity checks are evenly spaced at the end of a Trotter step. The circuit error rate $\lambda$ and CSP are computed without including the noise in the parity-check layers. The error bars represent a 1-$\sigma$ uncertainty due to the finite number of available shot and circuits.}
    \label{fig:6}
\end{figure*}

As expected, for noisy circuits (left sides of the plots in Fig.~\ref{fig:5}), the exponential cost of QEM favors the DK encoding, which contains the smallest number of gates and therefore the lowest error rate. In the left panel, the small number of circuits prevents a meaningful implementation of SNT and PEC and the QEM method of choice is thus SV. Encoding-wise, we observe that, as the CSP increases (see the black contour lines), it becomes convenient to exploit the slightly better detection properties of VC and subsequently also the higher distance of both HX and LE encodings, which enables more effective error detection, leading to better mitigation despite those encodings requiring more than twice the gates used by DK. In contrast, if more circuits are available (right panel), VC and DK outperform both higher-distance encodings, even in the high-fidelity region. The reason for this is that DK and VC+SNT have both a lower cost and a smaller bias compared to HX+SV. SNT therefore spans the region from 5\% to 90\% CSP, but ultimately in the regime where the exponential scaling of QEM is no longer an issue, the implementation of zero-bias PEC is the optimal strategy. PEC should always be combined with the encoding with the lowest gate count, making DK+PEC the only viable option. The presence of the VC+SNT phase on the right plot is due to the fact that the variance is limited by the number of available circuits rather than the number of available shots. More specifically, since VC is able to detect more errors (see Table~\ref{tab:encodings} and \cite{bausch2020mitigating}), it has a smaller bias (see Supplementary Fig.~S2) and relies less on PEC. Together with the improved sampling procedure described in the SI, the variance of the estimator for VC is similar or even smaller compared to the variance of the estimator with DK even when $C_\mathrm{DK+SNT} < C_\mathrm{VC+SNT}$ (see Supplementary Fig.~S3). The transition between SNT and PEC occurs at the point when the prefactor in the SNT cost (due to the PP) outweighs the considerably worse exponential scaling of the PEC cost. In such a scenario, an SNT variant comprising solely of PS and PEC would fare better and will likely extend the SNT region to even higher values of the CSP. Nonetheless, we choose to focus this work on the SNT variant with PP which is more efficient in the more practically relevant, low CSP regime.

We now focus more carefully on the region of small $\mathrm{CSP} \approx 0.1\% - 1\%$, as this is likely the most interesting region for attempting beyond-classical simulations on noisy, near-term hardware. So far, our analysis clearly indicated that the most convenient encoding is DK, owing to its low gate requirements. As for QEM, low CSP rules out costly techniques such as PEC. Fig.~\ref{fig:5} indicates that SV is the best option in this regime, given it has the lowest cost, but it is important to note that the resulting RMSE (see the inset) rapidly grows, quickly becoming too large to be of practical utility. This is due the large bias of SV and related to the high probability of the occurrence of multiple errors. To properly tackle the low CSP regime, beyond what is shown in Fig.~\ref{fig:5}, it is therefore necessary to refine the approach by making use of the possibility to implement multiple rounds of parity checks within the circuit. 

By performing PS more frequently, the probability of having multiple errors in between parity check rounds is drastically reduced. This significantly reduces the bias of SNT despite the additional errors associated with the noisy parity checks, but has a smaller effect on SV, which is limited by undetectable errors. This clearly emerges from the numerical analysis presented in Fig.~\ref{fig:6}, where the squared bias of SNT remains low at around $10^{-2}$ even for CSP as low as $0.05\%$. For such a high circuit error rate, the cost of PEC is impractical at $C^2_\mathrm{PEC} \sim 10^{13}$, and even more efficient QEM techniques with $\beta = 1$~\cite{OBrien_2021_EV,Huo_2022_EV,filippov_2023_TEM} are extremely challenging to implement in practice with $C^2_{\beta = 1} \sim 10^7$. Nonetheless, the introduction of the additional parity-checks will result in an increase in the cost according to Eq.~\ref{eq:meas_err_cost}, while the effect on the bias is expected to be small, especially at larger system sizes, as shown in Supplementary Fig.~S1. The errors on the data qubits can be mitigated using SNT and will contribute to the total circuit error rate $\lambda$.

Having demonstrated the potential of SNT with multiple parity check rounds in the low CSP regime with small-scale numerical simulations, we can extrapolate the performance of SNT to larger scales. The results are shown in Fig.~\ref{fig:1}. Specifically, the left panel in Fig.~\ref{fig:1} illustrates how SNT is able to obtain a lower error at larger infidelities compared to its constituents, whereas PEC and SV are limited by their cost and bias respectively. 

\section*{Discussion}

In this work we have introduced SNT, a novel QEM technique that combines error detection based on the stabilizers featured in low-distance fermionic encodings with tailored noise-shaping. Our method proved to be extremely cost-effective, with a scaling parameter as low as $\beta_\mathrm{SNT} \approx 0.6 - 0.8$ with a local noise model. This is a crucial feature allowing the mitigation of errors in very noisy regimes, with circuit error rates well above $\lambda = 1$, given that the statistical uncertainty contribution to the RMSE is exponentially large in $\beta \lambda$. At the same time, by leveraging multiple parity check rounds, SNT can provide a small bias that scales approximately as $\lambda^2$. As a result, our numerical simulations show that SNT can deliver results below $5\%$ RMSE also for circuits with an overall CSP below $0.1\%$, corresponding to a striking circuit error rate above $7$, while keeping the total run-time manageable. This unlocks the potential for quantum computers to rival state-of-the-art classical methods in fermionic simulations on 2D lattices, before the advent of fault-tolerance. Specifically, assuming the availability of high-quality QPUs with 158 qubits, whose main source of errors are noisy TQGs with fidelities of $99.95\%$, SNT may allow for the execution of a simulation of around $15$ Trotter steps of a $6\times6$ FHM, while still providing accurate results with less than 5\% RMSE, in a regime at the limit of current classical methods.

While ambitious, we consider these requirements to be achievable, given the rapid advancements in various quantum computing platforms and the fact that these conditions align closely with the anticipated needs and requirements for practical QEC development. To put the fidelity requirements into perspective, current large-scale processors, have been able to achieve a median two-qubit fidelity of 99.86\% with trapped ions~\cite{paetznick2024_trapped_ion_qec}, 99.67\% with superconducting qubits \cite{google_2024_qec} and 99.50\% with neutral atom platforms~\cite{Evered_2023_high_fid_neutral_atom}. Additionally, small-scale superconducting devices have demonstrated that fidelities of at least 99.9\% are achievable~\cite{ding_2023_highfid_fluxonium_TQG, Li_2024_doubletransmon_coupler}. A further halving of these error rates may thus be sufficient to push QPU capabilities close to or beyond classical reach. While the fidelities of single-qubit gates are typically an order of magnitude better than two-qubit gates, the opposite may apply to the readout fidelities~\cite{google_2024_qec}. Fortunately, for the specific example of four parity check rounds, even with currently achievable readout fidelities~\cite{google_2024_qec}, the increase in the cost $C_\mathrm{SNT}^2$ is limited by a factor of $\lesssim 7$, which could be further reduced to $\lesssim 1.5$ with a readout fidelity of $99.9\%$, according to Eq.~\ref{eq:meas_err_cost}. Note that this increase does not affect the results in Fig.~\ref{fig:1} where the reach is limited by the bias of the QEM. The effect on the bias is expected to be less significant according to the arguments presented in the text.

Another key aspect to consider is the connectivity of the QPU. If it is not possible to natively implement the chosen fermionic encoding, extra SWAP gates are required, leading to a decrease of the CSP for a fixed TQG fidelity. The presence of extra swaps due to limited QPU connectivity would result in an encoding-dependent rescaling of the CSP axis of Figs.~\ref{fig:5} and \ref{fig:1}, potentially affecting the choice of the best performing encoding. In this respect, we note that the DK encoding, which emerges from our current analysis as the best option to tackle large-scale problems, requires a maximal connectivity of $8$ to avoid the need for extra swaps. 

A successful implementation of QEM for very low CSP of the order of $0.01\%$ necessarily comes with a large sampling overhead. Even for the cost-effective SNT method we have $C^2\sim 10^6$, which requires the execution of a large number of shots. Importantly, given the probabilistic nature of the PEC part of SNT, the ability to execute a large number of (randomly sampled) circuits, ideally comparable with the number of shots, is crucial. In Supplementary Note 8 we analyze this point in detail, whilst proposing an optimized sampling strategy which reduces the number of required circuit executions. We show that it is particularly effective for SNT, taking advantage of the fact that most of the error mitigation is carried out by PS. Nevertheless, to keep the total run-time at reasonable levels, it is important to have fast circuit execution rates of the order of $1\,$kHz or higher, which is achievable with superconducting platforms given recent developments in control electronics \cite{fruitwala_2024_FPGA_RC}. 

The applicability and appeal of SNT is clearly not restricted to simulations of the FHM. Indeed, the main principle can be applied to any algorithm which can be rewritten as a product of multi-qubit parameterized Pauli rotations which commute with a set of stabilizer symmetries for arbitrary rotation angles, which also includes quantum simulations of bosonic or spin systems \cite{Macridin_2018, Sawaya_2020}. SNT can also be applied to completely general quantum algorithms by using low-distance quantum error detecting codes, which have demonstrated beyond break-even fidelities \cite{Self_2024, He_2025}. In general, it is important to stress that the best combination of encoding and mitigation strategy depends on the problem at hand, the noise profile, and the available number of shots and circuits that can be executed. This can be clearly seen from Fig.~\ref{fig:5}, where different budgets of shots and circuits have been explored. Additional \emph{state diagrams} are provided and discussed in Supplementary Note 9, where we consider the crossover between the 1D FHM, where the JW and LE encodings are clearly favored, and the square 2D FHM, dominated by the DK, VC and HX encodings.

As a general trend, with better HW performance and increasing resources, higher distance encodings are preferred, as their ability to detect a larger fraction of the errors outweighs the higher impact in terms of the number of qubits and number of required quantum operations. This holds true unless the noise level becomes low enough such that the mitigation cost ceases to be the limiting factor and more costly, \emph{bias-free} techniques like \emph{full} PEC can be implemented, without the need to perform any SV at all. In this respect, it is important to stress that in practice even noise-aware techniques suffer from imperfect noise characterization, stemming from parameter drifts \cite{govia_2024} and Pauli model violation with detrimental effects on the bias. Therefore, delegating a large fraction of the mitigation to noise-agnostic and more robust error detection, as in SNT, is a promising approach for practical scenarios. As described in the text, the SNT protocol can be applied in the presence of general Markovian noise acting after Clifford layers and partially also for the noise of the non-Clifford operations with the use of pseudo-twirling \cite{santos_2024}, as long as the noise characterization can be performed efficiently. While our results apply to the more physically motivated example of localized Pauli errors, high-weight spatially correlated errors can be treated with the same framework, however a decrease in the detectability is expected in such a scenario. In the presence of temporally correlated or non-Markovian errors the protocol can be integrated with the proposed non-Markovian PEC variants \cite{liu2024}, however in this case the main practical bottleneck is the scalable characterization of such noise processes. Nonetheless, temporal correlation can enhance the performance of SNT if the correlations are such that the presence of an error in one location reduces the probability for an error to occur in the same run at a later point in the circuit.

Finally, we note that the cost of SNT could be further reduced by using the recently introduced less costly TEM method~\cite{filippov_2023_TEM} instead of PEC to implement the noise-tailoring. A rough estimate based on Eq.~\ref{eq:cost_definition} and the assumption of a ratio of detectable noise of $R= 83\%$ would indeed indicate a reduction of the cost parameter from $\beta_{\rm SNT}\approx 0.77$ to $\beta_{\rm \overline{SNT}}\approx 0.58$. 
At a large circuit error rate of $\lambda = 7$, this would result in a cost for this SNT variant of $C_{\rm \overline{SNT}}^2\sim 4 \cdot 10^3$, to be compared with the costs of standard SNT at ${C}_{\rm SNT}^2\sim 5 \cdot 10^4$ and TEM at $C_{\rm TEM}^2 \sim 1 \cdot 10^6$. On the other hand, combining SNT with techniques such as Zero Noise Extrapolation \cite{Cai_2023_QEMreview} and variants thereof \cite{hosseinkhani_2025_NRE} is expected to further decrease the bias. The investigation of different encodings as well as further improvements of the SNT method will be the subject of future studies.

\section*{Methods}

\subsection*{Pauli Error Classification}
Here, we describe how it is possible to classify individual Pauli errors appearing at various locations in the circuit as detectable or undetectable, up to first order in the noise-strength. 

By using the unitary operator of the circuit from Eq.~\ref{eq:simple_circuit_unitary}, and the notation $\mathcal{U}[\bullet] = \mathsf{U}\bullet\mathsf{U}^\dagger$ we can therefore denote the noisy circuit implementation $\mathcal{U}_\mathrm{noisy}$ as 
\begin{equation}\label{eq:noisy_circuit}
    \mathcal{U}_\mathrm{noisy} = \prod_{k=1}^{N_L} \left[\mathcal{E}_k\mathcal{U}^\mathrm{C}_k \mathcal{R}_k(\theta_k) \right] \mathcal{E}_0\mathcal{U}^\mathrm{C}_0,
\end{equation}
where $\mathcal{E}_k$ is a Pauli channel as defined in Eq.~\ref{eq:pauli_noise_def} describing the noise of the $k$-th Clifford layer in the circuit.
If all $\theta_k = 0$, the circuit is Clifford and the Pauli error $\mathsf{P}_i$ is undetectable in the last layer $N_L$ if the Pauli operator 
\begin{equation}
    \mathsf{Q}_i^{(k:N_L)} \equiv  \mathcal{U}_{N_L}^\mathrm{C}...\mathcal{U}_{k+1}^\mathrm{C}[\mathsf{P}_i]
\end{equation}
commutes with all of the operators in the set $\mathbb{S}$. However, things are more complicated in the presence of non-Clifford rotations $\mathsf{R}_k(\theta_k)$.
For this, we consider the fact that for any two Pauli operators $\mathsf{P}$ and $\mathsf{Q}$ and an arbitrary angle $\theta$
\begin{equation}\label{eq:pauli_commutation_with_nonclifford}
    \exp\left[-i \frac{\theta}{2} \mathsf{P}\right] \mathsf{Q} = \mathsf{Q}\exp\left[-i(-1)^{\langle \mathsf{P} ,\mathsf{Q} \rangle} \frac{\theta}{2} \mathsf{P}\right], 
\end{equation}
where we have defined the symplectic inner product as: $\langle \mathsf{P} ,\mathsf{Q} \rangle = 0$ if $[\mathsf{P} ,\mathsf{Q}] = 0$ and $\langle \mathsf{P} ,\mathsf{Q} \rangle = 1$ if $\{\mathsf{P} ,\mathsf{Q}\} = 0$. The equality is a direct consequence of the fact that for any Pauli operator $ \exp\left[-i \frac{\theta}{2} \mathsf{P}\right]  = \cos\left({\frac{\theta}{2}}\right)\mathsf{I} - i \sin\left( \frac{\theta}{2}\right) \mathsf{P}$.

If a single Pauli error $\mathsf{P}_i$ occurs in layer $l$, the unitary evolution of that particular shot is modified from $\mathsf{U}$ (defined in Eq.~\ref{eq:simple_circuit_unitary}) to $\mathsf{U}_i^{(l)}$, which is given by
\begin{align}\label{eq:single_shot_unitary}
    \mathsf{U}_i^{(l)} &= \prod_{k=l+1}^{N_L} \left[\mathsf{U}^\mathrm{C}_k \mathsf{R}_k(\theta_k) \right]\mathsf{P}_i\mathsf{U}^\mathrm{C}_l \mathsf{R}_l(\theta_l) \prod_{k=1}^{l-1} \left[\mathsf{U}^\mathrm{C}_k \mathsf{R}_k(\theta_k) \right] \mathsf{U}^\mathrm{C}_0 \nonumber\\
    &= \mathsf{Q}_i^{(l:N_L)} \prod_{k=l+1}^{N_L} \left[\mathsf{U}^\mathrm{C}_k \mathsf{R}_k(\pm\theta_k) \right]  \prod_{k=1}^{l} \left[\mathsf{U}^\mathrm{C}_k \mathsf{R}_k(\theta_k) \right] \mathsf{U}^\mathrm{C}_0,\nonumber \\
    &= \mathsf{Q}_i^{(l:N_L)}\mathsf{U}_i^{(l:N_L)}
\end{align}
where the unitary $\mathsf{U}_i^{(l:N_L)}$ is determined by signs of the rotation angles $\pm\theta_k$ for $l < k \leq N_L$ which are in turn determined by the inner product of the Paulis $\langle \mathsf{Q}_i^{(l:k-1)},i\mathsf{R}_k(\pi)\rangle$.

Eq.~\ref{eq:single_shot_unitary} can be used to determine whether single errors in the circuit are detectable or undetectable. Applying SV effectively means that we apply the subspace projection operator $\mathsf{M}_\mathbb{S}$ to $\mathsf{U}_i^{(l)}$. For example, if we apply SV at the end of the circuit, for the initial state $|\psi\rangle=\mathsf{M}_\mathbb{S}|\psi\rangle$ the final state after the evolution and SV is given by Eq.~\ref{eq:single_shot_unitary}, or explicitly $\mathsf{M}_\mathbb{S}\mathsf{Q}_i^{(l:N_L)}\mathsf{U}_i^{(l:N_L)}|\psi\rangle $. Since every element of the stabilizer set commutes with the circuit unitary $\mathsf{U}$ for any value of $\theta_k$, it directly follows that also $\mathsf{U}_i^{(l:N_L)}$ commutes with the same set of stabilizer symmetries, and therefore $[\mathsf{M}_\mathbb{S},\mathsf{U}_i^{(l:N_L)}]=0$. If the Pauli $\mathsf{Q}_i^{(l:N_L)}$ commutes with all the elements in the stabilizer set $\mathbb{S}$, it also commutes with the projector $\mathsf{M}_\mathbb{S}$. The final state is then given by $\mathsf{Q}_i^{(l:N_L)}\mathsf{M}_\mathbb{S}\mathsf{U}_i^{(l:N_L)}|\psi\rangle = \mathsf{Q}_i^{(l:N_L)}\mathsf{U}_i^{(l:N_L)}|\psi\rangle $. This therefore means that the error $\mathsf{P}_i$ is \textit{undetectable}. In the opposite scenario, where $\mathsf{Q}_i^{(l:N_L)}$ anticommutes with at least one element in the stabilizer group, it is easy to see that the final state has completely left the subspace, since in this case $\mathsf{M}_\mathbb{S}\mathsf{Q}_i^{(l:N_L)}\mathsf{U}_i^{(l:N_L)}|\psi\rangle = 0$.

The above results demonstrate that the concept of  detectable (single) errors is well-defined also in the non-Clifford case. Moreover, since the Pauli $\mathsf{Q}_i^{(l:N_L)}$ is obtained by multiplying the original Pauli error $\mathsf{P}_i$ with a number of Clifford unitaries, the classification of the errors is computationally scalable.

The above analysis is valid in the high-fidelity regime, since the probability of observing more than one error per shot was neglected. Whether this approximation is justified or not can be estimated by considering the average number of errors per circuit run, which is given by $\lambda = \sum_k \sum_i p_i^{(k)}$\cite{Cai_2023_QEMreview}. We therefore require $\lambda \lessapprox 1$, however this estimate does not take into account the fact that in very large circuits, some errors may not appear in the causal-cone of the same stabilizers \cite{tran_2023_cones,eddins_2024_cones}. If this condition is not satisfied it is possible to perform PS more often, so that the probability of more than one error appearing in the causal-cone of a stabilizer check is negligible, as shown in Fig.~\ref{fig:6}.

By defining with $\lambda'$ the typical error rate within the causal cone of parity checks, which can be significantly smaller than $\lambda'<\lambda$ depending on the number of stabilizers and parity checks performed, we can set a (loose) upper bound on the bias
\begin{equation}\label{eq:SNT_approx_bias}
\mathrm{Bias}[\mathsf{O}_\mathrm{est.}]
\sim \mathcal{O}\left(  \lambda'^2 \right).    
\end{equation}
In practice, multiple detectable errors can still lead to a detectable error syndrome, reducing the bias well beyond Eq.~\ref{eq:SNT_approx_bias}, up to $\lambda \approx 2$, as indicated by the fits in Fig.~\ref{fig:4}.

\subsection*{\label{subsec:fermionic_encodings} Fermionic encodings for the FHM}

\begin{figure*}[t]
    \centering
    \includegraphics[width=0.9\textwidth]{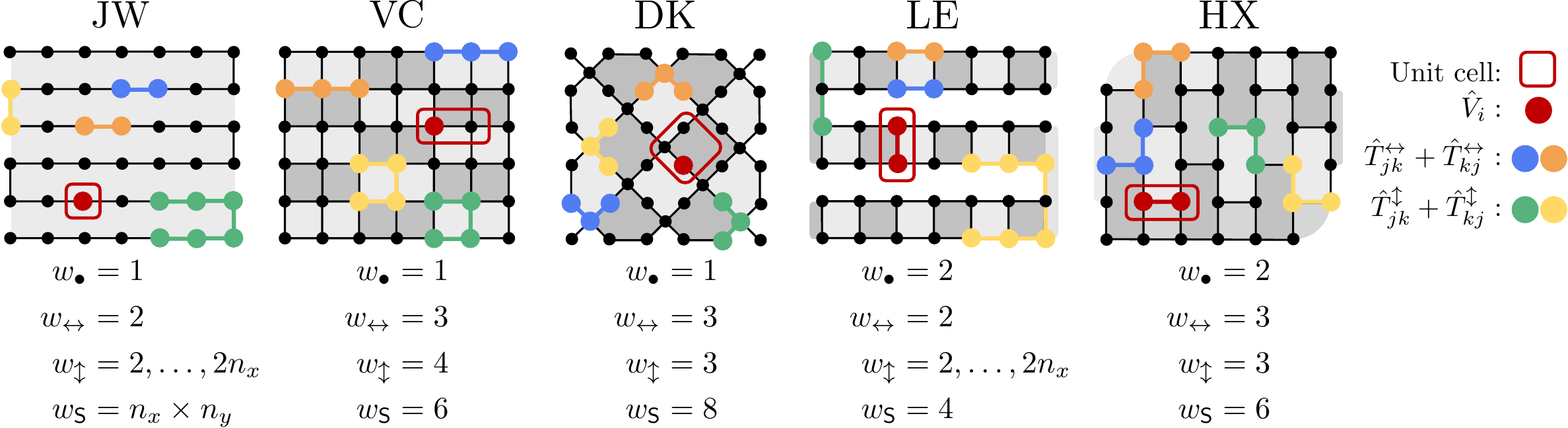}
    \caption{Pictorial representation of the different fermion-to-qubit encodings on an $n_x\times n_y$ qubit lattice studied in this work. Vertex operators are highlighted in red and the surrounding box represents the unit cell of the respective encoding. Orange and blue shapes correspond to the horizontal ($\leftrightarrow$) hopping operators, while the yellow and green shapes correspond to vertical ($\updownarrow$) hopping operators. The weights of the operators are summarized below each encoding. Differently shaded gray plaquettes represent possible stabilizer operators with a maximal weight $w_\mathsf{S}$.}
    \label{fig:7}
\end{figure*}

Our central application in this work is the simulation of the two-dimensional FHM model:
\begin{align}\label{eq:fhm}
\hat{H}_{\text{FH}} & =  - \sum_{\left<i,j\right>, \sigma} t_{ij} \hat{c}^\dagger_{i \sigma}\hat{c}_{j\sigma}^{\phantom{\dagger}}+U \sum_{i} \hat{n}_{i\uparrow}^{\phantom{\dagger}}\hat{n}_{i\downarrow}^{\phantom{\dagger}}
\end{align}
where $\hat{c}^{\dagger}_{i\sigma}$($\hat{c}_{i\sigma}$) creates(annihilates) a fermion with spin $\sigma$ on site $i$, $\hat{n}_{i\sigma} = \hat{c}^{\dagger}_{i\sigma} \hat{c}_{i\sigma}$ is the number operator, $t_{ij}\equiv t$ the nearest-neighbor hopping amplitude and $U$ is the on-site interaction. The total number of fermionic modes is set to be $N=N_x \times N_y$.

For the quantum simulation of fermionic models, the corresponding Hamiltonians must be mapped to spins. The most common such encoding is the Jordan-Wigner (JW) transformation\cite{Jordan1928}. The JW encoding, is one-dimensional by construction and leads to the formation of operators with long Pauli strings in higher dimensions. As an alternative, a number of so-called \emph{local} fermion-to-qubit encodings have been proposed in literature. These introduce additional \emph{ancillary} qubits with the aim of locally resolving anti-commutation relations between operators involved. Below, we will briefly introduce the main features of such encodings. More complete discussions on the subject can be found in Refs.~\cite{Algaba_2024,chien_2022,simkovic_2024}.

Any given local encoding can be defined in terms of a graph constructed using vertex ($\hat{V}_i$) and edge ($\hat{E}_{ij}$) operators defined for a fermionic mode $i$ and a pair of modes $(i,j)$, respectively. We can rewrite the operators from Eq.~\ref{eq:fhm} in terms of these new operators as follows:
\begin{align}
    \hat{n}_j^{\phantom{\dagger}} =& \frac{1}{2}(1-\hat{V}_j) \\
    \hat{c}^\dagger_j \hat{c}_k^{\phantom{\dagger}} + \hat{c}^\dagger_k \hat{c}_j^{\phantom{\dagger}} =& \frac{i}{2}(\hat{V}_k-\hat{V}_j)\hat{E}_{jk} \equiv \hat{T}_{jk} + \hat{T}_{kj} \nonumber
\end{align}

Edge and vertex operators are useful in deriving fermion-to-qubit encodings due to their fairly simple commutation relations, where two operators anticommute if they have exactly one common index, and commute otherwise. In an encoding, all vertex and edge operators that form a given graph must be assigned to Pauli operators acting on qubits in a way that satisfies their mutual commutation relations. It is worth noting, that not every required edge has to be directly defined, since it is possible to compose edges between further-apart vertices within the graph using the composite rule: 
    $\hat{E}_{jl}\hat{E}_{km}=-\hat{E}_{jm}\hat{E}_{kl}$.
Finally, every closed loop of edges defines a stabilizer $\hat{E}_{jl_1}\hat{E}_{jl_2}\dots\hat{E}_{l_M j} = \hat{S} \in \mathbb{S}$, which commutes with all other stabilizers and logical operators. As a consequence, they can be used for the purpose of quantum error mitigation and/or quantum error correction. In the one-dimensional JW encoding, it is possible to identify one such stabilizer by closing the loop of edges spanning the whole system. 

It has been shown that some local fermion-to-qubit encodings posses non-trivial code distance $d>1$, which allows for the detection of errors with weight up to $d-1$ and the correction of errors with weight up to $\lfloor(d-1)/2 \rfloor$. 

Besides JW, we will investigate four additional local encodings. The relative differences between all five encodings are summarized in Table~\ref{tab:encodings} and individual encodings are graphically represented in Fig.~\ref{fig:7}. The first, called ladder encoding (LE) \cite{LE} has a one-dimensional connectivity graph, similar to JW. The main difference to JW is the enlarged Hilbert space with a qubit-to-fermion ratio of $Q_r = 2$. This introduces weight-four stabilizers, keeps the weight of hopping terms ($\hat{T}_{jk}$, $\hat{T}_{kj}$) at two, and increases the weight of vertex operators from one to two. The upshot is that the code distance for this encoding is $2$ rather than $1$ for JW. 

Both these encodings, however, suffer from vertical hopping operator weights scaling with the linear system size, $N_x$. To address this, we also evaluate three additional encodings, VC (Ref.~\cite{verstraete2005mapping,Algaba_2024}), DK \cite{Derby_2021} and HX (Refs.~\cite{Kitaev_2006, chien_2022, simkovic_2024}) all of which have two-dimensional square-lattice edge-vertex graphs. This means that all Hamiltonian operators have constant weight, as illustrated in Fig.~\ref{fig:7}. The VC and HX encodings have stabilizer weights of $6$, and their main difference is their distance ($2$ for HX vs $1$ for VC) and the weights of their vertex and hopping operators. While both VC and DK are distance-$1$ encodings, they differ in the weight of the stabilizers $w_\mathsf{S}$, qubit-to-fermion ratio $Q_r$ as well as the weight of the hopping operators. It should be noted that, despite the distance being $1$, VC and DK contain a number of stabilizers which scale with the system size, and allows for the detection of a large fraction of all single-qubit errors (see Table~\ref{tab:encodings}). 

\subsection*{Numerical Simulations}
In this work we perform two types of numerical simulations: a Monte-Carlo based shot-by-shot simulation of a non-Clifford evolution, and a Monte-Carlo Clifford simulation of even larger systems. Additional data obtained from a density matrix simulation is presented in Supplementary Note 10.

In all simulations, we consider a local Pauli error noise model. The infidelity of a single layer $k$, associated with the unitary $\mathsf{U}_k^\mathrm{C}$, is given by $\varepsilon_k = 1 - \mathcal{F}_\mathrm{TQG}^{N_\mathrm{TQG}^{(k)}}$. Here, $\mathcal{F}_\mathrm{TQG}$ is the two-qubit gate (entanglement) fidelity and $N_\mathrm{TQG}^{(k)}$ is the number of two-qubit gates resulting from the decomposition of the unitary $\mathsf{U}_k^\mathrm{C}$ into the native gate set, consisting of CZ and arbitrary single-qubit rotations. The CSP is then varied by varying $\mathcal{F}_\mathrm{TQG}$. Since it is possible to measure the average gate fidelity in experiment via Randomized Benchmarking protocols \cite{Knill_2008}, we convert the entanglement fidelity to the average gate fidelity according to the formula presented in Ref.~\cite{Nielsen_2002} and present the results in terms of the latter.

The applied noise is local in the sense that it only acts on pairs of qubits between which a TQG is applied, and the probability $p_i^{(k)}$ of experiencing a two-qubit Pauli error compared to the probability of a single-qubit Pauli error is $0.8$. Notably, the same Pauli noise is also applied to the TQGs required to implement the parity checks. As for measurement errors, their impact on the parity checks is studied in Supplementary Note 4. 

As mentioned in the main text, we assume a native connectivity to the encoding, i.e. a connectivity specified by the logical operators illustrated in Fig.~\ref{fig:7} with an additional ancilla qubit used for a parity check in the center of each grey plaquette forming a stabilizer symmetry, which is connected to all the qubits on the edge of the plaquette. The maximal needed connectivity for each encoding is listed in Table~\ref{tab:encodings}.

As a benchmark problem of condensed matter physics, we consider the 1D Fermi-Hubbard Hamiltonian from Eq.~\ref{eq:fhm} with 2 (for the non-Clifford simulations) and 4 sites (for the Clifford simulations), with the parameters $U=4$ and $t = 1$. 

We evolve the initial states $|\!\!\uparrow\downarrow\,\rangle$ and $|\!\!\!\uparrow\downarrow\uparrow\downarrow\,\rangle$ for the non-Clifford and Clifford simulations respectively. Several methods can be used to prepare the desired initial state for a given encoding. Namely, dynamic circuits \cite{gottesman_1997_phd_thesis}, general unitary encodings \cite{gottesman_1997_phd_thesis} as well as ad-hoc strategies \cite{Higgott_2021}. The optimal choice of state preparation algorithm depends on the ability to perform dynamic circuits and their performance, and whether an ad-hoc strategy exists for the considered encoding. Determining the optimal state preparation procedure is outside of the scope of this work and we thus assume perfect state preparation in our simulations, focusing entirely on the effects of error happening during the subsequent time evolution.

In the non-Clifford simulations the state was evolved up to time $T=0.5$ with $N_\mathrm{Trotter} = 10$ steps. Similarly, the Clifford simulation circuits were obtained by rounding the angles of the non-Clifford single-qubit rotations in Eq.~\ref{eq:simple_circuit_unitary} to zero. As in the non-Clifford case, we perform the simulations with $10$ Trotter steps with a single-parity check round, and additionally with $12$ Trotter steps and mid-circuit parity check rounds. The parity check rounds are spaced evenly after a given number of Trotter steps. 

The obtained circuit depth is much larger compared to the number of qubits involved. This makes the \textit{total} circuit success probability a meaningful metric. Moreover, the specific choice of observables and related causal-cone arguments will not significantly affect the bias of the computed observables \cite{tran_2023_cones, eddins_2024_cones}. Throughout the paper, we consider the single-site occupations $\hat{n}_i^\sigma$ as the set of observables $\mathbb{O}$, i.e. 
\begin{equation}\label{eq:set_of_observables}
    \mathbb{O} = \{ \hat{n}_i^\sigma \, | \, \sigma \in \{\uparrow, \downarrow \},\,  i = 1,\dots,N  \},
\end{equation}
based on which the averaged (squared) bias is computed.
Nonetheless, further numerical results investigating the effect of the weight of the evaluated observable are provided in Supplementary Note 6.

All together, we perform simulations on 2 (JW, non-Cliff.) to 16 qubits (VC, LE or HX, Cliff.), not counting ancilla qubits used for stabilizer measurements. The non-Clifford data (in Fig.~\ref{fig:4}) was extracted from $1.5 \cdot 10^5$ shots, the (single parity check) Clifford data from $3.8 \cdot 10^5$ shots (in Supplementary Fig. S2 and ``Cost of SNT''), and the mid-circuit parity check Clifford data (in Fig.~\ref{fig:6}) from $4\cdot10^5$ shots. In all cases $N_\mathrm{shots} = N_\mathrm{circuits}$. Additionally, the PS and SV bias for the first column in Fig.~\ref{fig:5} were extracted from $6\cdot10^5$ shots, to better resolve the bias.

\subsection*{QEM Performance Measure}

In order to compare the performance of various QEM techniques, we employ the root-mean-squared error measure \cite{Cai_2023_QEMreview}, which takes into account the bias as well as the cost of a QEM technique. Indeed, assuming that the error-mitigated estimate $\mathsf{O}_\mathrm{est.}$ is normally distributed, with mean $\langle \mathsf{O}_\mathrm{est.} \rangle$ and variance $\mathrm{Var}[\mathsf{O}_\mathrm{est.}]$, the mean-squared-error is given by 
\begin{equation}
\begin{split}
\mathrm{MSE}[\mathsf{O}_\mathrm{est.}] &= \langle (\mathsf{O}_\mathrm{est.} - \mathsf{O})^2 \rangle \\&= (\langle \mathsf{O}_\mathrm{est.} \rangle - \langle \mathsf{O} \rangle)^2 +  \mathrm{Var}[\mathsf{O}_\mathrm{est.}] \\&= \mathrm{Bias}[\mathsf{O}_\mathrm{est.}]^2 + \mathrm{Var}[\mathsf{O}_\mathrm{est.}].   
\end{split}  
\end{equation}
Here, the variance $\mathrm{Var}[\mathsf{O}_\mathrm{est.}]$ is proportional to the squared cost $C^2$ of the QEM technique employed, which amplifies the statistical uncertainty due to the finite amount of shots $N_\mathrm{shots}$ and circuits $N_\mathrm{circuits}$ available. The exact expression of the variance in the general scenario $N_\mathrm{shots} \geq N_\mathrm{circuits}$ is derived and commented in Supplementary Note 7 and 8, where we also present a more efficient circuit sampling strategy which reduces $\mathrm{Var}[\mathsf{O}_\mathrm{est.}]$ for PEC and SNT. 

Given the set of observables $\mathbb{O}$, we then compute the RMSE-based metric 
\begin{align}\label{eq:RMSE_observable_avg}
    \mathrm{RMSE} &= \sqrt{\frac{1}{\mathbb{O}} \sum_{\mathsf{O}_{i} \in \mathbb{O}} \mathrm{MSE}[\mathsf{O}_{i,\mathrm{est.}}]  }.
\end{align}
to assess the QEM performance. 

\subsection*{Cost of SNT}

\begin{figure}[t]
    \centering
    \includegraphics[width=.49\textwidth]{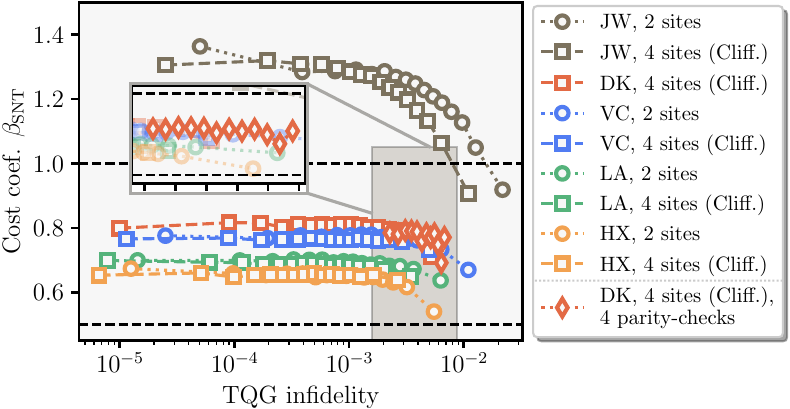}
    \caption{The coefficient $\beta_\mathrm{SNT}$ as defined in Eq.~\ref{eq:cost_fhm}, extracted by computing $\beta_\mathrm{SNT} = \log(C_\mathrm{SNT})/\lambda$, where the QEM cost $C$ is extracted from the simulations according to Eq.~\ref{eq:cost_fhm}. The plot includes three sets of data - a non-Clifford simulation (circles) is of 2 fermionic sites, while the Clifford simulation (squares) is of a 4-site chain with all the non-Clifford angles set to zero. Additionally, the diamonds (see inset), mark the cost obtained from simulations with three parity check rounds. The dashed lines indicate values of $\beta = 0.5$ and $\beta = 1$ corresponding to the techniques listed in Table~\ref{tab:qem_methods}. }
    \label{fig:8}
\end{figure}

Let us now present the numerical data supporting the scaling of the cost of SNT, as quoted in Table~\ref{tab:encodings}. For the JW encoding we have formally proven in Supplementary Note 11 that with the two stabilizers $\mathbb{S}_{\uparrow/\downarrow} = \{ \mathsf{S}_\uparrow, \mathsf{S}_\downarrow\}$, we expect $R \sim 67\% - 75\%$, where the lower limit applies in the local noise regime and the larger value in the limit of a global depolarizing channel. This results in the cost coefficient $\beta_\mathrm{JW+SNT} \approx 1.25-1.33$, which is consistent with the extracted numerical value from the simulations of $\beta_\mathrm{JW+SNT} \approx 1.3$, as seen in Fig.~\ref{fig:8}. 

Fig.~\ref{fig:8} displays $\beta_\mathrm{SNT}$ extracted from the simulations for two different system sizes. We can see that as soon as $\varepsilon_k \ll 1$, the values for $\beta_\mathrm{SNT} $ quickly stabilize to very similar values for both system sizes, and in both the Clifford and non-Clifford simulations. The reasons for lower values of $\beta_\mathrm{SNT}$ in the more noisy regime are the first order assumptions made in the derivation of the method. If the first order assumptions are valid, all the noise is perfectly mitigated and the bias is close to zero. Beyond this approximation, both SV and PEC will not compensate for all of the noise. SV, for example, will suffer from the combination of two detectable errors becoming undetectable. These effects result in a higher bias (see Fig.~\ref{fig:4}), but also in a lower cost. Additionally the pure exponential approximation $C \approx e^{\beta_\mathrm{SNT} \lambda}$ is also valid only for $\varepsilon_k \ll 1$. This is why $\beta_\mathrm{SNT}$ in the noiseless limit $\mathrm{CSP}\rightarrow 1$ is considered as a more truthful descriptor of the cost of SNT, even though lower values may be achievable in practice. 

More specifically we extract $\beta_\mathrm{VC+SNT} = 0.77$, $\beta_\mathrm{DK+SNT} = 0.81$, $\beta_\mathrm{LE+SNT} = 0.70$ and $\beta_\mathrm{HX+SNT} = 0.66$. If we only take into account the data points with $1 - \mathcal{F}_\mathrm{TQG}<10^{-3}$, the standard deviation of the points for the local encodings is on the order of $10^{-2}$ and no statistically significant difference between the two data sets is observed. Additionally, the average fraction of noise detected by the local stabilizers via PS and later PP can be extracted from the data for $\beta_\mathrm{PS}$ and $\beta_\mathrm{PP}$, by using the relations $C_\mathrm{PS} = \exp(R_\mathrm{PS}\lambda/2)$ and $C_\mathrm{SV} = C_\mathrm{SV}^{(\mathrm{PS})} C_\mathrm{SV}^{(\mathrm{PP})} = 1.5\exp(R_\mathrm{PS}\lambda/2)\exp(R_\mathrm{PP}\lambda)$. The contributions of SV and PEC to the total SNT cost were calculated based on the numerical values of $C_\mathrm{SNT}^{(\mathrm{PEC})} $ and $C_\mathrm{SNT}$, together with Eq.~\ref{eq:cost_fhm}. The results are listed in Table~\ref{tab:encodings}. In all cases (also when determining $\beta_\mathrm{SNT}$), the considered $\lambda$ does not include the errors in the parity checks.
Indeed, in Supplementary Note 4, we show that measurement errors on the ancillas do not contribute to $\beta_\mathrm{SNT}$, but act as a pre-factor in the final cost, and should be considered separately.

\section*{Data Availability}
The data supporting the findings of this article is available at {\tt https://doi.org/10.5281/zenodo.17660898}.

\section*{Code Availability}
The code supporting the findings of this article is available from the corresponding authors upon reasonable request.

\begin{acknowledgments}
No funding was received for this research. We thank all employees at IQM Quantum Computers for their insightful discussions, especially Francisco Revson F. Pereira, Stéphanie Cheylan and Martin Leib. We would also like to thank Sebastian Paeckel for the additional insights.
\end{acknowledgments}

\section*{Author Contributions}
M.P. developed the SNT algorithm, performed the circuit simulations, analyzed the resulting data and wrote the first manuscript draft. M.G.A. generated the circuits used in the simulations by implementing the presented encodings. E.G.-R. assisted in the simulation of the circuits. A.A. and I. de V. supervised the work. A.C and F.Š. conceptualized and supervised the project. All authors read, revised and approved the final manuscript.

\section*{Competing Interests}
The authors declare the following competing interests: The Subspace Noise Tailoring algorithm described in this manuscript is the subject of patent application FI20240044. All authors are listed as inventors on the application, which was filed by IQM Finland Oy and is pending at the time of the submission of this manuscript. The authors declare no further competing interests.

\clearpage
\newpage
\appendix
\begin{widetext}
\begin{center}
{\Large\bfseries Supplementary Information:}\\[0.5em]
{\Large Near-Term Fermionic Simulation with Subspace Noise Tailored Quantum Error Mitigation}\\[0.5em]

\end{center}
\end{widetext}
\newcounter{suppnote}
\newcommand{\suppsection}[1]{%
  \refstepcounter{suppnote}%
  \section*{Supplementary Note \thesuppnote:\\ #1}%
}

\renewcommand{\figurename}{SUPPLEMENTARY FIG.}

\setcounter{figure}{0}
\renewcommand{\thefigure}{S\arabic{figure}}
\renewcommand{\theHfigure}{S\arabic{figure}}
\suppsection{Classsical Simulations of the FHM}

\subsection*{Exact Diagonalization}

Exact Diagonalization of the FHM using GPUs was first demonstrated by Ref.~\cite{Siro_2012} on a $4\times 4$ lattice at half-filling. However, with the recent improvement in both GPU performance as well as the advent of GPU clusters, larger lattice sizes should be within reach. Nonetheless, an estimate of the memory requirements for the storage of a single Lancosz vector on a $5\times 5$ lattice at half-filling with 64 Bit precision is on the order of 100 TB, which is prohibitive under the assumption of 80 GB memory per GPU and 100s of available GPUs. Assuming a memory budget of 10 TB on a GPU cluster, 16 to 18 electrons on a $5\times 5$ lattice (with symmetry restrictions) could be stored on the cluster with multi-GPU communication and orthogonalization becoming the computational bottleneck. These estimates show that with a sufficient amount of computational resources and intermediate particle numbers below half-filling, exact diagonalization could be performed on lattice sizes of up to $5\times 5$.

\subsection*{Tensor Networks}
To date, the largest simulations of the time evolution of a spinful FHM were demonstrated using a circuit DMRG approach \cite{Schollwoeck_2011_dmrg,Paeckel_2019}. More specifically, Ref.~\cite{thompson2025_fermioniq} was able to evolve a $5\times6$ lattice encoded using the JW transformation up to a time $T=1$ by extrapolating the results from several noisy simulations. The simulations of Ref.~\cite{thompson2025_fermioniq} were performed on a single GPU and were limited by memory rather than compute time. The same simulation algorithm was later applied to a $6\times 6$ lattice with a local encoding closely related to the DK encoding \cite{Derby_2021}. Given access to a cluster of GPUs, the estimate of the CSP for the evolution up to time $T\approx 0.7$ (with 2 Trotter steps) was deemed too low to be able to perform any meaningful extrapolation at this lattice size \cite{granet2025}. Further estimates also revealed that even a hypothetical more efficient fermionic circuit DMRG algorithm is not expected to reach the required accuracy \cite{granet2025}. 

Time-dependent variational principle (TDVP), direct circuit contraction, and fermionic matrix product states were also benchmarked on the problem of the time evolution of a FHM in Ref.~\cite{alam2025}. With access to 8 CPUs with 12 GB of RAM each, the TDVP together with infinite bond dimension extrapolation was shown to be the most efficient. For the largest considered bond dimension, a FHM on a $4\times 7 $ lattice was evolved up to $T=2$ and the classical simulation was run for 14 days. Since the obtained results are in quantitative disagreement with exact methods for $U=0$, the circuit DMRG approach of \cite{thompson2025_fermioniq} is to date the most performant tensor-network based method for this type of problem documented in the literature. 

Additionally, 2D tensor-network methods have shown promising results in recent years, especially for the simulation of the time evolution of the Transverse Field Ising model (TFIM). Nonetheless, while the total number of qubits simulated is larger compared to the implementation of \cite{thompson2025_fermioniq} the success of the approaches of both \cite{liao2023} and \cite{Tindall_2024} was greatly aided by the low connectivity of the considered heavy hex TFIM lattice, which is significantly smaller compared to the standard square lattice we consider for the FHM. Furthermore, the TFIM evolution circuits from \cite{IBM_2023_utility} are relatively shallow, with observable causal cones covering a significantly smaller portion of the total circuit volume. Since no equivalent results regarding the time evolution of the FHM were published to the best of our knowledge, the ultimate reach of the approaches of \cite{liao2023} and \cite{Tindall_2024} cannot be determined. As a brute comparison we note that the quantum circuit classically simulated by \cite{liao2023} and \cite{Tindall_2024} contains 5760 entangling gates on 127 qubits, while a time evolution circuit of a $6\times 6$ FHM with 15 Trotter steps contains 7830 entangling gates on 108 qubits (using the DK encoding). The significantly larger number of entangling gates together with the lower number of qubits and higher connectivities suggest that the FHM evolution could be able to build up significantly more entanglement and thus prove to be less tractable for tensor-network methods.

Ref.~\cite{Yoshioka_2024} did compare the efficiency of a 2D and 1D tensor-network method for a related problem -- the calculation of the ground state energy of both the Heisenberg model and FHM. The authors considered Projected Entangled Pair States and DMRG, and DMRG was shown to be the more efficient choice for both models.

\subsection*{Path Truncation}

Recently several algorithms based on truncation methods \cite{rudolph2025} have emerged as promising candidates for the classical simulation of quantum dynamics and have successfully been used to simulate the TFIM time evolution on a heavy-hex lattice \cite{rudolph2023}.  

However, in the context of fermionic simulation, a similar, but more efficient approach referred to as Majorana propagation was recently proposed by Ref.~\cite{miller2025}. By performing the Heisenberg evolution of the observable in the native fermionic frame, without the need for a fermion-to-qubit encoding, the proposed algorithm is more efficient for two reasons: (a) by avoiding the use of a fermion-to-qubit encoding, there is no need to introduce ancilla qubits and (b) unlike the JW encoding, every term in the Hamiltonian will remain local, thus also reducing the overall number of entangling gates required. 

Majorana propagation was concurrently applied to the time evolution of the FHM by \cite{alam2025} and \cite{danna2025}. Ref.~\cite{danna2025} proves that Majorana propagation with a Gaussian Hamiltonian will preserve the weight of the observable $w$, which has two implications: (a) Majorana propagation is efficient for low- or very high-weight observables and (b) the non-Gaussianity of the Hamiltonian (rather than the physically less relevant Clifford structure) is a measure of the algorithm complexity. Ref.~\cite{danna2025} was also able to obtain qualitatively accurate results for a low-weight observable on a $7\times 7$ lattice -- the largest to date. While the computational resources were not disclosed, the simulation was performed by truncating Majorana strings with weights above $S > 6$ or prefactors below $\varepsilon < 10^{-6}$. These results promote Majorana propagation to one of the most efficient classical techniques for this computational task, however it is severely limited in the choice of observable. To illustrate this point, consider that even if the truncations in \cite{danna2025} were performed solely based on the weight of the Majorana string $S$ (and not its prefactor $\varepsilon$), the dimension of the explored Majorana-string space would be given by $\sum_{i = 0}^{6}\binom{2 \times2 \times 7\times 7}{i}\sim 10^{10}$. Conversely, even if the observable weight remained fixed, the dimension of the full Majorana-string space of a weight-8 observable on a $6\times 6$ lattice would be $ \binom{2 \times2 \times 6\times 6}{8} \sim 10^{12} $, implying that the estimation of an observable with a Majorana string weight of 8 on a $6\times 6$ lattice could prove to be more challenging, even in the Gaussian case, compared to a weight-4 observable on a $7\times 7$ lattice. In the worst case, even for Gaussian Hamiltonians, calculating the expectation value of an observable with weight $2N^2$ on an $N\times N$ lattice requires storing $\binom{4N^2}{2N^2} \sim 2^{4N^2}/N$ Majorana strings. This increase in computational complexity was numerically observed by \cite{alam2025} where the Majorana propagation algorithm was able to estimate single-site observables, but failed to converge for more complex observables at a lattice size of $4 \times 7$.

\subsection*{Neural Quantum States}

Additionally, time dependent neural quantum state based approaches have recently emerged as promising candidates for the simulation of correlated materials. Specifically, neural quantum states have recently been used to calculate the ground state energy of the FHM with the method reaching a lattice size of up to $4\times4$ \cite{Ibarra_Garcia_Padilla_2025}. Nonetheless, no calculations involving the time evolution of the FHM using neural quantum states have been reported in the literature to date. The main difficulty involved in such algorithms is the exponential scaling of the number of neural network parameters with time \cite{Lin_2022}. While time dependent neural quantum states have been used to evolve other 2D lattice models \cite{Lange_2024}, there have been no demonstrations of real time evolution of the Fermi-Hubbard model thus far. Moreover, the stability of the commonly used time-dependent Variational Monte Carlo method is known to be strongly dependent on the chosen variational ansatz thus hindering our ability to extrapolate the performance of such methods to the FHM.

\subsection*{Summary}

Based on the above presented arguments, the largest reported classical simulation of the time evolution of 2D FHM consist of extrapolated DMRG results for a $5\times 6$ lattice \cite{thompson2025_fermioniq} and the Majorana propagation results for a $7\times 7$ lattice \cite{danna2025}. Unlike the latter method which is severely limited in the weight of the observables that can be simulated, the MPS ansatz used by \cite{thompson2025_fermioniq} can be used to estimate arbitrary (unentangled) observables. 

In order to account for further optimizations of the algorithm presented in \cite{thompson2025_fermioniq} and potential access to even more GPUs, we will assume that a lattice size of $6\times 6$, while challenging, \emph{could} be within reach of classical computation in the future. While current quantum computers are limited in the number of Trotter steps $N_\mathrm{Trotter}$, tensor-network methods are restricted by the entanglement growth of the system \cite{Schollwoeck_2011_dmrg,Paeckel_2019}, which is related to the total evolution time $T$. Assuming that a classical simulation is able to evolve a FHM with $t, U \sim \mathcal{O}(1)$ up to time $T\sim 1$ a quantum computer would require a Trotter step size $\Delta t \ll T$ in order to avoid significant Trotter errors. In order to avoid Trotter errors significantly larger than the 5\% RMSE limit, we choose $\mathcal{O}(1)/5\% \sim 15$ Trotter steps in this setting. We have thus identified an optimistic limit of the classical simulability of the time evolution of a FHM in terms of the quantities plotted in Fig.~1. Since larger lattice sizes could also be simulated, albeit for shorter times, we note that the complexity of a tensor-network simulation (at a fixed bond dimension) is linear in the number of qubits thus implying an \emph{approximate} quadratic decay of the number of achievable Trotter steps beyond $6\times 6$.

\suppsection{Error Classification with General Decompositions}
The noise classification algorithm presented in the main text assumed a specific structure of the decomposed logical operator $\exp{(i\theta \mathsf{P})}$. The decompositions employed to obtain this form are completely general \cite{Cowtan_2020, Sriluckshmy_2023, Algaba_2024} and the presented noise classification algorithm has several favorable properties, namely: (a) The Pauli noise assumptions are justified with the use of Randomized Compiling, (b) the noise characterization can be performed in a scalable manner on a layer of gates.\\

For the sake of completeness, here we show that the proposed error classification can be performed without the need to invoke \emph{any} decomposition, but directly at the level of the circuit unitary, provided that the noise can be described as a Pauli channel acting before or after every logical operator. However in practice, this requirement can only be generally guaranteed in certain cases. \\

Some guarantees (similar to point (a)) can be obtained if the decomposition is symmetric, i.e. it has the form Clifford layer -- arbitrary unitary operator -- Clifford layer. Compared to the decomposition used in the main text, we no longer require that a Pauli error strictly commutes or anti-commutes with the central unitary. In this case the noise of both Clifford layers can be shaped and characterized identically as described in the main text. On the other hand the noise of the central arbitrary unitary must be either neglected or shaped using pseudo-twirling \cite{santos_2024}. Nonetheless, it must be noted that pseudo-twirling is not exact and the effective noise channel is not guaranteed to be completely Pauli, which could result in additional SNT bias. Under these conditions, the noise can be described as a Pauli channel acting before or after every logical operator, similarly to point (a).\\

As mentioned in the main text, cycle benchmarking and modifications thereof can be used to obtain a scalable noise model of a Clifford layer \cite{Erhard_2019_CB, flammia_2022_ACES, carignandugas2023, calzona_2024, van_den_Berg_2024,pelaez_2024_ACES, hockings_2024_ACES, chen_2025}. In this way we are able to efficiently characterize and mitigate both context and gate-dependent errors. However, if we want to characterize the noise of a non-Clifford operation, we are forced to utilize completely general characterization procedures (such as e.g. Gate Set Tomography \cite{Nielsen_2021}), which only characterize the noise on a single gate, rather than layer of gates. In this case we are therefore unable to fully characterize and mitigate any context-dependent errors induced by the non-Clifford central unitary. However, if such errors are not present in the device, the noise can be efficiently characterized (see point (b) in the first paragraph). \\

If we therefore assume that the noise in the circuit can be described as a Pauli channel acting before or after every logical operator and that these Pauli channels can be characterized, we can prove a more general version of the classification algorithm. We start with the completely general form of the unitary in Eq.~1 in the main text, before any decompositions have been applied
\begin{equation}
    \mathsf{U} = \prod_{k=1}^{N_\ell} e^{-i\theta_k \mathsf{M}_k},
\end{equation}
where $\mathsf{M}_k$ is a multi-qubit Pauli operator and $N_\ell$ is the number of logical operators in the algorithm. Assuming that a Pauli error $\mathsf{P}_i$ occurs after the logical operator indexed with $l$, or equivalently before the logical operator indexed with $l+1$, we can again use Eq.~10 in the main text to show that
\begin{align}
    \mathsf{U}_i^{(l)} &= \prod_{k = l+1}^{N_\ell} e^{-i\theta_k \mathsf{M}_k} \mathsf{P}_i \prod_{k = 1}^{l} e^{-i\theta_k \mathsf{M}_k} \nonumber \\
    &= \mathsf{P}_i \prod_{k = l+1}^{N_\ell} e^{\pm i\theta_k \mathsf{M}_k}  \prod_{k = 1}^{l} e^{-i\theta_k \mathsf{M}_k} \nonumber \\
    &= \mathsf{P}_i \mathsf{U}_i^{(l:N_\ell)}.
\end{align}
The above result has the exact same form as Eq.~11 in the main text meaning that all the related discussion still applies. Therefore, error $\mathsf{P}_i$ occurring after logical operator $l$ or before logical operator $l+1$ is \textit{un}detectable iff
\begin{equation}\label{eq:general_detectablity}
    [ \mathsf{P}_i , \mathsf{S}_j] = 0 \,\, \forall \mathsf{S}_j \in \mathbb{S}.
\end{equation}
Otherwise, the error $\mathsf{P}_i$ is \emph{detectable}. This condition is equivalent to Eq.~3 in the main text and provides a basis for the noise classification procedure required for the implementation of SNT in the more general scenario.

\suppsection{\label{app:sv_via_pp_cost} Cost of Symmetry Verification via Post-Processing}

In the main text, we make use of two additional global symmetries in the FHM, referred to as the spin-parity preservation, to perform an additional SV step via PP. Using the same notation, where $\mathsf{M}_{\mathbb{S}_{\uparrow / \downarrow}}$ is the projector onto the valid subspace and $\mathbb{S}_{\uparrow / \downarrow} = \{\mathsf{S}_\uparrow,  \mathsf{S}_\downarrow\}$ are the corresponding stabilizers, the QEM estimator is constructed as

\begin{align}\label{eq:PP_estimator}
    \langle \mathsf{O}_\mathrm{est.}\rangle &= \frac{\langle \mathsf{O} \mathsf{M}_{\mathbb{S}_{\uparrow / \downarrow}} \rangle}{\langle\mathsf{M}_{\mathbb{S}_{\uparrow / \downarrow}} \rangle} =  \frac{\langle \mathsf{A} \rangle }{\langle\mathsf{B}\rangle},
\end{align}

where we have denoted $\langle \mathsf{A} \rangle= \langle \mathsf{O} \mathsf{M}_{\mathbb{S}_{\uparrow / \downarrow}} \rangle = (\langle \mathsf{O} \rangle +\langle \mathsf{O}\mathsf{S}_\uparrow \rangle  + \langle \mathsf{O}\mathsf{S}_\downarrow \rangle  + \langle \mathsf{O}\mathsf{S}_\uparrow \mathsf{S}_\downarrow \rangle  )/4$ and $\langle \mathsf{B} \rangle= \langle  \mathsf{M}_{\mathbb{S}_{\uparrow / \downarrow}} \rangle = (1 +\langle \mathsf{S}_\uparrow \rangle  + \langle \mathsf{S}_\downarrow \rangle  + \langle \mathsf{S}_\uparrow \mathsf{S}_\downarrow \rangle  )/4$. It is evident that Eq.~\ref{eq:PP_estimator}, involves evaluating a number of observables, which in the most general case cannot be measured in a single circuit run. 

It is well established that the upper bound for the cost of SV via PP scales as $ e^{\lambda}$ \cite{Cai_2021,Cai_2023_QEMreview} with the bound being  saturated when all the noise is detectable. However, as we will demonstrate in the remainder of this section, the fact that several circuit runs are required will result in a constant multiplicative prefactor $\alpha$ in the cost, i.e. $C_\mathrm{PP}\leq \alpha e^{\lambda}$.

In order to derive the additional prefactor $\alpha$ we start with the variance of the estimator from Eq.~\ref{eq:PP_estimator}, which is given by \cite{Cai_2021}:

\begin{equation}\label{eq:var_O_est_PP}
    \mathrm{Var}[\mathsf{O}_\mathrm{est.}] = \frac{1}{\langle \mathsf{B}\rangle^2} \left(\mathrm{Var}[\mathsf{A}] + \langle \mathsf{O}_\mathrm{est.}\rangle^2 \mathrm{Var}[\mathsf{B}]\right),
\end{equation}

where the factor $1/\langle \mathsf{B}\rangle$ is upper bounded by $e^{\lambda}$ and the prefactor $\alpha$ that we are interested in is therefore related to the second term in the expression. We have omitted the term $\mathrm{Cov}[\mathsf{A},\mathsf{B}]$, since we are interested in the general case where $\mathsf{A}$ and $\mathsf{B}$ are not sampled from the same circuit runs. For any Pauli operator $\mathsf{O}$, we can bound the expression by

\begin{equation}
    \mathrm{Var}[\mathsf{O}_\mathrm{est.}] \leq e^{2\lambda} \left(\mathrm{Var}[\mathsf{A}] + \mathrm{Var}[\mathsf{B}]\right) = \alpha^2 e^{2\lambda}. 
\end{equation} 

In the specific case of $\mathbb{S}_{\uparrow / \downarrow} = \{\mathsf{S}_\uparrow,  \mathsf{S}_\downarrow\}$, the four expectation values needed for the estimation of $\mathsf{A}$ require four independent circuit runs. On the other hand, all the expectation values needed for the estimation of $\mathsf{B}$ can be sampled from the same circuit, since the stabilizers act non-trivially only on non-overlapping sets of qubits. By labeling the number of shots invested into each of the four required circuits needed to estimate $\langle \mathsf{A} \rangle$ as $n_\mathsf{A}$ and the number of invested shots for the single circuit run needed to estimate $\langle \mathsf{B} \rangle$ as $n_\mathsf{B}$, we can write
\begin{widetext}
\begin{align}
    \mathrm{Var}[\mathsf{A}] &= \frac{1}{16} \left( \mathrm{Var}[\mathsf{O}] + \mathrm{Var}[\mathsf{O}\mathsf{S}_\downarrow] + \mathrm{Var}[\mathsf{O}\mathsf{S}_\uparrow] + \mathrm{Var}[\mathsf{O}\mathsf{S}_\uparrow\mathsf{S}_\downarrow] \right)\leq \frac{1}{16}\frac{4}{n_\mathsf{A}}, \label{eq:varA_bound} \\
    \mathrm{Var}[\mathsf{B}] &= \frac{1}{16} \left( \mathrm{Var}[\mathsf{S}_\downarrow] + \mathrm{Var}[\mathsf{S}_\uparrow] + \mathrm{Var}[\mathsf{S}_\uparrow\mathsf{S}_\downarrow] + 2 \mathrm{Cov}[\mathsf{S}_\uparrow, \mathsf{S}_\uparrow \mathsf{S}_\downarrow]+ 2 \mathrm{Cov}[\mathsf{S}_\downarrow,\mathsf{S}_\uparrow \mathsf{S}_\downarrow ]+ 2 \mathrm{Cov}[\mathsf{S}_\uparrow,  \mathsf{S}_\downarrow]\right) \nonumber \\
    &= \frac{1}{16 n_\mathsf{B}} (1 - \langle \mathsf{S}_\downarrow\rangle^2 + 1- \langle \mathsf{S}_\uparrow\rangle^2 + 1- \langle \mathsf{S}_\uparrow\mathsf{S}_\downarrow\rangle^2 \label{eq:varB}\\
    &+ 2 \langle \mathsf{S}_\downarrow\rangle - 2 \langle \mathsf{S}_\uparrow\rangle\langle \mathsf{S}_\uparrow\mathsf{S}_\downarrow\rangle + 2 \langle \mathsf{S}_\uparrow\rangle - 2 \langle \mathsf{S}_\downarrow\rangle\langle \mathsf{S}_\uparrow\mathsf{S}_\downarrow\rangle + 2\langle \mathsf{S}_\uparrow\mathsf{S}_\downarrow\rangle - 2 \langle \mathsf{S}_\uparrow\rangle\langle\mathsf{S}_\downarrow\rangle ). \nonumber 
\end{align}
\end{widetext}
Moreover, $n_\mathsf{A}$ and $n_\mathsf{B}$ represent the effective number of shots defined in following sections (see Eq.~\ref{eq:N_eff_definition}), and are therefore related to the number of available circuits. The fact that a single run is sufficient to estimate $\mathsf{B}$ results in the additional covariance terms in the expression, since the samples are no longer uncorrelated. We have further used the following properties of the stabilizer operators: $\mathsf{S}_\updownarrow^2 = \mathsf{I}$ and $[\mathsf{S}_\uparrow, \mathsf{S}_\downarrow] = 0$. While we are able to straightforwardly obtain the bound for $\mathrm{Var}[\mathsf{A}]$ in Eq.~\ref{eq:varA_bound}, the bound for $\mathrm{Var}[\mathsf{B}]$ requires additional calculations. 

More specifically, the expression in the brackets of Eq.~\ref{eq:varB} can be rewritten as a multivariate polynomial
\begin{align}
    p(x,y,z) &= 3  - x^2 - y^2 - z^2 \nonumber \\
    &-2xy - 2 xz - 2yz + 2x + 2y + 2z .
\end{align}
Solving $\partial p(x,y,z)/\partial x = 0$, $\partial p(x,y,z)/\partial y = 0$, $\partial p(x,y,z)/\partial z = 0$, results in an underdetermined linear system of equations with the solutions specified by the plane $x + y + z = 1$. Plugging this result back into the original polynomial we obtain $p(x,y, 1- x-y) = 4$. Since the Hessian is negative semi-definite and all higher order derivatives are zero, the extremal value is considered a maximal surface, and therefore $\mathrm{Var}[\mathsf{B}] \leq 1/(4n_\mathsf{B})$

All in all, given an (effective) shot budget of $N_\mathrm{shots} = 4n_\mathsf{A} + n_\mathsf{B}$, we can minimize the upper bound for the variance of the estimator by solving the constrained optimization problem with the Lagrangian $\mathscr{L} = 1/(4 n_\mathrm{A}) + 1/(4 n_\mathrm{B}) - \lambda' (4n_\mathsf{A} + n_\mathsf{B} - N_\mathrm{shots})$, with $\lambda'$ being the Lagrange multiplier constant. Solving $\partial \mathscr{L}/\partial n_\mathsf{A}= 0$, $\partial \mathscr{L}/\partial n_\mathsf{B}= 0$ and $\partial \mathscr{L}/\partial \lambda' = 0$ yields the optimal values $n_\mathsf{A} = N_\mathrm{shots}/6$ and $n_\mathsf{B} = N_\mathrm{shots}/3$. Plugging these values back into Eq.~\ref{eq:var_O_est_PP} results in
\begin{equation}
    \mathrm{Var}[\mathsf{O}_\mathrm{est.}] \leq  \frac{9}{4} \frac{e^{2\lambda}}{N_\mathrm{shots}} 
\end{equation}
and therefore the constant prefactor associated with PP is $\alpha = 3/2$. The value $3/2$ is an upper bound, and could be further decreased if information about the variances of all the observables comprising $\langle \mathsf{O}_\mathrm{est.}\rangle$ is available. Nonetheless, the exact value of $\alpha$ is relevant only in the high-fidelity regime, where the exponential scaling of the cost is less detrimental.

\suppsection{Effect of Ancillary Qubit Measurement Errors}

\begin{figure}[h]
    \centering
    \includegraphics[width=0.47\textwidth]{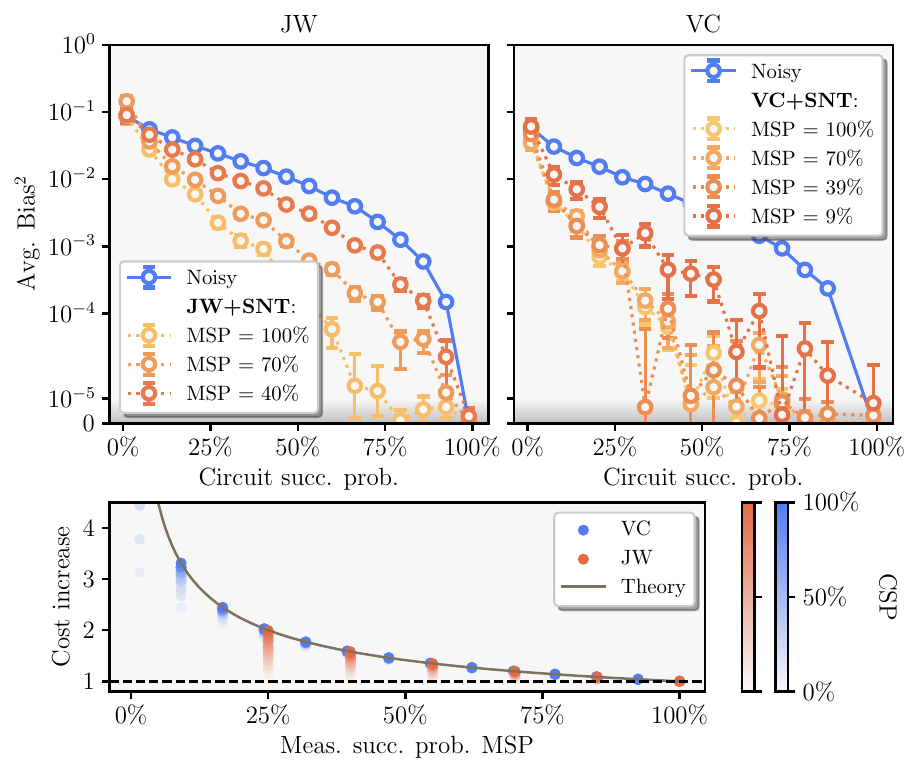}
    \caption{The effect of measurement errors on the bias and cost of SNT. The top row represents the average squared bias of SNT with varying (yellow to orange) measurement errors applied to the ancilla qubits and, for reference, the bias when no QEM is used (blue). The bottom row plots the increase in the cost of the error mitigation when measurement errors are present, i.e. we plot the quantity $C_\mathrm{SNT}^\mathrm{meas.} = C_\mathrm{SNT}[\mathrm{MSP}]/C_\mathrm{SNT}[100\%]$. The red points correspond to the JW encoding (also upper left) and the blue points to the VC (also upper right). The color intensity of the points is proportional to the CSP, i.e. darker points correspond to circuits will less errors. }
    \label{fig:meas_errors}
\end{figure}

While we have focused our attention on the effects of two-qubit gate errors in the main text, we now extend our analysis to readout errors. As mentioned in the main text, readout error mitigation \cite{Bravyi_2021_REM, Cai_2023_QEMreview} can be applied to the data qubits, meaning that the effects of readout errors on ancilla qubits is our primary concern.

In order to better understand the consequences of imperfect readout, we will look at the same problem as in Fig.~4 in the main text, but by performing SV exclusively via PS for the two distance-1 encodings. Measurement errors on the ancilla qubits are modeled by stochastically applying a bit-flip with probability $\varepsilon_\mathrm{meas.}$ before the measurement. Moreover, these results also apply to gate errors resulting in bit-flips after the TQGs used to implement the parity-checks before the measurement. For simplicity we make no distinction between the bit flip probabilities for $1\rightarrow 0$ or $0\rightarrow 1$, or between different ancilla qubits. We define the measurement success probability (MSP) analogously to the circuit success probability (CSP), i.e. as the probability of no measurement error occurring on any of the ancillas:
\begin{equation}\label{eq:MSP_definition}
    \mathrm{MSP} = (1 - \varepsilon_\mathrm{meas.})^{N_\mathrm{meas.}},
\end{equation}
with $N_\mathrm{meas.}$ being the number of ancilla measurements.

The measurement error (with varying $\varepsilon_\mathrm{meas.}$) was simulated with circuits of a varying CSP. The top row in Supplementary Fig.~\ref{fig:meas_errors} shows the estimator for the averaged squared bias with different measurement success probabilities. As seen in the top row of Supplementary Fig.~\ref{fig:meas_errors}, when the MSP is as low as a random guess (corresponding to 25\% for the JW and approx. 1.5\% for the VC encoding), the shots are being post-selected randomly and no bias improvement is observed. However, by comparing the JW and VC examples, we can see that with several low-weight stabilizers, a smaller MSP will have a smaller effect on the bias. More specifically, only when the MSP is as low as $9\%$ do we start to observe a clear degradation in the bias of the VC example, while the JW example has a significant bias reduction already at $\mathrm{MSP} = 70\%$. This can be intuitively understood by considering that a measurement error will contribute to the bias only if it occurs on the parity-check which would have otherwise detected an error. Therefore, at a fixed MSP, the bias degradation will be smaller if there are more stabilizers.

The effect of a measurement error on the bias is second-order since the bias will be affected only if an error has occurred in the circuit \textit{and} in the measurement. This is however not the case for the cost. Even if no error has occurred in the circuit, a noiseless shot will be discarded in the case of a measurement error. In the high CSP limit, where every measurement error results in an erroneously discarded shot, we therefore expect a cost increase due to measurement errors given by
\begin{equation}\label{eq:meas_err_cost_increase}
    C_\mathrm{SNT}^\mathrm{(meas.)} \leq 1/\sqrt{\mathrm{MSP}},
\end{equation}
where we have defined the cost increase as the quotient of the QEM cost with and without measurement errors. The above expression is an upper bound, since it neglects the possibility of a measurement error occurring and the shot being rightfully discarded due to a successful parity-check of a different stabilizer. The numerically extracted values for the cost increase, at varying MSP and CSP are plotted in the bottom row of Supplementary Fig.~\ref{fig:meas_errors}, for both JW and VC. We can clearly see that the formula from Eq.~\ref{eq:meas_err_cost_increase} accurately predicts the cost increase for high CSP. However, in the limit of lower CSP, the cost increase is observed to be always smaller compared to Eq.~\ref{eq:meas_err_cost_increase}. Note that in the high CSP regime where the effect of measurement errors is largest, cost increases are not as problematic as in the opposite case, due to the fact that the cost is already low.

\suppsection{\label{app:bias2_estimator} Squared Bias Estimation}

When attempting to compute the inherent bias of an error mitigation technique based on the results obtained from a simulation with a finite number of shots, the precision is fundamentally limited by $\mathrm{Var}[\mathsf{O}_\mathrm{est.}]$. In this section, we will show how to obtain an unbiased estimation of the squared bias from shot-based simulations and how to quantify the associated uncertainty due to a limited number of shots.

As described in the main text, the Root-Mean-Squared Error (RMSE) measure of QEM performance depends on the variance of the estimator, as well as on the \textit{squared} bias. We define the squared bias as the squared difference between the average of the expectation value of the error-mitigated estimator $\langle \mathsf{O}_{i,\mathrm{est.}} \rangle$ and the noiseless result $\langle \mathsf{O}_i \rangle$. Furthermore, we average the squared bias over a set of observables to minimize any dependence on the choice of observable, or more explicitly
\begin{align}\label{eq:sq_avg_bias_definition}
    \langle \mathrm{Bias}^2 \rangle_\mathbb{O} = \frac{1}{|\mathbb{O}|}\sum_{\mathsf{O}_i \in \mathbb{O}} \left(\langle \mathsf{O}_{i,\mathrm{est.}} \rangle - \langle \mathsf{O}_i \rangle \right)^2.
\end{align}

Unless a full density matrix simulation is feasible, we do not have access to $\langle \mathsf{O}_{i,\mathrm{est.}} \rangle$, but we can compute the finite sample mean, which we will denote with $\bar{\mathsf{O}}_{i,\mathrm{est.}}$ and assume it is normally distributed according to $\bar{\mathsf{O}}_{i,\mathrm{est.}} \sim \mathcal{N}(\langle \mathsf{O}_{i,\mathrm{est.}} \rangle, \mathrm{Var}[\mathsf{O}_{i,\mathrm{est.}}])$.

Since we are able to perform a noiseless simulation much more efficiently compared to a noisy simulation and since the noiseless simulation will not suffer from the variance increase due to QEM, we can neglect the difference between the sample mean and mean expectation value for the noiseless observable, i.e. $\bar{\mathsf{O}}_i \approx \langle\mathsf{O}_i\rangle$. Under this assumption, we can compute
\begin{equation}\label{eq:MSE_expression}
     \langle (\bar{\mathsf{O}}_{i,\mathrm{est.}} - \langle\mathsf{O}_i\rangle )^2 \rangle  \approx \mathrm{Var}[\mathsf{O}_{i,\mathrm{est.}}] + (\langle \mathsf{O}_{i,\mathrm{est.}} \rangle - \langle \mathsf{O}_{i} \rangle)^2.
\end{equation}
The second term in this expression is the squared bias of a single observable that we are interested in. However, while $ (\bar{\mathsf{O}}_{i,\mathrm{est.}} - \langle\mathsf{O}_i\rangle )^2$ can be easily computed from simulations, this quantity is not an approximation for the squared bias due to the presence of the term $\mathrm{Var}[\mathsf{O}_{i,\mathrm{est.}}]$ in Eq.~\ref{eq:MSE_expression}.  Eq.~\ref{eq:MSE_expression} therefore confirms our intuition, that the squared bias cannot be accurately estimated if the error mitigation cost is too high or equivalently the number of samples is too low, as both of these will increase $ \mathrm{Var}[\mathsf{O}_{i,\mathrm{est.}}]$. This becomes relevant as soon as $\mathrm{Var}[\mathsf{O}_{i,\mathrm{est.}}] \gtrapprox \left(\langle \mathsf{O}_{i,\mathrm{est.}} \rangle - \langle \mathsf{O}_i \rangle \right)^2$.

However, Eq.~\ref{eq:MSE_expression} provides a basis for the construction of an estimator of the quantity of interest. We can define
\begin{equation}
    \Theta_i = (\bar{\mathsf{O}}_{i,\mathrm{est.}} - \langle\mathsf{O}_i\rangle )^2 - \mathrm{\mathscr{S}^2}[\mathsf{O}_{i,\mathrm{est.}}],
\end{equation}
where $\mathscr{S}^2[\mathsf{O}_{i,\mathrm{est.}}]$ is the sample variance, which we define as $\mathscr{S}^2[\mathsf{X}] = 1/(N-1)\sum_{i=1}^N (x_i - \bar{\mathsf{X}})^2$, where $N$ is the total number of samples. Since, on average $\langle \Theta_i \rangle =  (\langle \mathsf{O}_{i,\mathrm{est.}} \rangle - \langle \mathsf{O}_{i} \rangle)^2$, we refer to $\Theta_i$ as an \textit{unbiased} estimator of the squared bias. Furthermore, it also follows that $\Theta = 1/|\mathbb{O}|\sum_{\mathsf{O}_i \in \mathbb{O}} \Theta_i$ is an unbiased estimator of the observable-averaged quantity.

While we have thus far determined that $\Theta_i$, on average, approximates the quantity of interest we must still quantify the uncertainty, or more specifically, we must derive expressions for $\mathrm{Var}[\Theta_i]$ and $\mathrm{Var}[\sum_{\mathsf{O}_i \in \mathbb{O}} \Theta_i]/|\mathbb{O} |$. In order to do so, we note that
\begin{align}\label{eq:var_thetas}
    \mathrm{Var}[\Theta] = \frac{1}{|\mathbb{O}|^2} \sum_{\mathsf{O}_i \in \mathbb{O}} \mathrm{Var}[\Theta_i],
\end{align}
since the estimators of all observables are uncorrelated. Furthermore, since the quantity $(\bar{\mathsf{O}}_{i,\mathrm{est.}} - \langle\mathsf{O}_i\rangle )^2$ is the square of a normally distributed random variable with a finite mean, it is distributed according to a non-central chi-squared distribution, i.e.
\begin{equation}\label{eq:theta_i_distribution}
    (\bar{\mathsf{O}}_{i,\mathrm{est.}} - \langle\mathsf{O}_i\rangle )^2 \sim \chi^2\left(1,\frac{(\langle \mathsf{O}_{i,\mathrm{est.}} \rangle - \langle\mathsf{O}_i\rangle )^2}{\mathrm{Var}[\mathsf{O}_{i,\mathrm{est.}}]}\right).
\end{equation}
This distribution has the following properties: if $\mathsf{X} \sim \chi^2(1,\lambda )$, then $\mathrm{Var}[\mathsf{X}] = 2 + 4 \lambda$. Additionally, the uncertainty in the sample variances, $\mathrm{Var}[\mathscr{S}^2[\mathsf{O}_{i,\mathrm{est.}}]] \propto \mathrm{Var}[\mathsf{O}_{i,\mathrm{est.}}]^2/N_\mathrm{eff}$, with $N_\mathrm{eff}$ defined in Eq.~\ref{eq:N_eff_definition} will also contribute to the variance of the estimator. However, in our simulations we typically consider $N_\mathrm{eff} \sim 10^{5}$ meaning that this contribution is much smaller compared to the other terms in the final expression. 

All together, Eqs.~\ref{eq:var_thetas}-\ref{eq:theta_i_distribution} imply 
\begin{align}\label{eq:var_theta_final}
    \mathrm{Var}[\Theta] &= \frac{1}{|\mathbb{O}|^2}\sum_{\mathsf{O}_i \in \mathbb{O}}  2\mathrm{Var}[\mathsf{O}_{i,\mathrm{est.}}]^2 \nonumber \\
    &+ 4 \mathrm{Var}[\mathsf{O}_{i,\mathrm{est.}}] (\langle \mathsf{O}_{i,\mathrm{est.}}\rangle - \langle\mathsf{O}_i\rangle )^2, \\
    &\approx \frac{1}{|\mathbb{O}|^2}\sum_{\mathsf{O}_i \in \mathbb{O}}  2\left(\mathscr{S}^2[\mathsf{O}_{i,\mathrm{est.}}]\right)^2 \nonumber \\
    &+ 4 \mathscr{S}^2[\mathsf{O}_{i,\mathrm{est.}}] (\bar{\mathsf{O}}_{i,\mathrm{est.}} - \langle\mathsf{O}_i\rangle )^2,   
\end{align}
Since the final estimator $\Theta$ is constructed as a weighted sum of non-central chi-squared distributed stochastic variables, the central limit theorem can be invoked when $|\mathbb{O}| \gg 1$. In this limit, the variance derived in Eq.~\ref{eq:var_theta_final} has a straightforward interpretation in terms of the percentiles of a normal distribution. Furthermore, it is evident that the precision in the estimation is always limited by the variances of the estimators of the individual observables $\mathrm{Var}[\mathsf{O}_{i,\mathrm{est.}}]$. To summarize, Eq.~\ref{eq:var_theta_final} is valid when the distribution of the error-mitigated observable is Gaussian with a variance that is much larger compared to the variance of the noiseless observable $\mathsf{O}_i$. We have additionally assumed that the sample mean and variance are close to the underlying mean values and variance. The latter assumption is only valid when a large number of shots is available.

\suppsection{\label{app:observable_weight} Effect of Observable Weight}

\begin{figure*}[t]
    \centering
    \includegraphics[width=0.95\textwidth]{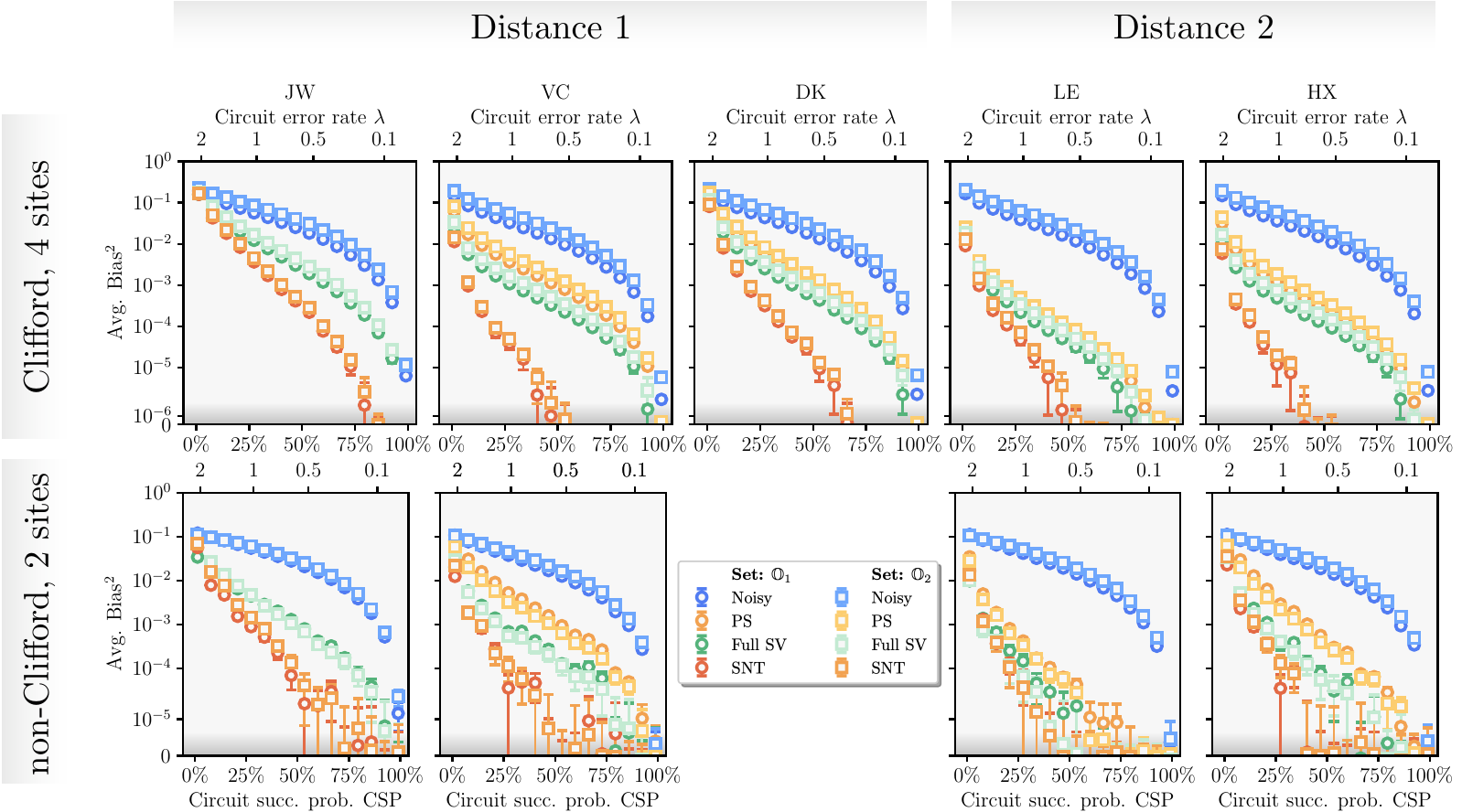}
    \caption{Squared bias (averaged over the site occupations) of the time evolution of a FHM with two (top row) and four (bottom row) sites after 10 Trotter steps as a function of CSP and the circuit error rate $\lambda$, for four different encodings and different mitigation schemes: no mitigation (blue), PS on local stabilizers (yellow), full SV including PP based on global stabilizers (green) and SNT (red). The two-site non-Clifford simulation was evolved up to time $T=0.5$. The error bars represent a 1-$\sigma$ uncertainty due to a finite number of shots/circuits, which starts to dominate in the gray shaded area. Darker marker colors correspond to the performance on observables from the set $\mathbb{O}_1$ and lighter to the set $\mathbb{O}_2$ (see legend). The noise level of the circuits is varied by changing the CZ gate fidelity.}
    \label{fig:high_weights}
\end{figure*}

Since the causal cone associated with a higher-weight observable is larger than a lower-weight observable, we have performed the bias analysis in the main text in the limit of a circuit with a significantly larger depth compared to the number of qubits. Here we provide additional data justifying this approach, along with the bias results obtained from a 4-site Clifford simulation. These results also provide evidence that the expected extrapolated performance of SNT in Figs. 1 and 5 of the main text is not degraded when considering high-weight observables.

The average squared bias plotted in Fig.~4 of the main text is defined as $\langle \mathrm{Bias}^2\rangle_\mathbb{O} = 1/|\mathbb{O}|\sum_{\mathsf{O}_i \in \mathbb{O}} \mathrm{Bias}[\mathsf{O}_i]^2$. We now distinguish between two sets of observables with different weights
\begin{align}
    \mathbb{O}_1 &= \{ (1 - \hat{V}_i)/2 \, | \,  i = 1,\dots,2N  \} \text{ and} \nonumber \\
    \mathbb{O}_2 &= \{ (1 - \hat{V}_i\hat{V}_j)/2 \, | \,  i,j = 1,\dots,2N; \, j>i  \}. 
\end{align}
The results in the main text are based on $\mathbb{O}_1$ which is the set of all single-site occupations $\hat{n}_i^\sigma$. For the JW and VC and DK encodings $\mathbb{O}_1$ contains weight-1 Pauli operators, while for the LE and HX $\mathbb{O}_1$ is comprised of weight-2 operators. The set $\mathbb{O}_2$ is constructed as a generalization of $\mathbb{O}_1$ to larger weights, in order to make the results directly comparable. Note that estimating the average value of a density-density term of the form $\langle \hat{n}_{i\updownarrow} \hat{n}_{j\updownarrow} \rangle$ would include measuring observables from both $\mathbb{O}_1$ and $\mathbb{O}_2$.

Comparing the results for $\mathbb{O}_1$ and $\mathbb{O}_2$ in Supplementary Fig.~\ref{fig:high_weights} we observe that, in the non-Clifford example, the difference between the squared bias is not statistically significant with the given number of shots. Note that the results in the top row of Supplementary Fig.~\ref{fig:high_weights} corresponding to the $\mathbb{O}_1$ set are a copy of Fig.~4 in the main text. We are able to discern the expected increase in the bias in the Clifford simulations in the bottom row of Supplementary Fig.~\ref{fig:high_weights} due to the smaller variance. The increase is most evident in the simulations without any error mitigation. All in all, the marginal differences in the performance of the two sets confirm that the depth of the circuits used in simulations is sufficient for the total circuit error rate (or success probability) to capture an adequate fraction of the noise affecting the observable. 

While we have benchmarked the bias performance in terms of the weights of the estimated observables, we note that for the presented SNT implementation as well as for the full SV approach, we still rely on the measurement of more global observables (e.g. $\mathsf{S}_{\uparrow/\downarrow}$) for the implementation of SV via PP. This means that the final result is composed from measurements of observables which have significantly larger causal cones than the weight of the observable we are trying to estimate would suggest, further justifying validity of the analysis based on single-site occupations for these two methods.

\suppsection{\label{app:PEC_variance} Variance of Probabilistic QEM}

In the derivation of the cost of probabilistic error cancellation (PEC) it is often implicitly assumed that we are sampling each shot from a different circuit, i.e. $N_\mathrm{shots} = N_\mathrm{circuits}$\cite{Endo_2018,Temme_2017_PEC}. However, this is often not the case in practice, since e.g. loading a new circuit into the electronics is significantly more time consuming \cite{fruitwala_2024_FPGA_RC} compared to sampling additional shots from an already loaded circuit\cite{van_den_Berg_2023_spl_pec, yoshioka_ibm_2024}. Due to the lack of an analytical expression, the uncertainty of the PEC result in Ref.~\cite{van_den_Berg_2023_spl_pec} was therefore evaluated using statistical bootstrapping. Here we provide an analytical expression for the uncertainty of a probabilistic estimator, thus simplifying the uncertainty analysis of QEM when circuit sampling methods are employed.

The estimator $\mathsf{O}_\mathrm{est.}$ of the observable of interest $\mathsf{O}$ is constructed as\cite{Endo_2018}
\begin{equation}
    \label{eq:O_est}
    \bar{\mathsf{O}}_\mathrm{est.} =  \frac{C}{N_\mathrm{c}}\sum_{\alpha = 1}^{N_\mathrm{c}} \frac{\mathrm{sign}_\alpha}{n_\mathrm{s}^{\alpha}} \sum_{i = 1}^{n_\mathrm{s}^\alpha} \Upsilon_i^\alpha.
\end{equation}
Here $\Upsilon_i^\alpha$ represents the measurement result of circuit $\alpha$ and shot $i$, and is a binary stochastic variable distributed according to a binomial distribution with a well-defined mean $\mu_\alpha$ and variance $\sigma_\alpha^2$. The estimator in Eq.~\ref{eq:O_est} is a sample average (denoted with $\bar{\bullet}$) of the results from $N_\mathrm{c}$ circuit executions, with $n_\mathrm{s}^{\alpha}$ shots per circuit $\alpha$ rescaled by the constant $C$ and term $\mathrm{sign}_\alpha \in \{-1,1\}$. We have shortened the notation compared to the main text, so that $N_\mathrm{c} = N_\mathrm{circuits}$. The term $\mathrm{sign}_\alpha \in \{-1,1\}$ can also be incorporated into the stochastic variable $\Upsilon_i^\alpha$ by redefining $\Upsilon_i^\alpha \rightarrow \mathrm{sign}_\alpha\Upsilon_i^\alpha$. For clarity, we will therefore omit $\mathrm{sign}_\alpha \in \{-1,1\}$ from the following expressions. Furthermore, we denote the average over the shots sampled from the same circuit with $\langle\bullet\rangle_\mathrm{s}$ and the average over different circuits with $\langle\bullet\rangle_\mathrm{c}$, e.g. $\mu_\alpha = \langle \Upsilon_i^\alpha \rangle_\mathrm{s}$. The same applies for the variance, such that e.g. $\mathrm{Var}_\mathrm{c}[\bullet] = \langle\bullet^2 \rangle_\mathrm{c} - \langle\bullet \rangle^2_\mathrm{c}$.

The uncertainty, characterized by the variance of the estimator is computed as
\begin{equation}\label{eq:var_O_est}
    \mathrm{Var}_\mathrm{s,c}[\bar{\mathsf{O}}_\mathrm{est.}] = \left\langle \left\langle \bar{\mathsf{O}}_\mathrm{est.}^2 \right \rangle_\mathrm{s} \right \rangle_\mathrm{c} - \left\langle \left\langle \bar{\mathsf{O}}_\mathrm{est.} \right \rangle_\mathrm{s} \right \rangle^2_\mathrm{c},
\end{equation}
where the double brackets represent the averaging over both the shots and circuits. While the averaging is interchangeable, we will average over the shots of a single circuit first and over all possible circuits later, as in an experimental implementation~\cite{van_den_Berg_2023_spl_pec}. Furthermore, in the experimental implementation presented in Ref.~\cite{van_den_Berg_2023_spl_pec}, each circuit was assigned a fixed number of shots so that $n_\mathrm{s}^{\alpha} \equiv n_\mathrm{s} $. We will first derive the expressions with this condition and list the general result afterwards.  

The second term of Eq.~\ref{eq:var_O_est} is given by
\begin{equation}
     \left\langle \left\langle \bar{\mathsf{O}}_\mathrm{est.} \right \rangle_\mathrm{s} \right \rangle_\mathrm{c}  = C \left\langle \left\langle \Upsilon_i^\alpha \right\rangle_\mathrm{s} \right \rangle_\mathrm{c} = C \langle \mu \rangle_\mathrm{c}.
\end{equation}
We have shortened the notation for the average over different circuits by omitting the circuit index $\alpha$ so that $\langle \mu \rangle_\mathrm{c} \equiv \langle\mu_\alpha\rangle_\mathrm{c} = \sum_{\alpha} p_\alpha \mu_\alpha$, where $p_\alpha$ is the probability to sample circuit $\alpha$ and is determined by the coefficients $p_i^{(k)}$ from Eq.~4 in the main text\cite{Endo_2018,Temme_2017_PEC,van_den_Berg_2023_spl_pec}. 

The first term of Eq.~\ref{eq:var_O_est} is less straightforward:
\begin{widetext}
\begin{align}
    \left\langle \left\langle\bar{\mathsf{O}}_\mathrm{est.}^2 \right \rangle_\mathrm{s} \right \rangle_\mathrm{c} &= \frac{C^2}{N_\mathrm{c}^2 n_\mathrm{s}^2} \left\langle \left\langle \sum_{\alpha, \beta =1}^{N_\mathrm{c}}\sum_{i,j =1}^{n_\mathrm{s}} \Upsilon_i^\alpha \Upsilon_j^\beta \right \rangle_\mathrm{s} \right \rangle_\mathrm{c} \\
    & = \frac{C^2}{N_\mathrm{c}^2 n_\mathrm{s}^2} \left\langle \left\langle \sum_{\alpha =1}^{N_\mathrm{c}}\sum_{i =1}^{n_\mathrm{s}} (\Upsilon_i^\alpha)^2 + \sum_{\alpha =1}^{N_\mathrm{c}}\sum_{\substack{i,j =1\\ i\neq j}}^{n_\mathrm{s}} \Upsilon_i^\alpha \Upsilon_j^\alpha + \sum_{\substack{\alpha, \beta =1\\ \alpha \neq \beta}}^{N_\mathrm{c}} \sum_{i =1}^{n_\mathrm{s}}  \Upsilon_i^\alpha \Upsilon_i^\beta +  \sum_{\substack{\alpha, \beta =1\\ \alpha \neq \beta}}^{N_\mathrm{c}} \sum_{\substack{i,j =1\\ i\neq j}}^{n_\mathrm{s}}  \Upsilon_i^\alpha \Upsilon_j^\beta \right \rangle_\mathrm{s} \right \rangle_\mathrm{c}  \\
    & = \frac{C^2}{N_\mathrm{c}^2 n_\mathrm{s}^2}  \left\langle \sum_{\alpha =1}^{N_\mathrm{c}}\sum_{i =1}^{n_\mathrm{s}} \langle(\Upsilon_i^\alpha)^2\rangle_\mathrm{s} + \sum_{\alpha =1}^{N_\mathrm{c}}\sum_{\substack{i,j =1\\ i\neq j}}^{n_\mathrm{s}} \langle \Upsilon_i^\alpha\rangle_\mathrm{s}\langle \Upsilon_j^\alpha \rangle_\mathrm{s} + \sum_{\substack{\alpha, \beta =1\\ \alpha \neq \beta}}^{N_\mathrm{c}} \sum_{i =1}^{n_\mathrm{s}} \langle \Upsilon_i^\alpha \rangle_\mathrm{s} \langle \Upsilon_i^\beta \rangle_\mathrm{s} +  \sum_{\substack{\alpha, \beta =1\\ \alpha \neq \beta}}^{N_\mathrm{c}} \sum_{\substack{i,j =1\\ i\neq j}}^{n_\mathrm{s}}  \langle \Upsilon_i^\alpha  \rangle_\mathrm{s} \langle \Upsilon_j^\beta \rangle_\mathrm{s} \right \rangle_\mathrm{c}  \\
    &= \frac{C^2}{N_\mathrm{c}^2 n_\mathrm{s}^2}  \left\langle \sum_{\alpha =1}^{N_\mathrm{c}}n_\mathrm{s} (\mu_\alpha^2 + \sigma_\alpha^2) +  \sum_{\alpha =1}^{N_\mathrm{c}} n_\mathrm{s}(n_\mathrm{s} - 1) \mu_\alpha^2 
    + \sum_{\substack{\alpha, \beta =1\\ \alpha \neq \beta}}^{N_\mathrm{c}} n_\mathrm{s}^2 \mu_\alpha \mu_\beta  \right \rangle_\mathrm{c} . 
\end{align}
\end{widetext}
In the calculation we have split the sum over all indices over four sums, based on whether we have summed over the same or strictly different circuits or shot indices and used the fact that for any set of samples $\{ x_i \}$, where $x_i$ was sampled from a distribution with a finite average $\mu$, $\langle \sum_{i=1}^{N_X} x_i \rangle = \sum_{i=1}^{N_X}\langle x_i \rangle = {N_X}\mu$. Additionally, we have acknowledged that the outcome of each shot and circuit execution is uncorrelated, meaning that $\langle \Upsilon_i^\alpha \Upsilon_j^\beta \rangle = \langle \Upsilon_i^\alpha  \rangle \langle \Upsilon_j^\beta \rangle$, unless $\alpha = \beta$ and $i = j$. We will now proceed with evaluating the average over the circuits using the same reasoning as before
\begin{align}
    \left\langle \left\langle \bar{\mathsf{O}}_\mathrm{est.}^2 \right \rangle_\mathrm{s} \right \rangle_\mathrm{c} &= \frac{C^2}{N_\mathrm{c}^2 n_\mathrm{s}^2}  ( N_\mathrm{c} n_\mathrm{s} \left[\langle \mu^2 \rangle_\mathrm{c} +\langle \sigma^2 \rangle_\mathrm{c} \right] \\ \nonumber
    &+  N_\mathrm{c} n_\mathrm{s}(n_\mathrm{s} - 1) \langle \mu^2 \rangle_\mathrm{c} \\
    &+ N_\mathrm{c}(N_\mathrm{c}-1) n_\mathrm{s}^2 \langle \mu\rangle_\mathrm{c}^2  ).\nonumber
\end{align}
Further calculations result in
\begin{align}\label{eq:var_O_est_final}
    \mathrm{Var}_\mathrm{s,c}[\bar{\mathsf{O}}_\mathrm{est.}] = \frac{C^2}{N_\mathrm{c}} \left( \frac{\langle \sigma^2 \rangle_\mathrm{c}}{ n_\mathrm{s} } + \mathrm{Var}_\mathrm{c}[\mu] \right).
\end{align}
Here we have defined $\mathrm{Var}_\mathrm{c}[\mu] = \langle \mu^2\rangle_\mathrm{c} -\langle \mu \rangle_\mathrm{c}^2 $ as the variance of the individual circuit expectation values. The expression in Eq.~\ref{eq:var_O_est_final} has two distinct contributions: the first term ($\frac{\langle \sigma^2 \rangle_\mathrm{c}}{ n_\mathrm{s} }$) represents the average shot noise of the circuits, while the second term ($\mathrm{Var}_\mathrm{c}[\mu] $) represents the uncertainty due to the finite circuit sampling. The whole derivation can be repeated for the more general case where the number of shots per circuit is circuit dependent, i.e. $n^\alpha_\mathrm{s}$, in which case we obtain
\begin{align}\label{eq:var_O_est_final_SNT}
    \mathrm{Var}_\mathrm{s,c}[\bar{\mathsf{O}}_\mathrm{est.}] = \frac{C^2}{N_\mathrm{c}} \left( \frac{\langle \sigma^2 \rangle_\mathrm{c}}{ \langle n_\mathrm{s} \rangle_\mathrm{c} } + \mathrm{Var}_\mathrm{c}[\mu] \left\{ 1 + \frac{\mathrm{Var}_\mathrm{c}[n_\mathrm{s}]}{\langle n_\mathrm{s} \rangle_\mathrm{c}^2}\right\} \right).
\end{align}
The latter expression is also valid for SNT where due to the PS, $n_\mathrm{s}^\alpha$ may vary between the sampled circuits. Nonetheless, as long as the shot rejection rate is similar between all sampled circuits this additional term does not contribute to the variance significantly.

In order to obtain a bound on the expression in Eq.~\ref{eq:var_O_est_final}, we can use the fact that estimating any Pauli observable as $\Upsilon_{i}^\alpha \in \{-1,1 \} $, implies $ \langle \sigma^2 \rangle_\mathrm{c}$, $\langle \mu^2 \rangle_\mathrm{c} \leq 1$ and $\mathrm{Var}_\mathrm{c}[\mu]\leq 1$. The expression for the uncertainty of the estimator is therefore bounded by
\begin{align}\label{eq:var_O_est_upper_lim}
    \mathrm{Var}_\mathrm{s,c}[\bar{\mathsf{O}}_\mathrm{est.}] &\leq \frac{C^2}{N_\mathrm{c}} \left[ \frac{1}{ n_\mathrm{s} } + 1 \right].
\end{align}
It is therefore evident that a large $n_\mathrm{s}$ is not able to compensate for a small number of circuits $N_\mathrm{c}$. 

Furthermore, we can also define an effective number of samples
\begin{equation}\label{eq:N_eff_definition}
    \frac{1}{N_\mathrm{eff}} = \frac{1}{N_\mathrm{c}} + \frac{1}{N_\mathrm{c} n_\mathrm{s}},
\end{equation}
so that $\mathrm{Var}_\mathrm{s,c}[\mathsf{O}_\mathrm{est.}] \leq C^2/N_\mathrm{eff}$.
We are thus guaranteed that $N_\mathrm{eff} < N_\mathrm{c}$, meaning that the precision of a Pauli observable will be always limited by the number of circuits rather than the number of shots per circuit. It also means that investing a small number of shots per each circuit, possibly as low as, $n_\mathrm{s} \sim 10 $ or $ 100$ is sufficient.

\suppsection{Improved Circuit Sampling}\label{app:circuit_sampling}

While the expressions derived in Eqs.~\ref{eq:var_O_est_final} and ~\ref{eq:var_O_est_final_SNT} describe the variance in a practical scenario for both PEC and SNT, they can also be used to construct a more optimal circuit sampling strategy.

We have previously assumed that the circuits are sampled from the discrete distribution specified by the coefficients $p_\alpha$. We will label the probability of sampling the $0$-circuit (i.e. the original circuit without any added operations) as $p_0$. As long as $p_0 \gg \mathrm{max}_{\alpha \neq 0} p_\alpha$, i.e. the probability of sampling the 0-circuit is much larger compared to any other circuit, which can be exploited for an improved sampling scheme requiring less circuits. A practical PEC implementation in this case should be performed by 1. sampling the circuits, 2. grouping and counting the number of times the $0$-circuit has been sampled, 3. running the $0$-circuit with $n_\mathrm{s}^0$ shots and 4. running the rest of the sampled circuits, each with a smaller number of shots $n_\mathrm{s}'< n_\mathrm{s}^0$. However, at this point, the question of what is the optimal choice of $n_\mathrm{s}^0$ and $n_\mathrm{s}'$ arises.

The observable estimator in this case is comprised of two parts
\begin{align}
    \bar{\mathsf{O}}'_\mathrm{est.} &=  \frac{C\,p_0}{n_\mathrm{s}^0} \sum_{i = 1}^{n_\mathrm{s}^0} \Upsilon_i^0 + \frac{C(1-p_0)}{(N_\mathrm{c}-1) n_\mathrm{s}'}\sum_{\alpha = 1}^{N_\mathrm{c}-1} \sum_{i = 1}^{n_\mathrm{s}'} \Upsilon_i^\alpha,
\end{align}
and it is possible to check that it is still unbiased, i.e. $\langle\langle \bar{\mathsf{O}}'_\mathrm{est.} \rangle_\mathrm{s}\rangle_\mathrm{c} = C\langle \mu \rangle_\mathrm{c}$. To clarify, the sampled circuits in the second part of the expression do not include the $0$-circuit, meaning that the probability of sampling circuit $\alpha \neq 0$ is given by $p_\alpha/(1 - p_0)$. Since the samples from the $0$-circuit are uncorrelated with the samples from the $\alpha\neq 0$ circuits, the variance of this new estimator is the weighted sum of the two parts, as derived in Eq.~\ref{eq:var_O_est_final}
\begin{align}
    \mathrm{Var}[\bar{\mathsf{O}}'_\mathrm{est.}]/C^2 = p_0^2 \frac{\sigma_0^2}{n_\mathrm{s}^0} + \frac{(1 - p_0)^2 }{N_\mathrm{c}-1}\left( \frac{\langle \sigma^2 \rangle_{\mathrm{c}|\alpha\neq0}}{ n_\mathrm{s}' } + \mathrm{Var}_{\mathrm{c}|\alpha\neq0}[\mu] \right).
\end{align}
The optimal choice for $n_\mathrm{s}^0$ and $n_\mathrm{s}'$ with a restricted budget $n_\mathrm{s}^0 + (N_\mathrm{c} - 1)n_\mathrm{s}' = N_\mathrm{s} $, where $N_\mathrm{s}$ is the shortened notation for the total available number of shots $N_\mathrm{shots}$, is obtained by minimizing the Lagrangian
\begin{equation}
    \mathscr{L}[n_\mathrm{s}^0,n_\mathrm{s}',\lambda'] = \mathrm{Var}[\bar{\mathsf{O}}'_\mathrm{est.}] - \lambda' (N_\mathrm{s} - n_\mathrm{s}^0 - (N_\mathrm{c} - 1)n_\mathrm{s}').
\end{equation}
Solving $\partial \mathscr{L}/\partial n_\mathrm{s}^0 = 0$, $\partial \mathscr{L}/\partial n_\mathrm{s}' = 0$ and $\partial \mathscr{L}/\partial \lambda' = 0$ results in
\begin{align}
    n_\mathrm{s}^0 &= N_\mathrm{s} \,p_0 \frac{s}{1 + (s-1)\, p_0} \\
    n_\mathrm{s}' &= \frac{N_\mathrm{s}}{N_\mathrm{c}-1}  \frac{1 - p_0}{1 + (s-1)\, p_0},
\end{align}
where $s = \sqrt{\sigma_0^2/\langle \sigma^2 \rangle_{\mathrm{c}|\alpha\neq0}}$. Since $s$ is not known before the circuits are executed, we further simplify the results in the limit $s\rightarrow 1$, and obtain for the final optimized variance
\begin{align}\label{eq:var_O_opt}
    \mathrm{Var}_\mathrm{s,c}[\bar{\mathsf{O}}'_\mathrm{est.}]/C^2 &= p_0 \frac{\sigma_0^2}{N_\mathrm{s}} + \frac{1 - p_0 }{N_\mathrm{s}} \langle \sigma^2 \rangle_{\mathrm{c}|\alpha\neq0}\nonumber \\
    &+ \frac{(1 - p_0  )^2}{N_\mathrm{c} - 1} \mathrm{Var}_{\mathrm{c}|\alpha\neq0}[\mu] \\
    &\approx \frac{(1 - p_0  )^2}{N_\mathrm{c} - 1} \mathrm{Var}_{\mathrm{c}|\alpha\neq0}[\mu]. \label{eq:var_O_opt_approx}
\end{align}
Comparing Eq.~\ref{eq:var_O_opt} with the original expression in Eq.~\ref{eq:var_O_est_final}, we can see that the first two terms in Eq.~\ref{eq:var_O_opt} are now $\propto 1/N_\mathrm{s}$ and are no longer significant if $N_\mathrm{s} \gg N_\mathrm{c}  $. Even though the last term is still proportional to $1/N_\mathrm{c}$ the prefactor of $(1 - p_0)^2$ will mitigate a smaller $N_\mathrm{c}$. For example, at $p_0\approx 68\%$, the required number of circuits for a similar variance compared to Eq.~\ref{eq:var_O_est_final} is $10\times$ smaller.

For standard PEC, $p_0 =\prod_k (1 - \varepsilon_k)=\mathrm{CSP} = \exp(-\lambda)$. However, when partial PEC is performed, such as is the case for SNT, the value of $p_0$ for a given CSP can be significantly smaller, thus extending the range of $\lambda$ values where the improved circuit sampling leads to a significant reduction in the required circuit executions. More specifically, if $R$ denotes the ratio of detectable errors, the variance of SNT (when $N_\mathrm{c}\ll N_\mathrm{s}$) is reduced by a factor of $(1 - \exp[-(1-R)\lambda])^2$. In the very noisy regime, where $\lambda \approx 7$, with a typical $R\approx 0.85$ (as observed for the VC encoding), the variance is suppressed by a factor of 0.4. The limit for a 10-fold reduction in the required $N_\mathrm{c}$ for SNT is $\lambda \approx 2.5$, corresponding to $\mathrm{CSP}\approx 8\%$.

\begin{figure}[t]
    \centering
    \includegraphics[width=0.45\textwidth]{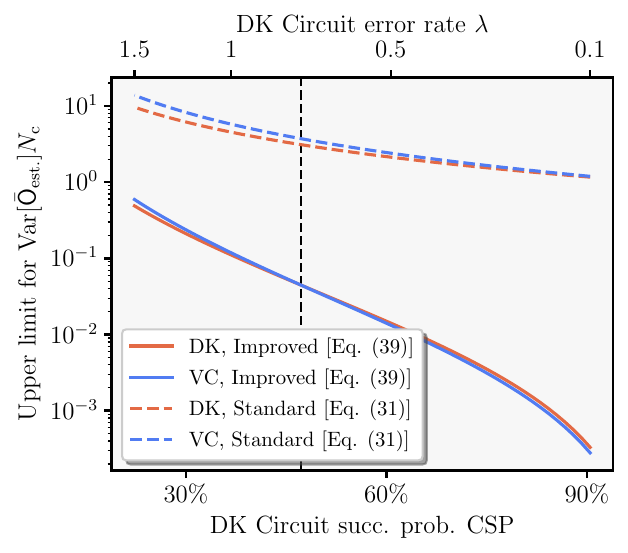}
    \caption{
    The upper limit for the QEM estimator variance $\mathrm{Var}[\bar{\mathsf{O}}_\mathrm{est.}] N_\mathrm{c}$ with the improved sampling strategy (Eq.~\ref{eq:var_O_opt_approx}) and the standard approach (Eq.~\ref{eq:var_O_est_upper_lim}) in the low-circuit regime of $N_\mathrm{s} \gg N_\mathrm{c}$ for the DK and VC encodings. We have chosen the ratio of $\lambda_\mathrm{VC}/\lambda_\mathrm{DK} = 1.3 $, corresponding to a FHM lattice size of $6\times 6$ along with the values for $R = R_\mathrm{PS} + R_\mathrm{PP}$ listed in Table I in the main text. The dashed vertical line indicates the point beyond which the variance limit for the VC is smaller compared to that of DK for the improved sampling strategy. 
    }
    \label{fig:improved_sampling}
\end{figure}

Eq.~\ref{eq:var_O_opt_approx} also shows that the upper limit for the variance of the estimator does not depend \emph{only} on the cost of the error mitigation due to the prefactor of $(1 - p_0)^2$. This becomes especially relevant when comparing the variance limit of encodings with the same distance, such as DK and VC. More specifically the variance still increases with $\lambda$, favoring DK, and decreases with $R$ which is beneficial for VC. Supplementary Fig.~\ref{fig:improved_sampling} clearly illustrates these two competing effects and shows that even though $C_\mathrm{VC+SNT} > C_\mathrm{DK+SNT}$, after the improved sampling strategy is applied, the upper variance limit for both encodings is very similar, and even slightly favors VC in the high-fidelity regime. The value of CSP below which the variance of VC is smaller compared to DK (black vertical line) will shift to lower values of the CSP at larger lattice sizes, since $\lim_{N\times N\rightarrow \infty }[\lambda_\mathrm{VC}/\lambda_\mathrm{DK}] < 1.3 $. Supplementary Fig.~\ref{fig:improved_sampling} also highlights the orders-of-magnitude improvement in the variance of the improved sampling strategy compared to the standard approach.

Overall, using the improved sampling strategy, SNT will require significantly less circuits for a similar variance compared to PEC. This strategy can also be further extended to higher-orders, by considering circuits with 1, 2, etc. additional operations.

\suppsection{\label{app:rectangular_lattice} Optimal Strategies in 1D}

\begin{figure*}[t]
    \centering
    \includegraphics[width=0.9\textwidth]{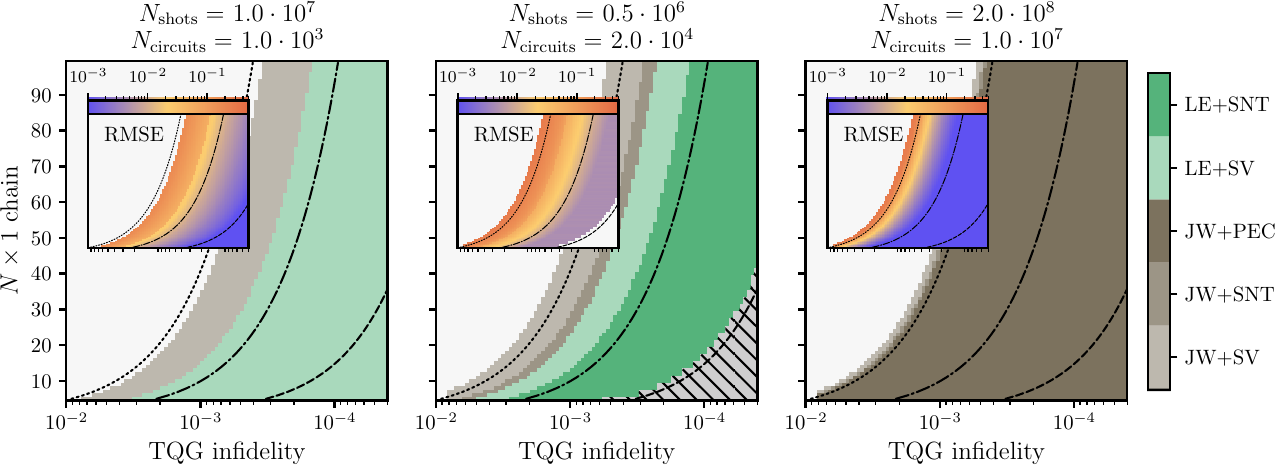}
    \caption{
     Optimal combinations of encoding and QEM, for the time evolution ($10$ Trotter steps) of a 1D FHM. The three black contour lines represent a CSP of 5\%, 50\% and 90\% (left to right) of the circuit generated by the encoding with the smallest number of TQGs, depending on the system size. The white region in the top left corresponds to low circuit success probabilities, where a single parity check is not sufficient for a significant bias reduction, and the cost of PEC $C_\mathrm{PEC}^2 \gtrsim 10^6$ is too large for the given shot budget. The hatched region in the lower right corner represents the region where the optimal combination cannot be determined due to the finite number of shots in the bias estimation. Insets show the RMSE for optimal combinations of QEM and encoding.
    }
    \label{fig:best_strategy_1D}
\end{figure*}

\begin{figure*}[t]
    \centering
    \includegraphics[width=0.9\textwidth]{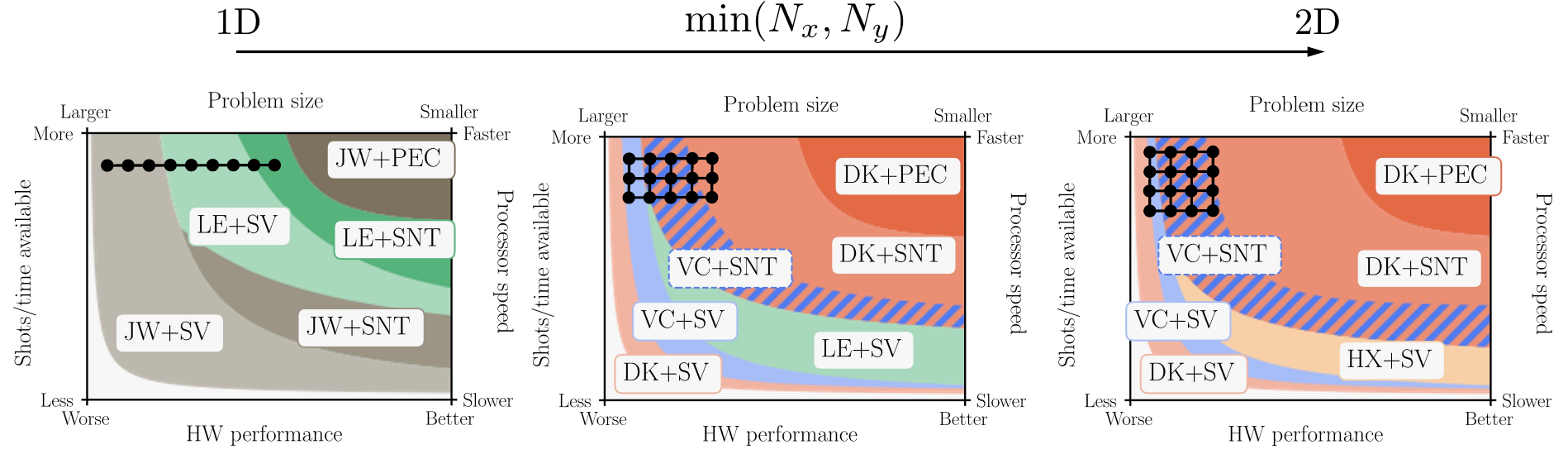}
    \caption{Qualitative behavior of the optimal QEM + encoding combination based on the results from Fig.~5 in the main text and Supplementary Fig.~\ref{fig:best_strategy_1D}. The plots from left to right correspond to an interpolation between a 1D chain (left) and a 2D square lattice (right), in the limit of a large number of fermionic sites. The center plot corresponds to systems on rectangular or quasi-1D lattices where $N_x \gg N_y$. The hatched blue region indicates the possible presence of VC+SNT, which depends on the exact circuit to shot ratio, as seen in Fig.~5 in the main text.
    }
    \label{fig:phase_diagram_sketch}
\end{figure*}

The results in the main text are focused on the implementation of the time evolution of a two-dimensional FHM, due to the high complexity of performing this task using classical methods. We now expand this analysis to the 1D case, and interpolate the results for the optimal encoding and QEM combination to general rectangular lattices.

Fig.~\ref{fig:best_strategy_1D} shows the optimal choice of encoding and QEM for the time evolution of a FHM on a linear chain, using the same methodology as for the square lattice examples in the main text. The required number of two-qubit gates per Trotter step for each encoding is listed in Table~I of the main text. We consider the same shot and circuit budgets as in Fig.~5 of the main text, with an additional panel highlighting the emergence of PEC as the optimal QEM approach in the large shot and circuit budget limit. In the case of a 1D FHM, only 1D encodings (JW and LE) appear on the phase diagram, as the super-linear scaling in the number of TQGs is no longer present (see Table I in the main text). JW still appears at lower circuit fidelities, where the QEM cost is crucial, with LE appearing at higher fidelities, where a low bias is preferential. Identically as in the 2D example we observe a threshold between a low and high-distance encoding in the left panel of Supplementary Fig.~\ref{fig:best_strategy_1D}. The middle panel of Supplementary Fig.~\ref{fig:best_strategy_1D} also displays the same transition between the two encodings, which contrasts with the results of the 2D model, where only distance-1 encodings with different QEM combinations are observed. This difference is attributed to the fact that the cost associated with JW+SNT is significantly larger compared to VC or DK+SNT. The lack of local stabilizers in the JW encoding therefore favors the distance-2 LE encoding. The encoding hierarchy is further broken in the high-circuit limit (right plot of Supplementary Fig.~\ref{fig:best_strategy_1D}) when the number of available circuits is large enough for the implementation of PEC, which is symmetry agnostic.

Having obtained an understanding of the phase diagrams for one- and two-dimensional examples, we will now interpolate between these results to obtain results for the case of a rectangular fermionic lattice, to further illustrate the effect of dimensionality of the system under consideration. A qualitative sketch of the best encoding and QEM combinations is presented in Supplementary Fig.~\ref{fig:phase_diagram_sketch}. Note that the general trend in the plots in Supplementary Fig.~\ref{fig:phase_diagram_sketch} is largely determined by the QEM cost (first SV, second SNT and lastly PEC) and secondly by the choice of fermion-to-qubit encoding. The exception to this rule is the purely 1D case, as explained in the previous paragraph.

In all panels of Supplementary Fig.~\ref{fig:phase_diagram_sketch} the lower left corner is populated by the combination of a low-distance encoding and symmetry verification, due to the comparably low QEM cost. Conversely, the upper right corner will always feature the lowest-distance encoding together with PEC. The intermediate region for the 1D panel on the left is based on Supplementary Fig.~\ref{fig:best_strategy_1D}, while the right-most plot is based on Fig.~5 in the main text. The intermediate panel is again based on the results from Fig.~5, where both VC and DK outperform JW, and the 1D LE encoding outperforms the 2D HX for intermediate system sizes. All together, these phase diagrams illustrate the rich behavior of the combinations of fermionic encodings and QEM.

\suppsection{\label{subsec:JW_example} Density Matrix Simulations}

\begin{figure}[h]
    \includegraphics[width=0.45\textwidth]{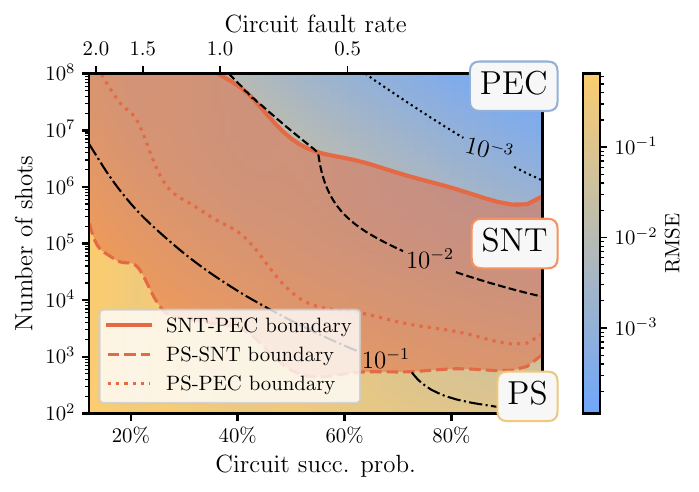}
    \caption{RMSE of SNT, PS and PEC computed via density matrix simulation of the spinless $t$-$J$ model. The number of different circuits used for PEC and SNT is $N_\mathrm{circ} = N_\mathrm{shots}/10$, with a layer infidelity of $\varepsilon = 0.5\%$. The region where SNT is the optimal choice is colored in orange, with orange lines representing the boundaries between the three considered QEM techniques. The black lines correspond to contours of constant RMSE.}
    \label{fig:JW_exact}
\end{figure}

To complement the Fermi-Hubbard model simulation results in the main text, we also apply SNT to a related problem of the time evolution of a spinless $t$-$J$ model with the Hamiltonian
\begin{equation}\label{eq:t-j_model}
    \hat{H}=t\sum_i \hat{c}_i^\dagger\hat{c}_{i+1} + \hat{c}_i \hat{c}^\dagger_{i+1} + J \sum_i \hat{c}^\dagger_i  \hat{c}_i \hat{c}^\dagger_{i+1} \hat{c}_{i+1}.
\end{equation}
We consider a 1D chain with 4 sites, set $t=1$ and $J=2$ and employ the JW encoding, which features a single global stabilizer of the form $\mathbb{S}=\{ \mathsf{ZZZZ}\}$. As in the main text, we choose the single-site occupations as the set of observables over which to average the squared bias, so that $\mathbb{O} = \{\hat{n}_i,\, i = 1,2,3,4 \}$. 

In this case, we perform exact density matrix simulations meaning that the bias can be extracted exactly, and without precision limitations. The symmetry verification step is performed by applying the operator $\mathsf{M}_\mathbb{S} = (\mathsf{I} + \mathsf{S})/2$ and subsequently normalizing the density matrix before the final measurement. For SNT the first-order noise cancellation is also performed by applying the corresponding noise inverse. Since the RMSE of PEC depends only on the cost, which in turn depends on the CSP, the RMSE of PEC can be computed without having to resort to any simulation. We consider the sampling strategy described in Sec.~\ref{app:circuit_sampling}.

In these simulations we have fixed the layer infidelity to $\varepsilon_k \equiv \varepsilon$. Since each Trotter step of the circuit is comprised of 6 layers of gates, the CSP is given by $\mathrm{CSP} = (1 - \varepsilon)^{6N_\mathrm{Trotter}}$, i.e. $N_L = 6N_\mathrm{Trotter} $. Unlike the main text, where the CSP is varied by varying the layer infidelity, here we vary the CSP by changing the number of Trotter steps. A parity-check is added, i.e we apply $\mathsf{M}_\mathbb{S}$ after each Trotter step, and we assume that SV is performed via PS.

We compute the RMSE identically as in the main text and plot the values for the best performing QEM method in Supplementary Fig.~\ref{fig:JW_exact}. It reveals a large region (in orange) where SNT achieves the lowest RMSE, outperforming both PS and PEC across several orders of magnitude in the number of shots. Notably, this region includes the range of $10^5-10^6$ shots per observable, which is currently used in QEM experiments on superconducting hardware \cite{IBM_2023_utility,yoshioka_ibm_2024,Google_2020_hartreefock}. PEC has a low bias but is limited by its high cost, making it the preferred choice only when the circuit fault rate is low and the sampling budget is high, thereby keeping the variance under control. In contrast, PS is characterized by a large bias but a small cost, rendering it the best option when the shot budget is severely limited and the circuits are noisy, i.e., in the bottom right part of the plot. However, even in this region where PS is the best choice, its overall performance in terms of RMSE is poor, with errors well above $10\%$, which might hinder the usefulness of the results. These results therefore confirm the data obtained from the larger Fermi-Hubbard simulations in the main text.

\suppsection{\label{app:JW_noise_ratio}Ratio of detectable noise in the JW encoding}

In this section we will analytically derive the ratio of detectable errors $R$ for the JW encoding where the only preserved stabilizer symmetries are the number parity, given by $\mathsf{S}_\text{\o} = \mathsf{Z}^{\otimes N_\mathrm{Q}}$ and the two spin parities (for spin-$1/2$ systems), of the form $\mathsf{S}_\uparrow = \mathsf{Z}^{\otimes N_\mathrm{Q}/2} \otimes \mathsf{I}^{\otimes N_\mathrm{Q}/2}$ and vice versa for $\mathsf{S}_\downarrow$, where we have denoted the total number of qubits with $N_\mathrm{Q}$. We will analyze two limits - the local limit, with the noise restricted to single-qubit errors, and the global limit where the noise is described by a global depolarizing channel. An analysis of both limits of the noise model is necessary to understand the performance for a more realistic scenario which lies in between the two extremes.

\subsection*{\label{subsec:local_errors} Local noise}

More specifically, $R$ can be interpreted as the conditional probability that an error has been detected, provided that an error has occurred. In terms of probabilities (denoted with Pr), we therefore define
\begin{align}
    R &\equiv \mathrm{Pr}[\text{error detected }|\text{ error occurred}] \\
    &= \frac{\mathrm{Pr}[\text{error detected and occurred}]}{\mathrm{Pr}[\text{error occurred}]}.
\end{align}

In the localized error model, we assume that the noise channel before post-selection is comprised solely of uncorrelated single-qubit errors. We denote the probability of observing a Pauli $\mathsf{X}$, $\mathsf{Y}$ or $\mathsf{Z}$ error with $p_\mathsf{X}$, $p_\mathsf{Y}$ and $p_\mathsf{Z}$ respectively, so that the error channel applied to \textit{each} qubit is given by
\begin{equation}
    \mathcal{E}[\bullet] = (1 - \sum_{i \in \{\mathsf{X},\mathsf{Y},\mathsf{Z} \}} p_i) \bullet + p_\mathsf{X} \mathsf{X} \bullet  \mathsf{X}  + p_\mathsf{Y} \mathsf{Y} \bullet  \mathsf{Y}  + p_\mathsf{Z} \mathsf{Z} \bullet  \mathsf{Z}.
\end{equation}

The stabilizers of the JW encoding will be able to detect any odd number of $\mathsf{X}$ or $\mathsf{Y}$ errors occurring on their support. For simplicity, we assume the probability of observing either an $\mathsf{X}$ or $\mathsf{Y}$ error on a single qubit is given by $p_\mathsf{X/Y} = p_\mathsf{X} + p_\mathsf{Y}$. In this case the probability of observing an odd number of $\mathsf{X}$ or $\mathsf{Y}$ errors can be computed by using the relation
\begin{align}
    \mathrm{Pr}[\text{odd \# of $\mathsf{X},\mathsf{Y}$}] + \mathrm{Pr}[\text{even \# of $\mathsf{X},\mathsf{Y}$}] &= 1, \label{eq:even_plus_odd_number_of_errors}\\
    \mathrm{Pr}[\text{even \# of $\mathsf{X},\mathsf{Y}$}] - \mathrm{Pr}[\text{odd \# of $\mathsf{X},\mathsf{Y}$}] &= \nonumber \\
     ((1 - p_\mathsf{X/Y}) - p_\mathsf{X/Y})^{w_\mathsf{S}},\label{eq:even_minus_odd_number_of_errors} &
\end{align}
where the latter expression is derived from the properties of a binomial distribution, and $w_\mathsf{S}$ is the weight of the stabilizer. For the spinless example where $w_\mathsf{S} = N_\mathrm{Q}$, we can use the above result to express the fraction of detectable noise $R_\text{JW,\o}^\text{local}$:
\begin{align}
    R_\text{JW,\o}^\text{local} &= \frac{\mathrm{Pr}[\text{odd \# of $\mathsf{X},\mathsf{Y}$}]}{\mathrm{Pr}[\text{error occurred}]} \\
    &= \frac{1}{2}\frac{1 -  \left( 1 - 2p_\mathsf{X/Y} \right)^{N_\mathrm{Q}} }{1 - \left( 1 - p_\mathsf{X/Y} - p_\mathsf{Z}\right)^{N_\mathrm{Q}}}.
\end{align}
This result converges to the value $R_\text{JW,\o}^\text{local} = 1/2$, in the increasingly noisy limit of ${N_\mathrm{Q} \rightarrow \infty} $ with fixed $p_\mathsf{X,Y,Z}$, and also with in the case of a fixed circuit success probability $\text{CSP} = (1 - \sum_{i \in \{\mathsf{X},\mathsf{Y},\mathsf{Z} \}} p_i)^{N_\mathrm{Q}} $. On the other hand, in the low-noise limit of $p_\mathsf{X,Y,Z} \rightarrow 0$, $R_\text{JW,\o}^\text{local}$ converges to $p_\mathsf{X/Y}/(p_\mathsf{Z} + p_\mathsf{X/Y})$. Assuming the probabilities $p_\mathsf{X}$, $p_\mathsf{Y}$ and $p_\mathsf{Z}$  are equal, we therefore expect $R_\text{JW,\o}^\text{local} \sim 50\% -  67 \%$.

When two spin symmetries are preserved, we will detect an error provided that an odd number of $\mathsf{X}$ or $\mathsf{Y}$ errors have occurred on the support of \textit{either} of the two stabilizers. Equivalently, an error will be detected \textit{unless} an even number of $\mathsf{X}$ or $\mathsf{Y}$ errors has occurred on the supports of \textit{both} $\mathsf{S}_\uparrow$ and $\mathsf{S}_\downarrow$. Relying on the single-qubit error model, we note that the ratio of detectable noise in this case is given by
\begin{align}
    R_{\text{JW},\uparrow/\downarrow}^\text{local} &= \frac{1 - \mathrm{Pr}[\text{even \# of $\mathsf{X},\mathsf{Y}$ on $N_\mathrm{Q}/2$ qubits}]^2 }{\mathrm{Pr}[\text{error occurred}]}. 
\end{align}
Using Eqs.~\ref{eq:even_plus_odd_number_of_errors} and \ref{eq:even_minus_odd_number_of_errors}  to evaluate the expression in the numerator, we obtain 
\begin{equation}
   R_{\text{JW},\uparrow/\downarrow}^\text{local} = \frac{1 - \frac{1}{4}\left(1 -  [1 - 2 p_\mathsf{X/Y}]^{N_\mathrm{Q}/2}\right)^2 }{1 -  \left( 1 - p_\mathsf{X/Y} - p_\mathsf{Z}\right)^{N_\mathrm{Q}} } .
\end{equation}
As in the spinless example, the ratio of detectable errors in the high-fidelity limit is given by $p_\mathsf{X/Y}/(p_\mathsf{Z} + p_\mathsf{X/Y})$, while in the large $N_\mathrm{Q}$ and low-fidelity limits $R_{\text{JW},\uparrow/\downarrow}^\text{local}$ converges to the value of $3/4$. The expected range for $R_{\text{JW},\uparrow/\downarrow}^\text{local}$ is therefore $67\% -  75 \%$. 

\subsection*{\label{subsec:global_errors} Global Depolarizing noise}

On the other hand, provided the circuits are sufficiently deep, we can approximate the noise at the end of the circuit by a global depolarizing noise channel - an approximation which is valid for deep, random circuits \cite{Tsubouchi_2023}. The global depolarizing noise channel acting on $N_\mathrm{Q}$ qubits is defined as 
\begin{equation}
\mathcal{E}[\bullet] = (1 - p)\bullet  +\, p \sum_{i|\mathsf{P}_i \in \mathbb{P}^{\otimes N_\mathrm{Q}} \setminus \{ \mathsf{I}^{\otimes N_\mathrm{Q}} \} } \mathsf{P}_i \bullet \mathsf{P}_i .
\end{equation}
In this case the ratio of undetectable errors is determined by the number of all possible Pauli operators $ \{ \mathsf{I},\mathsf{X},\mathsf{Y},\mathsf{Z}\}^{\otimes N_\mathrm{Q}}$ on $N_\mathrm{Q}$ qubits, which commute with all the elements of the stabilizer set $\mathbb{S}$. We will denote the size of the set $|\mathbb{S}|$ as the number of distinct Pauli operators in the set (ignoring the phases). Under this definition, the size of the set is determined by the number of generators $g$ so that $|\mathbb{S}| = 2^g$. For JW, $g=1$ if no spin-symmetry is present in the Hamiltonian, and $g=2$ otherwise.

The set of Paulis that commute with the stabilizer group are known as the centralizer of the stabilizer group, which we will denote with $\mathbb{C}(\mathbb{S}) = \{\mathsf{P} \in \mathbb{P}^{\otimes N_\mathrm{Q}} | [\mathsf{P},\mathsf{S}] = 0, \, \forall \mathsf{S} \in \mathbb{S} \}$. The size of the set $|\mathbb{C}(\mathbb{S})|$ is therefore directly related to the number of undetectable errors and can be computed by noting
\begin{align}
    |\mathbb{C}(\mathbb{S})| &= |\mathbb{C}(\mathbb{S}) / \mathbb{S} | \cdot | \mathbb{S} | = 4^{N_\mathrm{Q} - g} \cdot 2^g,
\end{align}
where we have denoted the quotient of two groups with a $/$ and used the fact that the quotient group is isomorphic to the Pauli group, i.e. $\mathbb{C}(\mathbb{S})/\mathbb{S} \simeq \mathbb{P}^{\otimes (N_\mathrm{Q} - g)} $, which is a result of $ |\mathbb{P}^{\otimes N_\mathrm{Q}}| = |\mathbb{P}^{\otimes N_\mathrm{Q}}/\mathbb{C}(\mathbb{S})| \cdot|\mathbb{C}(\mathbb{S}) | $. We have therefore established that for the global depolarizing channel case, there are $ |\mathbb{C}(\mathbb{S})| = 4^{N_\mathrm{Q} - g/2}$ Paulis which are not detectable by symmetry verification.

The ratio $R_\mathrm{JW}^\text{global}$ of all detectable Pauli errors in the global depolarizing noise limit is therefore given by $R_\mathrm{JW}^\text{global} = 4^{N_\mathrm{Q} }(1 - 2^{-g})/(4^{N_\mathrm{Q}} - 1)$, where the -1 is added to account for the identity. In the limit $N_\mathrm{Q} \gg 1$, $R \approx  1 - 2^{-g}$, suggesting that even the JW encoding will be able to detect 50\% of all errors in the most general case (just with the number parity conservation, $g=1$) and 75\% of errors with spin symmetry ($g=2$). Together with the previous results for local noise $R_\text{JW,\o}^\text{local} \sim 50\% -  67 \%$ and $R_{\text{JW},\uparrow/\downarrow}^\text{local} \sim 67\% -  75 \%$, we therefore expect that with a more realstic noise model composed mostly of single- and two-qubit errors $R_{\text{JW},\uparrow/\downarrow} \sim 50\% - 67 \%$, while for the case of two spin-parity stabilizers $R_{\text{JW},\uparrow/\downarrow} \sim 67\%-75\%$. These numbers can be translated into a SNT cost coefficient by using the relation $\beta_\mathrm{JW+SNT} = R/2  + 2 (1-R)$ (with post-selection) or alternatively $\beta_\mathrm{JW+SNT} = R  + 2 (1-R)$ (with post-processing).

\bibliography{bibliography}

@article{simkovic_2024,
  title = {Low-weight high-distance error-correcting fermionic encodings},
  author = {Fedor \v{S}imkovic IV and Martin Leib and Francisco Revson F. Pereira},
  journal = {Phys. Rev. Res.},
  volume = {6},
  issue = {4},
  pages = {043123},
  numpages = {18},
  year = {2024},
  month = {Nov},
  publisher = {American Physical Society},
  doi = {10.1103/PhysRevResearch.6.043123},
  url = {https://link.aps.org/doi/10.1103/PhysRevResearch.6.043123}
}

@article{Jordan1928,
	doi = {10.1007/bf01331938},
	url = {https://doi.org/10.1007%2Fbf01331938},
	year = 1928,
	month = {sep},
	publisher = {Springer Science and Business Media {LLC}},
	volume = {47},
	number = {9-10},
	pages = {631--651},
	author = {P. Jordan and E. Wigner},
	title = {Über das Paulische Äquivalenzverbot},
	journal = {Zeitschrift für Physik}
}

@article{bausch2020mitigating,
  title={Mitigating errors in local fermionic encodings},
  author={Bausch, Johannes and Cubitt, Toby and Derby, Charles and Klassen, Joel},
  journal={arXiv preprint arXiv:2003.07125},
  year={2020}
}

@article{Derby_2021,
   title={Compact fermion to qubit mappings},
   volume={104},
   ISSN={2469-9969},
   url={http://dx.doi.org/10.1103/PhysRevB.104.035118},
   DOI={10.1103/physrevb.104.035118},
   number={3},
   journal={Physical Review B},
   publisher={American Physical Society (APS)},
   author={Derby, Charles and Klassen, Joel and Bausch, Johannes and Cubitt, Toby},
   year={2021},
   month=jul }

@article{Algaba_2024,
   title={Low-depth simulations of fermionic systems on square-grid quantum hardware},
   volume={8},
   ISSN={2521-327X},
   url={http://dx.doi.org/10.22331/q-2024-04-30-1327},
   DOI={10.22331/q-2024-04-30-1327},
   journal={Quantum},
   publisher={Verein zur Forderung des Open Access Publizierens in den Quantenwissenschaften},
   author={Algaba, Manuel G. and Sriluckshmy, P. V. and Leib, Martin and \v{S}imkovic IV, Fedor},
   year={2024},
   month=apr, pages={1327} }

@article{Sriluckshmy_2023,
   title={Optimal, hardware native decomposition of parameterized multi-qubit Pauli gates},
   volume={8},
   ISSN={2058-9565},
   url={http://dx.doi.org/10.1088/2058-9565/acfa20},
   DOI={10.1088/2058-9565/acfa20},
   number={4},
   journal={Quantum Science and Technology},
   publisher={IOP Publishing},
   author={Sriluckshmy, P. V. and Pina-Canelles, Vicente and Ponce, Mario and Algaba, Manuel G. and IV, Fedor \v{S}imkovic and Leib, Martin},
   year={2023},
   month=sep, pages={045029} }

@article{Cowtan_2020,
   title={Phase Gadget Synthesis for Shallow Circuits},
   volume={318},
   ISSN={2075-2180},
   url={http://dx.doi.org/10.4204/EPTCS.318.13},
   DOI={10.4204/eptcs.318.13},
   journal={Electronic Proceedings in Theoretical Computer Science},
   publisher={Open Publishing Association},
   author={Cowtan, Alexander and Dilkes, Silas and Duncan, Ross and Simmons, Will and Sivarajah, Seyon},
   year={2020},
   month=may, pages={213–228} }

@article{BonetMonroig_2018,
  title = {Low-cost error mitigation by symmetry verification},
  author = {Bonet-Monroig, X. and Sagastizabal, R. and Singh, M. and O'Brien, T. E.},
  journal = {Phys. Rev. A},
  volume = {98},
  issue = {6},
  pages = {062339},
  numpages = {10},
  year = {2018},
  month = {Dec},
  publisher = {American Physical Society},
  doi = {10.1103/PhysRevA.98.062339},
  url = {https://link.aps.org/doi/10.1103/PhysRevA.98.062339}
}

@article{Cai_2023_QEMreview,
  title = {Quantum error mitigation},
  author = {Cai, Zhenyu and Babbush, Ryan and Benjamin, Simon C. and Endo, Suguru and Huggins, William J. and Li, Ying and McClean, Jarrod R. and O'Brien, Thomas E.},
  journal = {Rev. Mod. Phys.},
  volume = {95},
  issue = {4},
  pages = {045005},
  numpages = {37},
  year = {2023},
  month = {Dec},
  publisher = {American Physical Society},
  doi = {10.1103/RevModPhys.95.045005},
  url = {https://link.aps.org/doi/10.1103/RevModPhys.95.045005}
}

@article{McArdle_2019,
  title = {Error-Mitigated Digital Quantum Simulation},
  author = {McArdle, Sam and Yuan, Xiao and Benjamin, Simon},
  journal = {Phys. Rev. Lett.},
  volume = {122},
  issue = {18},
  pages = {180501},
  numpages = {6},
  year = {2019},
  month = {May},
  publisher = {American Physical Society},
  doi = {10.1103/PhysRevLett.122.180501},
  url = {https://link.aps.org/doi/10.1103/PhysRevLett.122.180501}
}

@article{verstraete2005mapping,
  title={Mapping local Hamiltonians of fermions to local Hamiltonians of spins},
  author={Verstraete, Frank and Cirac, J Ignacio},
  journal={Journal of Statistical Mechanics: Theory and Experiment},
  volume={2005},
  number={09},
  pages={P09012},
  year={2005},
  publisher={IOP Publishing}
}

@article{chien_2022,
  title={Optimizing fermionic encodings for both Hamiltonian and hardware},
  author={Chien, Riley W and Klassen, Joel},
  journal={arXiv preprint arXiv:2210.05652},
  year={2022}
}

@article{setia2019superfast,
  title={Superfast encodings for fermionic quantum simulation},
  author={Setia, Kanav and Bravyi, Sergey and Mezzacapo, Antonio and Whitfield, James D},
  journal={Physical Review Research},
  volume={1},
  number={3},
  pages={033033},
  year={2019},
  publisher={APS}
}

@article{jiang2019majorana,
  title={Majorana loop stabilizer codes for error mitigation in fermionic quantum simulations},
  author={Jiang, Zhang and McClean, Jarrod and Babbush, Ryan and Neven, Hartmut},
  journal={Physical Review Applied},
  volume={12},
  number={6},
  pages={064041},
  year={2019},
  publisher={APS}
}

@article{chen2024error,
  title={Error-correcting codes for fermionic quantum simulation},
  author={Chen, Yu-An and Gorshkov, Alexey V and Xu, Yijia},
  journal={SciPost Physics},
  volume={16},
  number={1},
  pages={033},
  year={2024}
}

@article{Google_2020_hartreefock,
   title={Hartree-Fock on a superconducting qubit quantum computer},
   volume={369},
   ISSN={1095-9203},
   url={http://dx.doi.org/10.1126/science.abb9811},
   DOI={10.1126/science.abb9811},
   number={6507},
   journal={Science},
   publisher={American Association for the Advancement of Science (AAAS)},
   author={Arute, Frank and Arya, Kunal and Babbush, Ryan and Bacon, Dave and Bardin, Joseph C. and Barends, Rami and Boixo, Sergio and Broughton, Michael and Buckley, Bob B. and Buell, David A. and Burkett, Brian and Bushnell, Nicholas and Chen, Yu and Chen, Zijun and Chiaro, Benjamin and Collins, Roberto and Courtney, William and Demura, Sean and Dunsworth, Andrew and Farhi, Edward and Fowler, Austin and Foxen, Brooks and Gidney, Craig and Giustina, Marissa and Graff, Rob and Habegger, Steve and Harrigan, Matthew P. and Ho, Alan and Hong, Sabrina and Huang, Trent and Huggins, William J. and Ioffe, Lev and Isakov, Sergei V. and Jeffrey, Evan and Jiang, Zhang and Jones, Cody and Kafri, Dvir and Kechedzhi, Kostyantyn and Kelly, Julian and Kim, Seon and Klimov, Paul V. and Korotkov, Alexander and Kostritsa, Fedor and Landhuis, David and Laptev, Pavel and Lindmark, Mike and Lucero, Erik and Martin, Orion and Martinis, John M. and McClean, Jarrod R. and McEwen, Matt and Megrant, Anthony and Mi, Xiao and Mohseni, Masoud and Mruczkiewicz, Wojciech and Mutus, Josh and Naaman, Ofer and Neeley, Matthew and Neill, Charles and Neven, Hartmut and Niu, Murphy Yuezhen and O’Brien, Thomas E. and Ostby, Eric and Petukhov, Andre and Putterman, Harald and Quintana, Chris and Roushan, Pedram and Rubin, Nicholas C. and Sank, Daniel and Satzinger, Kevin J. and Smelyanskiy, Vadim and Strain, Doug and Sung, Kevin J. and Szalay, Marco and Takeshita, Tyler Y. and Vainsencher, Amit and White, Theodore and Wiebe, Nathan and Yao, Z. Jamie and Yeh, Ping and Zalcman, Adam},
   year={2020},
   month=aug, pages={1084–1089} }

@article{Chen_2023,
   title={Error-mitigated quantum simulation of interacting fermions with trapped ions},
   volume={9},
   ISSN={2056-6387},
   url={http://dx.doi.org/10.1038/s41534-023-00784-8},
   DOI={10.1038/s41534-023-00784-8},
   number={1},
   journal={npj Quantum Information},
   publisher={Springer Science and Business Media LLC},
   author={Chen, Wentao and Zhang, Shuaining and Zhang, Jialiang and Su, Xiaolu and Lu, Yao and Zhang, Kuan and Qiao, Mu and Li, Ying and Zhang, Jing-Ning and Kim, Kihwan},
   year={2023},
   month=dec }

@misc{chien_2023,
      title={Simulating quantum error mitigation in fermionic encodings}, 
      author={Riley W. Chien and Kanav Setia and Xavier Bonet-Monroig and Mark Steudtner and James D. Whitfield},
      year={2023},
      eprint={2303.02270},
      archivePrefix={arXiv},
      primaryClass={quant-ph},
      url={https://arxiv.org/abs/2303.02270}, 
}

@article{guimaraes_2024,
  title = {Noise-Assisted Digital Quantum Simulation of Open Systems Using Partial Probabilistic Error Cancellation},
  author = {Guimar\~aes, Jos\'e D. and Lim, James and Vasilevskiy, Mikhail I. and Huelga, Susana F. and Plenio, Martin B.},
  journal = {PRX Quantum},
  volume = {4},
  issue = {4},
  pages = {040329},
  numpages = {19},
  year = {2023},
  month = {Nov},
  publisher = {American Physical Society},
  doi = {10.1103/PRXQuantum.4.040329},
  url = {https://link.aps.org/doi/10.1103/PRXQuantum.4.040329}
}

@article{van_den_Berg_2023_spl_pec,
   title={Probabilistic error cancellation with sparse Pauli–Lindblad models on noisy quantum processors},
   volume={19},
   ISSN={1745-2481},
   url={http://dx.doi.org/10.1038/s41567-023-02042-2},
   DOI={10.1038/s41567-023-02042-2},
   number={8},
   journal={Nature Physics},
   publisher={Springer Science and Business Media LLC},
   author={van den Berg, Ewout and Minev, Zlatko K. and Kandala, Abhinav and Temme, Kristan},
   year={2023},
   month=may, pages={1116–1121} }

@article{hashim_2021_rc,
  title = {Randomized Compiling for Scalable Quantum Computing on a Noisy Superconducting Quantum Processor},
  author = {Hashim, Akel and Naik, Ravi K. and Morvan, Alexis and Ville, Jean-Loup and Mitchell, Bradley and Kreikebaum, John Mark and Davis, Marc and Smith, Ethan and Iancu, Costin and O'Brien, Kevin P. and Hincks, Ian and Wallman, Joel J. and Emerson, Joseph and Siddiqi, Irfan},
  journal = {Phys. Rev. X},
  volume = {11},
  issue = {4},
  pages = {041039},
  numpages = {12},
  year = {2021},
  month = {Nov},
  publisher = {American Physical Society},
  doi = {10.1103/PhysRevX.11.041039},
  url = {https://link.aps.org/doi/10.1103/PhysRevX.11.041039}
}

@article{Cai_2021,
   title={Multi-exponential error extrapolation and combining error mitigation techniques for NISQ applications},
   volume={7},
   ISSN={2056-6387},
   url={http://dx.doi.org/10.1038/s41534-021-00404-3},
   DOI={10.1038/s41534-021-00404-3},
   number={1},
   journal={npj Quantum Information},
   publisher={Springer Science and Business Media LLC},
   author={Cai, Zhenyu},
   year={2021},
   month=may }

@article{McClean_2020,
   title={Decoding quantum errors with subspace expansions},
   volume={11},
   ISSN={2041-1723},
   url={http://dx.doi.org/10.1038/s41467-020-14341-w},
   DOI={10.1038/s41467-020-14341-w},
   number={1},
   journal={Nature Communications},
   publisher={Springer Science and Business Media LLC},
   author={McClean, Jarrod R. and Jiang, Zhang and Rubin, Nicholas C. and Babbush, Ryan and Neven, Hartmut},
   year={2020},
   month=jan }

@article{Huggins_2021,
   title={Efficient and noise resilient measurements for quantum chemistry on near-term quantum computers},
   volume={7},
   ISSN={2056-6387},
   url={http://dx.doi.org/10.1038/s41534-020-00341-7},
   DOI={10.1038/s41534-020-00341-7},
   number={1},
   journal={npj Quantum Information},
   publisher={Springer Science and Business Media LLC},
   author={Huggins, William J. and McClean, Jarrod R. and Rubin, Nicholas C. and Jiang, Zhang and Wiebe, Nathan and Whaley, K. Birgitta and Babbush, Ryan},
   year={2021},
   month=feb }

@article{Terhal_2015,
   title={Quantum error correction for quantum memories},
   volume={87},
   ISSN={1539-0756},
   url={http://dx.doi.org/10.1103/RevModPhys.87.307},
   DOI={10.1103/revmodphys.87.307},
   number={2},
   journal={Reviews of Modern Physics},
   publisher={American Physical Society (APS)},
   author={Terhal, Barbara M.},
   year={2015},
   month=apr, pages={307–346} }

@article{Sagastizabal_2019,
   title={Experimental error mitigation via symmetry verification in a variational quantum eigensolver},
   volume={100},
   ISSN={2469-9934},
   url={http://dx.doi.org/10.1103/PhysRevA.100.010302},
   DOI={10.1103/physreva.100.010302},
   number={1},
   journal={Physical Review A},
   publisher={American Physical Society (APS)},
   author={Sagastizabal, R. and Bonet-Monroig, X. and Singh, M. and Rol, M. A. and Bultink, C. C. and Fu, X. and Price, C. H. and Ostroukh, V. P. and Muthusubramanian, N. and Bruno, A. and Beekman, M. and Haider, N. and O’Brien, T. E. and DiCarlo, L.},
   year={2019},
   month=jul }

@article{Stanisic_2022,
   title={Observing ground-state properties of the Fermi-Hubbard model using a scalable algorithm on a quantum computer},
   volume={13},
   ISSN={2041-1723},
   url={http://dx.doi.org/10.1038/s41467-022-33335-4},
   DOI={10.1038/s41467-022-33335-4},
   number={1},
   journal={Nature Communications},
   publisher={Springer Science and Business Media LLC},
   author={Stanisic, Stasja and Bosse, Jan Lukas and Gambetta, Filippo Maria and Santos, Raul A. and Mruczkiewicz, Wojciech and O’Brien, Thomas E. and Ostby, Eric and Montanaro, Ashley},
   year={2022},
   month=oct }

@misc{google_2020b_chargespin,
      title={Observation of separated dynamics of charge and spin in the Fermi-Hubbard model}, 
      author={Frank Arute and Kunal Arya and Ryan Babbush and Dave Bacon and Joseph C. Bardin and Rami Barends and Andreas Bengtsson and Sergio Boixo and Michael Broughton and Bob B. Buckley and David A. Buell and Brian Burkett and Nicholas Bushnell and Yu Chen and Zijun Chen and Yu-An Chen and Ben Chiaro and Roberto Collins and Stephen J. Cotton and William Courtney and Sean Demura and Alan Derk and Andrew Dunsworth and Daniel Eppens and Thomas Eckl and Catherine Erickson and Edward Farhi and Austin Fowler and Brooks Foxen and Craig Gidney and Marissa Giustina and Rob Graff and Jonathan A. Gross and Steve Habegger and Matthew P. Harrigan and Alan Ho and Sabrina Hong and Trent Huang and William Huggins and Lev B. Ioffe and Sergei V. Isakov and Evan Jeffrey and Zhang Jiang and Cody Jones and Dvir Kafri and Kostyantyn Kechedzhi and Julian Kelly and Seon Kim and Paul V. Klimov and Alexander N. Korotkov and Fedor Kostritsa and David Landhuis and Pavel Laptev and Mike Lindmark and Erik Lucero and Michael Marthaler and Orion Martin and John M. Martinis and Anika Marusczyk and Sam McArdle and Jarrod R. McClean and Trevor McCourt and Matt McEwen and Anthony Megrant and Carlos Mejuto-Zaera and Xiao Mi and Masoud Mohseni and Wojciech Mruczkiewicz and Josh Mutus and Ofer Naaman and Matthew Neeley and Charles Neill and Hartmut Neven and Michael Newman and Murphy Yuezhen Niu and Thomas E. O'Brien and Eric Ostby and Bálint Pató and Andre Petukhov and Harald Putterman and Chris Quintana and Jan-Michael Reiner and Pedram Roushan and Nicholas C. Rubin and Daniel Sank and Kevin J. Satzinger and Vadim Smelyanskiy and Doug Strain and Kevin J. Sung and Peter Schmitteckert and Marco Szalay and Norm M. Tubman and Amit Vainsencher and Theodore White and Nicolas Vogt and Z. Jamie Yao and Ping Yeh and Adam Zalcman and Sebastian Zanker},
      year={2020},
      eprint={2010.07965},
      archivePrefix={arXiv},
      primaryClass={quant-ph},
      url={https://arxiv.org/abs/2010.07965}, 
}

@article{Dborin_2022,
   title={Simulating groundstate and dynamical quantum phase transitions on a superconducting quantum computer},
   volume={13},
   ISSN={2041-1723},
   url={http://dx.doi.org/10.1038/s41467-022-33737-4},
   DOI={10.1038/s41467-022-33737-4},
   number={1},
   journal={Nature Communications},
   publisher={Springer Science and Business Media LLC},
   author={Dborin, James and Wimalaweera, Vinul and Barratt, F. and Ostby, Eric and O’Brien, Thomas E. and Green, A. G.},
   year={2022},
   month=oct }

@misc{yoshioka_ibm_2024,
      title={Diagonalization of large many-body Hamiltonians on a quantum processor}, 
      author={Nobuyuki Yoshioka and Mirko Amico and William Kirby and Petar Jurcevic and Arkopal Dutt and Bryce Fuller and Shelly Garion and Holger Haas and Ikko Hamamura and Alexander Ivrii and Ritajit Majumdar and Zlatko Minev and Mario Motta and Bibek Pokharel and Pedro Rivero and Kunal Sharma and Christopher J. Wood and Ali Javadi-Abhari and Antonio Mezzacapo},
      year={2024},
      eprint={2407.14431},
      archivePrefix={arXiv},
      primaryClass={quant-ph},
      url={https://arxiv.org/abs/2407.14431}, 
}

@article{Erhard_2019_CB,
   title={Characterizing large-scale quantum computers via cycle benchmarking},
   volume={10},
   ISSN={2041-1723},
   url={http://dx.doi.org/10.1038/s41467-019-13068-7},
   DOI={10.1038/s41467-019-13068-7},
   number={1},
   journal={Nature Communications},
   publisher={Springer Science and Business Media LLC},
   author={Erhard, Alexander and Wallman, Joel J. and Postler, Lukas and Meth, Michael and Stricker, Roman and Martinez, Esteban A. and Schindler, Philipp and Monz, Thomas and Emerson, Joseph and Blatt, Rainer},
   year={2019},
   month=nov }

@inproceedings{flammia_2022_ACES,
  doi = {10.4230/LIPICS.TQC.2022.4},
  url = {https://drops.dagstuhl.de/entities/document/10.4230/LIPIcs.TQC.2022.4},
  author = {Flammia, Steven T.},  
  title = {Averaged Circuit Eigenvalue Sampling},
  publisher = {Schloss Dagstuhl – Leibniz-Zentrum f\"{u}r Informatik},
  year = {2022},
  copyright = {Creative Commons Attribution 4.0 International license}
}

@misc{calzona_2024,
      title={Multi-Layer Cycle Benchmarking for high-accuracy error characterization}, 
      author={Alessio Calzona and Miha Papič and Pedro Figueroa-Romero and Adrian Auer},
      year={2024},
      eprint={2412.09332},
      archivePrefix={arXiv},
      primaryClass={quant-ph},
      url={https://arxiv.org/abs/2412.09332}, 
}

@article{Chen2023Jan,
	author = {Chen, Senrui and Liu, Yunchao and Otten, Matthew and Seif, Alireza and Fefferman, Bill and Jiang, Liang},
	title = {{The learnability of Pauli noise}},
	journal = {Nat. Commun.},
	volume = {14},
	number = {52},
	pages = {1--8},
	year = {2023},
	month = jan,
	issn = {2041-1723},
	publisher = {Nature Publishing Group},
	doi = {10.1038/s41467-022-35759-4}
}

@article{van_den_Berg_2024,
   title={Techniques for learning sparse Pauli-Lindblad noise models},
   volume={8},
   ISSN={2521-327X},
   url={http://dx.doi.org/10.22331/q-2024-12-10-1556},
   DOI={10.22331/q-2024-12-10-1556},
   journal={Quantum},
   publisher={Verein zur Forderung des Open Access Publizierens in den Quantenwissenschaften},
   author={van den Berg, Ewout and Wocjan, Pawel},
   year={2024},
   month=dec, pages={1556} }

@article{Temme_2017_PEC,
   title={Error Mitigation for Short-Depth Quantum Circuits},
   volume={119},
   ISSN={1079-7114},
   url={http://dx.doi.org/10.1103/PhysRevLett.119.180509},
   DOI={10.1103/physrevlett.119.180509},
   number={18},
   journal={Physical Review Letters},
   publisher={American Physical Society (APS)},
   author={Temme, Kristan and Bravyi, Sergey and Gambetta, Jay M.},
   year={2017},
   month=nov }

@article{Li_2017_PEC,
   title={Efficient Variational Quantum Simulator Incorporating Active Error Minimization},
   volume={7},
   ISSN={2160-3308},
   url={http://dx.doi.org/10.1103/PhysRevX.7.021050},
   DOI={10.1103/physrevx.7.021050},
   number={2},
   journal={Physical Review X},
   publisher={American Physical Society (APS)},
   author={Li, Ying and Benjamin, Simon C.},
   year={2017},
   month=jun }

@article{IBM_2023_utility,
	author = {Kim, Youngseok and Eddins, Andrew and Anand, Sajant and Wei, Ken Xuan and van den Berg, Ewout and Rosenblatt, Sami and Nayfeh, Hasan and Wu, Yantao and Zaletel, Michael and Temme, Kristan and Kandala, Abhinav},
	date = {2023/06/01},
	doi = {10.1038/s41586-023-06096-3},
	id = {Kim2023},
	isbn = {1476-4687},
	journal = {Nature},
	number = {7965},
	pages = {500--505},
	title = {Evidence for the utility of quantum computing before fault tolerance},
	url = {https://doi.org/10.1038/s41586-023-06096-3},
	volume = {618},
	year = {2023}}

@article{Huggins_2021_VD,
   title={Virtual Distillation for Quantum Error Mitigation},
   volume={11},
   ISSN={2160-3308},
   url={http://dx.doi.org/10.1103/PhysRevX.11.041036},
   DOI={10.1103/physrevx.11.041036},
   number={4},
   journal={Physical Review X},
   publisher={American Physical Society (APS)},
   author={Huggins, William J. and McArdle, Sam and O’Brien, Thomas E. and Lee, Joonho and Rubin, Nicholas C. and Boixo, Sergio and Whaley, K. Birgitta and Babbush, Ryan and McClean, Jarrod R.},
   year={2021},
   month=nov }

@misc{czarnik_2021_VD,
      title={Qubit-efficient exponential suppression of errors}, 
      author={Piotr Czarnik and Andrew Arrasmith and Lukasz Cincio and Patrick J. Coles},
      year={2021},
      eprint={2102.06056},
      archivePrefix={arXiv},
      primaryClass={quant-ph},
      url={https://arxiv.org/abs/2102.06056}, 
}

@misc{filippov_2023_TEM,
      title={Scalable tensor-network error mitigation for near-term quantum computing}, 
      author={Sergei Filippov and Matea Leahy and Matteo A. C. Rossi and Guillermo García-Pérez},
      year={2023},
      eprint={2307.11740},
      archivePrefix={arXiv},
      primaryClass={quant-ph},
      url={https://arxiv.org/abs/2307.11740}, 
}

@article{OBrien_2021_EV,
   title={Error Mitigation via Verified Phase Estimation},
   volume={2},
   ISSN={2691-3399},
   url={http://dx.doi.org/10.1103/PRXQuantum.2.020317},
   DOI={10.1103/prxquantum.2.020317},
   number={2},
   journal={PRX Quantum},
   publisher={American Physical Society (APS)},
   author={O’Brien, Thomas E. and Polla, Stefano and Rubin, Nicholas C. and Huggins, William J. and McArdle, Sam and Boixo, Sergio and McClean, Jarrod R. and Babbush, Ryan},
   year={2021},
   month=may }

@article{Huo_2022_EV,
   title={Dual-state purification for practical quantum error mitigation},
   volume={105},
   ISSN={2469-9934},
   url={http://dx.doi.org/10.1103/PhysRevA.105.022427},
   DOI={10.1103/physreva.105.022427},
   number={2},
   journal={Physical Review A},
   publisher={American Physical Society (APS)},
   author={Huo, Mingxia and Li, Ying},
   year={2022},
   month=feb }

@misc{hockings_2024_ACES,
      title={Scalable noise characterisation of syndrome extraction circuits with averaged circuit eigenvalue sampling}, 
      author={Evan T. Hockings and Andrew C. Doherty and Robin Harper},
      year={2024},
      eprint={2404.06545},
      archivePrefix={arXiv},
      primaryClass={quant-ph},
      url={https://arxiv.org/abs/2404.06545}, 
}

@misc{pelaez_2024_ACES,
      title={Average circuit eigenvalue sampling on NISQ devices}, 
      author={Emilio Pelaez and Victory Omole and Pranav Gokhale and Rich Rines and Kaitlin N. Smith and Michael A. Perlin and Akel Hashim},
      year={2024},
      eprint={2403.12857},
      archivePrefix={arXiv},
      primaryClass={quant-ph},
      url={https://arxiv.org/abs/2403.12857}, 
}

@misc{tran_2023_cones,
      title={Locality and Error Mitigation of Quantum Circuits}, 
      author={Minh C. Tran and Kunal Sharma and Kristan Temme},
      year={2023},
      eprint={2303.06496},
      archivePrefix={arXiv},
      primaryClass={quant-ph},
      url={https://arxiv.org/abs/2303.06496}, 
}

@article{bravyi2002fermionic,
  title={Fermionic quantum computation},
  author={Bravyi, Sergey B and Kitaev, Alexei Yu},
  journal={Annals of Physics},
  volume={298},
  number={1},
  pages={210--226},
  year={2002},
  publisher={Elsevier}
}

@misc{eddins_2024_cones,
      title={Lightcone shading for classically accelerated quantum error mitigation}, 
      author={Andrew Eddins and Minh C. Tran and Patrick Rall},
      year={2024},
      eprint={2409.04401},
      archivePrefix={arXiv},
      primaryClass={quant-ph},
      url={https://arxiv.org/abs/2409.04401}, 
}

@article{Kandala_2017,
   title={Hardware-efficient variational quantum eigensolver for small molecules and quantum magnets},
   volume={549},
   ISSN={1476-4687},
   url={http://dx.doi.org/10.1038/nature23879},
   DOI={10.1038/nature23879},
   number={7671},
   journal={Nature},
   publisher={Springer Science and Business Media LLC},
   author={Kandala, Abhinav and Mezzacapo, Antonio and Temme, Kristan and Takita, Maika and Brink, Markus and Chow, Jerry M. and Gambetta, Jay M.},
   year={2017},
   month=sep, pages={242–246} }

@article{Bravyi_2021_REM,
   title={Mitigating measurement errors in multiqubit experiments},
   volume={103},
   ISSN={2469-9934},
   url={http://dx.doi.org/10.1103/PhysRevA.103.042605},
   DOI={10.1103/physreva.103.042605},
   number={4},
   journal={Physical Review A},
   publisher={American Physical Society (APS)},
   author={Bravyi, Sergey and Sheldon, Sarah and Kandala, Abhinav and Mckay, David C. and Gambetta, Jay M.},
   year={2021},
   month=apr }

@misc{fruitwala_2024_FPGA_RC,
      title={Hardware-Efficient Randomized Compiling}, 
      author={Neelay Fruitwala and Akel Hashim and Abhi D. Rajagopala and Yilun Xu and Jordan Hines and Ravi K. Naik and Irfan Siddiqi and Katherine Klymko and Gang Huang and Kasra Nowrouzi},
      year={2024},
      eprint={2406.13967},
      archivePrefix={arXiv},
      primaryClass={quant-ph},
      url={https://arxiv.org/abs/2406.13967}, 
}

@article{google_2024_qec,
	author = {Acharya, Rajeev and Abanin, Dmitry A. and Aghababaie-Beni, Laleh and Aleiner, Igor and Andersen, Trond I. and Ansmann, Markus and Arute, Frank and Arya, Kunal and Asfaw, Abraham and Astrakhantsev, Nikita and Atalaya, Juan and Babbush, Ryan and Bacon, Dave and Ballard, Brian and Bardin, Joseph C. and Bausch, Johannes and Bengtsson, Andreas and Bilmes, Alexander and Blackwell, Sam and Boixo, Sergio and Bortoli, Gina and Bourassa, Alexandre and Bovaird, Jenna and Brill, Leon and Broughton, Michael and Browne, David A. and Buchea, Brett and Buckley, Bob B. and Buell, David A. and Burger, Tim and Burkett, Brian and Bushnell, Nicholas and Cabrera, Anthony and Campero, Juan and Chang, Hung-Shen and Chen, Yu and Chen, Zijun and Chiaro, Ben and Chik, Desmond and Chou, Charina and Claes, Jahan and Cleland, Agnetta Y. and Cogan, Josh and Collins, Roberto and Conner, Paul and Courtney, William and Crook, Alexander L. and Curtin, Ben and Das, Sayan and Davies, Alex and De Lorenzo, Laura and Debroy, Dripto M. and Demura, Sean and Devoret, Michel and Di Paolo, Agustin and Donohoe, Paul and Drozdov, Ilya and Dunsworth, Andrew and Earle, Clint and Edlich, Thomas and Eickbusch, Alec and Elbag, Aviv Moshe and Elzouka, Mahmoud and Erickson, Catherine and Faoro, Lara and Farhi, Edward and Ferreira, Vinicius S. and Burgos, Leslie Flores and Forati, Ebrahim and Fowler, Austin G. and Foxen, Brooks and Ganjam, Suhas and Garcia, Gonzalo and Gasca, Robert and Genois, {\'E}lie and Giang, William and Gidney, Craig and Gilboa, Dar and Gosula, Raja and Dau, Alejandro Grajales and Graumann, Dietrich and Greene, Alex and Gross, Jonathan A. and Habegger, Steve and Hall, John and Hamilton, Michael C. and Hansen, Monica and Harrigan, Matthew P. and Harrington, Sean D. and Heras, Francisco J. H. and Heslin, Stephen and Heu, Paula and Higgott, Oscar and Hill, Gordon and Hilton, Jeremy and Holland, George and Hong, Sabrina and Huang, Hsin-Yuan and Huff, Ashley and Huggins, William J. and Ioffe, Lev B. and Isakov, Sergei V. and Iveland, Justin and Jeffrey, Evan and Jiang, Zhang and Jones, Cody and Jordan, Stephen and Joshi, Chaitali and Juhas, Pavol and Kafri, Dvir and Kang, Hui and Karamlou, Amir H. and Kechedzhi, Kostyantyn and Kelly, Julian and Khaire, Trupti and Khattar, Tanuj and Khezri, Mostafa and Kim, Seon and Klimov, Paul V. and Klots, Andrey R. and Kobrin, Bryce and Kohli, Pushmeet and Korotkov, Alexander N. and Kostritsa, Fedor and Kothari, Robin and Kozlovskii, Borislav and Kreikebaum, John Mark and Kurilovich, Vladislav D. and Lacroix, Nathan and Landhuis, David and Lange-Dei, Tiano and Langley, Brandon W. and Laptev, Pavel and Lau, Kim-Ming and Le Guevel, Lo{\"\i}ck and Ledford, Justin and Lee, Joonho and Lee, Kenny and Lensky, Yuri D. and Leon, Shannon and Lester, Brian J. and Li, Wing Yan and Li, Yin and Lill, Alexander T. and Liu, Wayne and Livingston, William P. and Locharla, Aditya and Lucero, Erik and Lundahl, Daniel and Lunt, Aaron and Madhuk, Sid and Malone, Fionn D. and Maloney, Ashley and Mandr{\`a}, Salvatore and Manyika, James and Martin, Leigh S. and Martin, Orion and Martin, Steven and Maxfield, Cameron and McClean, Jarrod R. and McEwen, Matt and Meeks, Seneca and Megrant, Anthony and Mi, Xiao and Miao, Kevin C. and Mieszala, Amanda and Molavi, Reza and Molina, Sebastian and Montazeri, Shirin and Morvan, Alexis and Movassagh, Ramis and Mruczkiewicz, Wojciech and Naaman, Ofer and Neeley, Matthew and Neill, Charles and Nersisyan, Ani and Neven, Hartmut and Newman, Michael and Ng, Jiun How and Nguyen, Anthony and Nguyen, Murray and Ni, Chia-Hung and Niu, Murphy Yuezhen and O'Brien, Thomas E. and Oliver, William D. and Opremcak, Alex and Ottosson, Kristoffer and Petukhov, Andre and Pizzuto, Alex and Platt, John and Potter, Rebecca and Pritchard, Orion and Pryadko, Leonid P. and Quintana, Chris and Ramachandran, Ganesh and Reagor, Matthew J. and Redding, John and Rhodes, David M. and Roberts, Gabrielle and Rosenberg, Eliott and Rosenfeld, Emma and Roushan, Pedram and Rubin, Nicholas C. and Saei, Negar and Sank, Daniel and Sankaragomathi, Kannan and Satzinger, Kevin J. and Schurkus, Henry F. and Schuster, Christopher and Senior, Andrew W. and Shearn, Michael J. and Shorter, Aaron and Shutty, Noah and Shvarts, Vladimir and Singh, Shraddha and Sivak, Volodymyr and Skruzny, Jindra and Small, Spencer and Smelyanskiy, Vadim and Smith, W. Clarke and Somma, Rolando D. and Springer, Sofia and Sterling, George and Strain, Doug and Suchard, Jordan and Szasz, Aaron and Sztein, Alex and Thor, Douglas and Torres, Alfredo and Torunbalci, M. Mert and Vaishnav, Abeer and Vargas, Justin and Vdovichev, Sergey and Vidal, Guifre and Villalonga, Benjamin and Heidweiller, Catherine Vollgraff and Waltman, Steven and Wang, Shannon X. and Ware, Brayden and Weber, Kate and Weidel, Travis and White, Theodore and Wong, Kristi and Woo, Bryan W. K. and Xing, Cheng and Yao, Z. Jamie and Yeh, Ping and Ying, Bicheng and Yoo, Juhwan and Yosri, Noureldin and Young, Grayson and Zalcman, Adam and Zhang, Yaxing and Zhu, Ningfeng and Zobrist, Nicholas and Google Quantum AI and Collaborators},
	date = {2024/12/09},
	doi = {10.1038/s41586-024-08449-y},
	id = {Acharya2024},
	isbn = {1476-4687},
	journal = {Nature},
	title = {Quantum error correction below the surface code threshold},
	url = {https://doi.org/10.1038/s41586-024-08449-y},
	year = {2024}}

@article{Schollwoeck_2011_dmrg,
   title={The density-matrix renormalization group in the age of matrix product states},
   volume={326},
   ISSN={0003-4916},
   url={http://dx.doi.org/10.1016/j.aop.2010.09.012},
   DOI={10.1016/j.aop.2010.09.012},
   number={1},
   journal={Annals of Physics},
   publisher={Elsevier BV},
   author={Schollwöck, Ulrich},
   year={2011},
   month=jan, pages={96–192} }

@article{Mildenberger_2025,
   title={Confinement in a ${{\mathbb{Z}
}}_{2}$ lattice gauge theory on a quantum computer},
   ISSN={1745-2481},
   url={http://dx.doi.org/10.1038/s41567-024-02723-6},
   DOI={10.1038/s41567-024-02723-6},
   journal={Nature Physics},
   publisher={Springer Science and Business Media LLC},
   author={Mildenberger, Julius and Mruczkiewicz, Wojciech and Halimeh, Jad C. and Jiang, Zhang and Hauke, Philipp},
   year={2025},
   month=jan }

@article{Georgescu_2014,
   title={Quantum simulation},
   volume={86},
   ISSN={1539-0756},
   url={http://dx.doi.org/10.1103/RevModPhys.86.153},
   DOI={10.1103/revmodphys.86.153},
   number={1},
   journal={Reviews of Modern Physics},
   publisher={American Physical Society (APS)},
   author={Georgescu, I. M. and Ashhab, S. and Nori, Franco},
   year={2014},
   month=mar, pages={153–185} }

@misc{hoefler2023,
      title={Disentangling Hype from Practicality: On Realistically Achieving Quantum Advantage}, 
      author={Torsten Hoefler and Thomas Haener and Matthias Troyer},
      year={2023},
      eprint={2307.00523},
      archivePrefix={arXiv},
      primaryClass={quant-ph},
      url={https://arxiv.org/abs/2307.00523}, 
}

@article{Fauseweh_2024,
	author = {Fauseweh, Benedikt},
	date = {2024/03/08},
	doi = {10.1038/s41467-024-46402-9},
	id = {Fauseweh2024},
	isbn = {2041-1723},
	journal = {Nature Communications},
	number = {1},
	pages = {2123},
	title = {Quantum many-body simulations on digital quantum computers: State-of-the-art and future challenges},
	url = {https://doi.org/10.1038/s41467-024-46402-9},
	volume = {15},
	year = {2024}}

@article{smith_2019,
	author = {Smith, Adam and Kim, M. S. and Pollmann, Frank and Knolle, Johannes},
	date = {2019/11/28},
	doi = {10.1038/s41534-019-0217-0},
	id = {Smith2019},
	isbn = {2056-6387},
	journal = {npj Quantum Information},
	number = {1},
	pages = {106},
	title = {Simulating quantum many-body dynamics on a current digital quantum computer},
	url = {https://doi.org/10.1038/s41534-019-0217-0},
	volume = {5},
	year = {2019}}

@article{Wintersperger_2023,
   title={Neutral atom quantum computing hardware: performance and end-user perspective},
   volume={10},
   ISSN={2196-0763},
   url={http://dx.doi.org/10.1140/epjqt/s40507-023-00190-1},
   DOI={10.1140/epjqt/s40507-023-00190-1},
   number={1},
   journal={EPJ Quantum Technology},
   publisher={Springer Science and Business Media LLC},
   author={Wintersperger, Karen and Dommert, Florian and Ehmer, Thomas and Hoursanov, Andrey and Klepsch, Johannes and Mauerer, Wolfgang and Reuber, Georg and Strohm, Thomas and Yin, Ming and Luber, Sebastian},
   year={2023},
   month=aug }

@misc{govia_2024,
      title={Bounding the systematic error in quantum error mitigation due to model violation}, 
      author={L. C. G. Govia and S. Majumder and S. V. Barron and B. Mitchell and A. Seif and Y. Kim and C. J. Wood and E. J. Pritchett and S. T. Merkel and D. C. McKay},
      year={2024},
      eprint={2408.10985},
      archivePrefix={arXiv},
      primaryClass={quant-ph},
      url={https://arxiv.org/abs/2408.10985}, 
}

@misc{paetznick2024_trapped_ion_qec,
      title={Demonstration of logical qubits and repeated error correction with better-than-physical error rates}, 
      author={A. Paetznick and M. P. da Silva and C. Ryan-Anderson and J. M. Bello-Rivas and J. P. Campora III and A. Chernoguzov and J. M. Dreiling and C. Foltz and F. Frachon and J. P. Gaebler and T. M. Gatterman and L. Grans-Samuelsson and D. Gresh and D. Hayes and N. Hewitt and C. Holliman and C. V. Horst and J. Johansen and D. Lucchetti and Y. Matsuoka and M. Mills and S. A. Moses and B. Neyenhuis and A. Paz and J. Pino and P. Siegfried and A. Sundaram and D. Tom and S. J. Wernli and M. Zanner and R. P. Stutz and K. M. Svore},
      year={2024},
      eprint={2404.02280},
      archivePrefix={arXiv},
      primaryClass={quant-ph},
      url={https://arxiv.org/abs/2404.02280}, 
}

@article{Evered_2023_high_fid_neutral_atom,
	author = {Evered, Simon J. and Bluvstein, Dolev and Kalinowski, Marcin and Ebadi, Sepehr and Manovitz, Tom and Zhou, Hengyun and Li, Sophie H. and Geim, Alexandra A. and Wang, Tout T. and Maskara, Nishad and Levine, Harry and Semeghini, Giulia and Greiner, Markus and Vuleti{\'c}, Vladan and Lukin, Mikhail D.},
	journal = {Nature},
	number = {7982},
	pages = {268--272},
	title = {High-fidelity parallel entangling gates on a neutral-atom quantum computer},
	volume = {622},
	year = {2023}}

@article{ding_2023_highfid_fluxonium_TQG,
  title = {High-Fidelity, Frequency-Flexible Two-Qubit Fluxonium Gates with a Transmon Coupler},
  author = {Ding, Leon and Hays, Max and Sung, Youngkyu and Kannan, Bharath and An, Junyoung and Di Paolo, Agustin and Karamlou, Amir H. and Hazard, Thomas M. and Azar, Kate and Kim, David K. and Niedzielski, Bethany M. and Melville, Alexander and Schwartz, Mollie E. and Yoder, Jonilyn L. and Orlando, Terry P. and Gustavsson, Simon and Grover, Jeffrey A. and Serniak, Kyle and Oliver, William D.},
  journal = {Phys. Rev. X},
  volume = {13},
  issue = {3},
  pages = {031035},
  numpages = {24},
  year = {2023},
  month = {Sep},
  publisher = {American Physical Society},
  doi = {10.1103/PhysRevX.13.031035},
  url = {https://link.aps.org/doi/10.1103/PhysRevX.13.031035}
}

@article{Li_2024_doubletransmon_coupler,
  title = {Realization of High-Fidelity CZ Gate Based on a Double-Transmon Coupler},
  author = {Li, Rui and Kubo, Kentaro and Ho, Yinghao and Yan, Zhiguang and Nakamura, Yasunobu and Goto, Hayato},
  journal = {Phys. Rev. X},
  volume = {14},
  issue = {4},
  pages = {041050},
  numpages = {30},
  year = {2024},
  month = {Nov},
  publisher = {American Physical Society},
  doi = {10.1103/PhysRevX.14.041050},
  url = {https://link.aps.org/doi/10.1103/PhysRevX.14.041050}
}

@article{Clinton_2021,
	author = {Clinton, Laura and Bausch, Johannes and Cubitt, Toby},
	date = {2021/08/17},
	doi = {10.1038/s41467-021-25196-0},
	id = {Clinton2021},
	isbn = {2041-1723},
	journal = {Nature Communications},
	number = {1},
	pages = {4989},
	title = {Hamiltonian simulation algorithms for near-term quantum hardware},
	url = {https://doi.org/10.1038/s41467-021-25196-0},
	volume = {12},
	year = {2021}}

@article{Cade_2020,
  title = {Strategies for solving the Fermi-Hubbard model on near-term quantum computers},
  author = {Cade, Chris and Mineh, Lana and Montanaro, Ashley and Stanisic, Stasja},
  journal = {Phys. Rev. B},
  volume = {102},
  issue = {23},
  pages = {235122},
  numpages = {25},
  year = {2020},
  month = {Dec},
  publisher = {American Physical Society},
  doi = {10.1103/PhysRevB.102.235122},
  url = {https://link.aps.org/doi/10.1103/PhysRevB.102.235122}
}

@article{Paeckel_2019,
   title={Time-evolution methods for matrix-product states},
   volume={411},
   ISSN={0003-4916},
   url={http://dx.doi.org/10.1016/j.aop.2019.167998},
   DOI={10.1016/j.aop.2019.167998},
   journal={Annals of Physics},
   publisher={Elsevier BV},
   author={Paeckel, Sebastian and Köhler, Thomas and Swoboda, Andreas and Manmana, Salvatore R. and Schollwöck, Ulrich and Hubig, Claudius},
   year={2019},
   month=dec, pages={167998} }

@misc{jafarizadeh2024,
      title={A recipe for local simulation of strongly-correlated fermionic matter on quantum computers: the 2D Fermi-Hubbard model}, 
      author={Arash Jafarizadeh and Frank Pollmann and Adam Gammon-Smith},
      year={2024},
      eprint={2408.14543},
      archivePrefix={arXiv},
      primaryClass={quant-ph},
      url={https://arxiv.org/abs/2408.14543}, 
}

@article{O_Brien_2023,
   title={Purification-based quantum error mitigation of pair-correlated electron simulations},
   volume={19},
   ISSN={1745-2481},
   url={http://dx.doi.org/10.1038/s41567-023-02240-y},
   DOI={10.1038/s41567-023-02240-y},
   number={12},
   journal={Nature Physics},
   publisher={Springer Science and Business Media LLC},
   author={O’Brien, T. E. and Anselmetti, G. and Gkritsis, F. and Elfving, V. E. and Polla, S. and Huggins, W. J. and Oumarou, O. and Kechedzhi, K. and Abanin, D. and Acharya, R. and Aleiner, I. and Allen, R. and Andersen, T. I. and Anderson, K. and Ansmann, M. and Arute, F. and Arya, K. and Asfaw, A. and Atalaya, J. and Bardin, J. C. and Bengtsson, A. and Bortoli, G. and Bourassa, A. and Bovaird, J. and Brill, L. and Broughton, M. and Buckley, B. and Buell, D. A. and Burger, T. and Burkett, B. and Bushnell, N. and Campero, J. and Chen, Z. and Chiaro, B. and Chik, D. and Cogan, J. and Collins, R. and Conner, P. and Courtney, W. and Crook, A. L. and Curtin, B. and Debroy, D. M. and Demura, S. and Drozdov, I. and Dunsworth, A. and Erickson, C. and Faoro, L. and Farhi, E. and Fatemi, R. and Ferreira, V. S. and Flores Burgos, L. and Forati, E. and Fowler, A. G. and Foxen, B. and Giang, W. and Gidney, C. and Gilboa, D. and Giustina, M. and Gosula, R. and Grajales Dau, A. and Gross, J. A. and Habegger, S. and Hamilton, M. C. and Hansen, M. and Harrigan, M. P. and Harrington, S. D. and Heu, P. and Hoffmann, M. R. and Hong, S. and Huang, T. and Huff, A. and Ioffe, L. B. and Isakov, S. V. and Iveland, J. and Jeffrey, E. and Jiang, Z. and Jones, C. and Juhas, P. and Kafri, D. and Khattar, T. and Khezri, M. and Kieferová, M. and Kim, S. and Klimov, P. V. and Klots, A. R. and Korotkov, A. N. and Kostritsa, F. and Kreikebaum, J. M. and Landhuis, D. and Laptev, P. and Lau, K.-M. and Laws, L. and Lee, J. and Lee, K. and Lester, B. J. and Lill, A. T. and Liu, W. and Livingston, W. P. and Locharla, A. and Malone, F. D. and Mandrà, S. and Martin, O. and Martin, S. and McClean, J. R. and McCourt, T. and McEwen, M. and Mi, X. and Mieszala, A. and Miao, K. C. and Mohseni, M. and Montazeri, S. and Morvan, A. and Movassagh, R. and Mruczkiewicz, W. and Naaman, O. and Neeley, M. and Neill, C. and Nersisyan, A. and Newman, M. and Ng, J. H. and Nguyen, A. and Nguyen, M. and Niu, M. Y. and Omonije, S. and Opremcak, A. and Petukhov, A. and Potter, R. and Pryadko, L. P. and Quintana, C. and Rocque, C. and Roushan, P. and Saei, N. and Sank, D. and Sankaragomathi, K. and Satzinger, K. J. and Schurkus, H. F. and Schuster, C. and Shearn, M. J. and Shorter, A. and Shutty, N. and Shvarts, V. and Skruzny, J. and Smith, W. C. and Somma, R. D. and Sterling, G. and Strain, D. and Szalay, M. and Thor, D. and Torres, A. and Vidal, G. and Villalonga, B. and Vollgraff Heidweiller, C. and White, T. and Woo, B. W. K. and Xing, C. and Yao, Z. J. and Yeh, P. and Yoo, J. and Young, G. and Zalcman, A. and Zhang, Y. and Zhu, N. and Zobrist, N. and Bacon, D. and Boixo, S. and Chen, Y. and Hilton, J. and Kelly, J. and Lucero, E. and Megrant, A. and Neven, H. and Smelyanskiy, V. and Gogolin, C. and Babbush, R. and Rubin, N. C.},
   year={2023},
   month=oct, pages={1787–1792} }

@article{Sawaya_2020,
   title={Resource-efficient digital quantum simulation of d-level systems for photonic, vibrational, and spin-s Hamiltonians},
   volume={6},
   ISSN={2056-6387},
   url={http://dx.doi.org/10.1038/s41534-020-0278-0},
   DOI={10.1038/s41534-020-0278-0},
   number={1},
   journal={npj Quantum Information},
   publisher={Springer Science and Business Media LLC},
   author={Sawaya, Nicolas P. D. and Menke, Tim and Kyaw, Thi Ha and Johri, Sonika and Aspuru-Guzik, Alán and Guerreschi, Gian Giacomo},
   year={2020},
   month=jun }

@misc{koyluoglu_2024,
      title={Measuring central charge on a universal quantum processor}, 
      author={Nazlı Uğur Köylüoğlu and Swarndeep Majumder and Mirko Amico and Sarah Mostame and Ewout van den Berg and M. A. Rajabpour and Zlatko Minev and Khadijeh Najafi},
      year={2024},
      eprint={2408.06342},
      archivePrefix={arXiv},
      primaryClass={quant-ph},
      url={https://arxiv.org/abs/2408.06342}, 
}

@misc{nigmatullin2024,
      title={Experimental Demonstration of Break-Even for the Compact Fermionic Encoding}, 
      author={Ramil Nigmatullin and Kevin Hemery and Khaldoon Ghanem and Steven Moses and Dan Gresh and Peter Siegfried and Michael Mills and Thomas Gatterman and Nathan Hewitt and Etienne Granet and Henrik Dreyer},
      year={2024},
      eprint={2409.06789},
      archivePrefix={arXiv},
      primaryClass={quant-ph},
      url={https://arxiv.org/abs/2409.06789}, 
}

@misc{evered2025,
      title={Probing topological matter and fermion dynamics on a neutral-atom quantum computer}, 
      author={Simon J. Evered and Marcin Kalinowski and Alexandra A. Geim and Tom Manovitz and Dolev Bluvstein and Sophie H. Li and Nishad Maskara and Hengyun Zhou and Sepehr Ebadi and Muqing Xu and Joseph Campo and Madelyn Cain and Stefan Ostermann and Susanne F. Yelin and Subir Sachdev and Markus Greiner and Vladan Vuletić and Mikhail D. Lukin},
      year={2025},
      eprint={2501.18554},
      archivePrefix={arXiv},
      primaryClass={quant-ph},
      url={https://arxiv.org/abs/2501.18554}, 
}

@misc{will2025,
      title={Probing Non-Equilibrium Topological Order on a Quantum Processor}, 
      author={M. Will and T. A. Cochran and E. Rosenberg and B. Jobst and N. M Eassa and P. Roushan and M. Knap and A. Gammon-Smith and F. Pollmann},
      year={2025},
      eprint={2501.18461},
      archivePrefix={arXiv},
      primaryClass={quant-ph},
      url={https://arxiv.org/abs/2501.18461}, 
}

@article{Endo_2018,
   title={Practical Quantum Error Mitigation for Near-Future Applications},
   volume={8},
   ISSN={2160-3308},
   url={http://dx.doi.org/10.1103/PhysRevX.8.031027},
   DOI={10.1103/physrevx.8.031027},
   number={3},
   journal={Physical Review X},
   publisher={American Physical Society (APS)},
   author={Endo, Suguru and Benjamin, Simon C. and Li, Ying},
   year={2018},
   month=jul }

@misc{gottesman_1997_phd_thesis,
      title={Stabilizer Codes and Quantum Error Correction}, 
      author={Daniel Gottesman},
      year={1997},
      eprint={quant-ph/9705052},
      archivePrefix={arXiv},
      primaryClass={quant-ph},
      url={https://arxiv.org/abs/quant-ph/9705052}, 
}

@article{Higgott_2021,
   title={Optimal local unitary encoding circuits for the surface code},
   volume={5},
   ISSN={2521-327X},
   url={http://dx.doi.org/10.22331/q-2021-08-05-517},
   DOI={10.22331/q-2021-08-05-517},
   journal={Quantum},
   publisher={Verein zur Forderung des Open Access Publizierens in den Quantenwissenschaften},
   author={Higgott, Oscar and Wilson, Matthew and Hefford, James and Dborin, James and Hanif, Farhan and Burton, Simon and Browne, Dan E.},
   year={2021},
   month=aug, pages={517} }

@misc{hosseinkhani_2025_NRE,
      title={Noise-Robust Estimation of Quantum Observables in Noisy Hardware}, 
      author={Amin Hosseinkhani and Fedor Šimkovic and Alessio Calzona and Tianhan Liu and Adrian Auer and Inés de Vega},
      year={2025},
      eprint={2503.06695},
      archivePrefix={arXiv},
      primaryClass={quant-ph},
      url={https://arxiv.org/abs/2503.06695}, 
}

@misc{LE,
      title={Fermion-to-qubit encodings with arbitrary code distance}, 
      author={Manuel G. Algaba and Miha Papič and Inés de Vega and Alessio Calzona and Fedor Šimkovic IV},
      year={2025},
      eprint={2505.02916},
      archivePrefix={arXiv},
      primaryClass={quant-ph},
      url={https://arxiv.org/abs/2505.02916}
    }

@article{Macridin_2018,
   title={Digital quantum computation of fermion-boson interacting systems},
   volume={98},
   ISSN={2469-9934},
   url={http://dx.doi.org/10.1103/PhysRevA.98.042312},
   DOI={10.1103/physreva.98.042312},
   number={4},
   journal={Physical Review A},
   publisher={American Physical Society (APS)},
   author={Macridin, Alexandru and Spentzouris, Panagiotis and Amundson, James and Harnik, Roni},
   year={2018},
   month=oct }

@article{Nielsen_2002,
   title={A simple formula for the average gate fidelity of a quantum dynamical operation},
   volume={303},
   ISSN={0375-9601},
   url={http://dx.doi.org/10.1016/S0375-9601(02)01272-0},
   DOI={10.1016/s0375-9601(02)01272-0},
   number={4},
   journal={Physics Letters A},
   publisher={Elsevier BV},
   author={Nielsen, Michael A},
   year={2002},
   month=oct, pages={249–252} }

@article{Knill_2008,
   title={Randomized benchmarking of quantum gates},
   volume={77},
   ISSN={1094-1622},
   url={http://dx.doi.org/10.1103/PhysRevA.77.012307},
   DOI={10.1103/physreva.77.012307},
   number={1},
   journal={Physical Review A},
   publisher={American Physical Society (APS)},
   author={Knill, E. and Leibfried, D. and Reichle, R. and Britton, J. and Blakestad, R. B. and Jost, J. D. and Langer, C. and Ozeri, R. and Seidelin, S. and Wineland, D. J.},
   year={2008},
   month=jan }

@misc{thompson2025_fermioniq,
      title={Non-zero noise extrapolation: accurately simulating noisy quantum circuits with tensor networks}, 
      author={Anthony P. Thompson and Arie Soeteman and Chris Cade and Ido Niesen},
      year={2025},
      eprint={2501.13237},
      archivePrefix={arXiv},
      primaryClass={quant-ph},
      url={https://arxiv.org/abs/2501.13237}, 
}

@article{Tsubouchi_2023,
   title={Universal Cost Bound of Quantum Error Mitigation Based on Quantum Estimation Theory},
   volume={131},
   ISSN={1079-7114},
   url={http://dx.doi.org/10.1103/PhysRevLett.131.210601},
   DOI={10.1103/physrevlett.131.210601},
   number={21},
   journal={Physical Review Letters},
   publisher={American Physical Society (APS)},
   author={Tsubouchi, Kento and Sagawa, Takahiro and Yoshioka, Nobuyuki},
   year={2023},
   month=nov }

@misc{vilchezestevez2025,
      title={Extracting the spin excitation spectrum of a fermionic system using a quantum processor}, 
      author={Lucia Vilchez-Estevez and Raul A. Santos and Sabrina Wang and Filippo Maria Gambetta},
      year={2025},
      eprint={2501.04649},
      archivePrefix={arXiv},
      primaryClass={quant-ph},
      url={https://arxiv.org/abs/2501.04649}, 
}

@article{Kitaev_2006,
   title={Anyons in an exactly solved model and beyond},
   volume={321},
   ISSN={0003-4916},
   url={http://dx.doi.org/10.1016/j.aop.2005.10.005},
   DOI={10.1016/j.aop.2005.10.005},
   number={1},
   journal={Annals of Physics},
   publisher={Elsevier BV},
   author={Kitaev, Alexei},
   year={2006},
   month=jan, pages={2–111} }

@misc{chen_2025,
      title={Disambiguating Pauli noise in quantum computers}, 
      author={Edward H. Chen and Senrui Chen and Laurin E. Fischer and Andrew Eddins and Luke C. G. Govia and Brad Mitchell and Andre He and Youngseok Kim and Liang Jiang and Alireza Seif},
      year={2025},
      eprint={2505.22629},
      archivePrefix={arXiv},
      primaryClass={quant-ph},
      url={https://arxiv.org/abs/2505.22629}, 
}

@misc{chen_2024,
      title={Efficient self-consistent learning of gate set Pauli noise}, 
      author={Senrui Chen and Zhihan Zhang and Liang Jiang and Steven T. Flammia},
      year={2025},
      eprint={2410.03906},
      archivePrefix={arXiv},
      primaryClass={quant-ph},
      url={https://arxiv.org/abs/2410.03906}, 
}

@misc{carignandugas2023,
      title={The Error Reconstruction and Compiled Calibration of Quantum Computing Cycles}, 
      author={Arnaud Carignan-Dugas and Dar Dahlen and Ian Hincks and Egor Ospadov and Stefanie J. Beale and Samuele Ferracin and Joshua Skanes-Norman and Joseph Emerson and Joel J. Wallman},
      year={2023},
      eprint={2303.17714},
      archivePrefix={arXiv},
      primaryClass={quant-ph},
      url={https://arxiv.org/abs/2303.17714}, 
}

@misc{granet2025,
      title={Superconducting pairing correlations on a trapped-ion quantum computer}, 
      author={Etienne Granet and Sheng-Hsuan Lin and Kevin Hémery and Reza Hagshenas and Pablo Andres-Martinez and David T. Stephen and Anthony Ransford and Jake Arkinstall and M. S. Allman and Pete Campora and Samuel F. Cooper and Robert D. Delaney and Joan M. Dreiling and Brian Estey and Caroline Figgatt and Cameron Foltz and John P. Gaebler and Alex Hall and Ali Husain and Akhil Isanaka and Colin J. Kennedy and Nikhil Kotibhaskar and Michael Mills and Alistair R. Milne and Annie J. Park and Adam P. Reed and Brian Neyenhuis and Justin G. Bohnet and Michael Foss-Feig and Andrew C. Potter and Ramil Nigmatullin and Mohsin Iqbal and Henrik Dreyer},
      year={2025},
      eprint={2511.02125},
      archivePrefix={arXiv},
      primaryClass={quant-ph},
      url={https://arxiv.org/abs/2511.02125}, 
}

@misc{alam2025,
      title={Fermionic dynamics on a trapped-ion quantum computer beyond exact classical simulation}, 
      author={Faisal Alam and Jan Lukas Bosse and Ieva Čepaitė and Adrian Chapman and Laura Clinton and Marcos Crichigno and Elizabeth Crosson and Toby Cubitt and Charles Derby and Oliver Dowinton and Paul K. Faehrmann and Steve Flammia and Brian Flynn and Filippo Maria Gambetta and Raúl García-Patrón and Max Hunter-Gordon and Glenn Jones and Abhishek Khedkar and Joel Klassen and Michael Kreshchuk and Edward Harry McMullan and Lana Mineh and Ashley Montanaro and Caterina Mora and John J. L. Morton and Dhrumil Patel and Pete Rolph and Raul A. Santos and James R. Seddon and Evan Sheridan and Wilfrid Somogyi and Marika Svensson and Niam Vaishnav and Sabrina Yue Wang and Gethin Wright},
      year={2025},
      eprint={2510.26300},
      archivePrefix={arXiv},
      primaryClass={quant-ph},
      url={https://arxiv.org/abs/2510.26300}, 
}

@article{Tindall_2024,
   title={Efficient Tensor Network Simulation of IBM’s Eagle Kicked Ising Experiment},
   volume={5},
   ISSN={2691-3399},
   url={http://dx.doi.org/10.1103/PRXQuantum.5.010308},
   DOI={10.1103/prxquantum.5.010308},
   number={1},
   journal={PRX Quantum},
   publisher={American Physical Society (APS)},
   author={Tindall, Joseph and Fishman, Matthew and Stoudenmire, E. Miles and Sels, Dries},
   year={2024},
   month=jan }

@misc{liao2023,
      title={Simulation of IBM's kicked Ising experiment with Projected Entangled Pair Operator}, 
      author={Hai-Jun Liao and Kang Wang and Zong-Sheng Zhou and Pan Zhang and Tao Xiang},
      year={2023},
      eprint={2308.03082},
      archivePrefix={arXiv},
      primaryClass={quant-ph},
      url={https://arxiv.org/abs/2308.03082}, 
}

@article{Yoshioka_2024,
   title={Hunting for quantum-classical crossover in condensed matter problems},
   volume={10},
   ISSN={2056-6387},
   url={http://dx.doi.org/10.1038/s41534-024-00839-4},
   DOI={10.1038/s41534-024-00839-4},
   number={1},
   journal={npj Quantum Information},
   publisher={Springer Science and Business Media LLC},
   author={Yoshioka, Nobuyuki and Okubo, Tsuyoshi and Suzuki, Yasunari and Koizumi, Yuki and Mizukami, Wataru},
   year={2024},
   month=apr }

@misc{rudolph2023,
      title={Classical surrogate simulation of quantum systems with LOWESA}, 
      author={Manuel S. Rudolph and Enrico Fontana and Zoë Holmes and Lukasz Cincio},
      year={2023},
      eprint={2308.09109},
      archivePrefix={arXiv},
      primaryClass={quant-ph},
      url={https://arxiv.org/abs/2308.09109}, 
}

@misc{rudolph2025,
      title={Pauli Propagation: A Computational Framework for Simulating Quantum Systems}, 
      author={Manuel S. Rudolph and Tyson Jones and Yanting Teng and Armando Angrisani and Zoë Holmes},
      year={2025},
      eprint={2505.21606},
      archivePrefix={arXiv},
      primaryClass={quant-ph},
      url={https://arxiv.org/abs/2505.21606}, 
}

@misc{miller2025,
      title={Simulation of Fermionic circuits using Majorana Propagation}, 
      author={Aaron Miller and Zoë Holmes and Özlem Salehi and Rahul Chakraborty and Anton Nykänen and Zoltán Zimborás and Adam Glos and Guillermo García-Pérez},
      year={2025},
      eprint={2503.18939},
      archivePrefix={arXiv},
      primaryClass={quant-ph},
      url={https://arxiv.org/abs/2503.18939}, 
}

@misc{danna2025,
      title={Majorana string simulation of nonequilibrium dynamics in two-dimensional lattice fermion systems}, 
      author={Matteo D'Anna and Jannes Nys and Juan Carrasquilla},
      year={2025},
      eprint={2511.02809},
      archivePrefix={arXiv},
      primaryClass={cond-mat.quant-gas},
      url={https://arxiv.org/abs/2511.02809}, 
}

@article{Lin_2022,
   title={Scaling of Neural‐Network Quantum States for Time Evolution},
   volume={259},
   ISSN={1521-3951},
   url={http://dx.doi.org/10.1002/pssb.202100172},
   DOI={10.1002/pssb.202100172},
   number={5},
   journal={physica status solidi (b)},
   publisher={Wiley},
   author={Lin, Sheng-Hsuan and Pollmann, Frank},
   year={2022},
   month=jan }

@article{Lange_2024,
doi = {10.1088/2058-9565/ad7168},
url = {https://doi.org/10.1088/2058-9565/ad7168},
year = {2024},
month = {sep},
publisher = {IOP Publishing},
volume = {9},
number = {4},
pages = {040501},
author = {Lange, Hannah and Van de Walle, Anka and Abedinnia, Atiye and Bohrdt, Annabelle},
title = {From architectures to applications: a review of neural quantum states},
journal = {Quantum Science and Technology}
}

@article{Ibarra_Garcia_Padilla_2025,
   title={Autoregressive neural quantum states of Fermi Hubbard models},
   volume={7},
   ISSN={2643-1564},
   url={http://dx.doi.org/10.1103/PhysRevResearch.7.013122},
   DOI={10.1103/physrevresearch.7.013122},
   number={1},
   journal={Physical Review Research},
   publisher={American Physical Society (APS)},
   author={Ibarra-García-Padilla, Eduardo and Lange, Hannah and Melko, Roger G. and Scalettar, Richard T. and Carrasquilla, Juan and Bohrdt, Annabelle and Khatami, Ehsan},
   year={2025},
   month=feb }

@article{Siro_2012,
   title={Exact diagonalization of the Hubbard model on graphics processing units},
   volume={183},
   ISSN={0010-4655},
   url={http://dx.doi.org/10.1016/j.cpc.2012.04.006},
   DOI={10.1016/j.cpc.2012.04.006},
   number={9},
   journal={Computer Physics Communications},
   publisher={Elsevier BV},
   author={Siro, T. and Harju, A.},
   year={2012},
   month=sep, pages={1884–1889} }

@article{He_2025,
   title={Performance of quantum approximate optimization with quantum error detection},
   volume={8},
   ISSN={2399-3650},
   url={http://dx.doi.org/10.1038/s42005-025-02136-8},
   DOI={10.1038/s42005-025-02136-8},
   number={1},
   journal={Communications Physics},
   publisher={Springer Science and Business Media LLC},
   author={He, Zichang and Amaro, David and Shaydulin, Ruslan and Pistoia, Marco},
   year={2025},
   month=may }

@article{santos_2024,
	author = {P. Santos, Jader and Bar, Ben and Uzdin, Raam},
	journal = {npj Quantum Information},
	number = {1},
	pages = {100},
	title = {Pseudo twirling mitigation of coherent errors in non-Clifford gates},
	volume = {10},
	year = {2024}}

@misc{liu2024,
      title={Non-Markovian Noise Suppression Simplified through Channel Representation}, 
      author={Zhenhuan Liu and Yunlong Xiao and Zhenyu Cai},
      year={2024},
      eprint={2412.11220},
      archivePrefix={arXiv},
      primaryClass={quant-ph},
      url={https://arxiv.org/abs/2412.11220}, 
}

@article{Nielsen_2021,
   title={Gate Set Tomography},
   volume={5},
   ISSN={2521-327X},
   url={http://dx.doi.org/10.22331/q-2021-10-05-557},
   DOI={10.22331/q-2021-10-05-557},
   journal={Quantum},
   publisher={Verein zur Forderung des Open Access Publizierens in den Quantenwissenschaften},
   author={Nielsen, Erik and Gamble, John King and Rudinger, Kenneth and Scholten, Travis and Young, Kevin and Blume-Kohout, Robin},
   year={2021},
   month=oct, pages={557} }

@article{Self_2024,
   title={Protecting expressive circuits with a quantum error detection code},
   volume={20},
   ISSN={1745-2481},
   url={http://dx.doi.org/10.1038/s41567-023-02282-2},
   DOI={10.1038/s41567-023-02282-2},
   number={2},
   journal={Nature Physics},
   publisher={Springer Science and Business Media LLC},
   author={Self, Chris N. and Benedetti, Marcello and Amaro, David},
   year={2024},
   month=jan, pages={219–224} }

\end{document}